\documentclass[twocolumn,pre,superscriptaddress,notitlepage]{revtex4-2}

\usepackage{amsmath, amssymb}
\usepackage{graphicx}
\usepackage{textcomp}
\usepackage{float}
\usepackage{tabularx}
\usepackage{nicefrac}
\usepackage{color}
\usepackage{booktabs} 
\usepackage{chngcntr}

\usepackage{enumitem}

\usepackage{fancyvrb}

\usepackage{makecell}
\usepackage[table]{xcolor} 
\usepackage[most]{tcolorbox}
\tcbset{colback=gray!10, 
colframe=gray!50, 
boxrule=0.4pt, 
arc=1mm, 
boxsep=1pt,      
  left=2pt,        
  right=2pt,       
  top=2pt,         
  bottom=2pt       
  }
  
\usepackage{times}  
\usepackage{helvet}  
\usepackage{courier}  

\usepackage{natbib}  
\usepackage[justification=justified,singlelinecheck=false]{caption}
\makeatletter
\long\def\@makecaption#1#2{%
  \vskip\abovecaptionskip
  \begingroup
    \small
    \leftskip=\z@skip
    \rightskip=\z@skip
    \parfillskip=\z@\@plus 1fil\relax
    \parindent=\z@
    \noindent\textbf{#1:}~#2\par
  \endgroup
  \vskip\belowcaptionskip}
\makeatother

\definecolor{quotegray}{HTML}{008000}
\usepackage{xcolor}
\newcommand{\homa}[1]{\textcolor{black}{#1}}

\newcommand{\yt}{\ensuremath{\text{YT}}}
\newcommand{\emb}{\ensuremath{\text{emb}}}
\newcommand{\bllm}{\ensuremath{\text{bLLM}}}
\newcommand{\bllmp}{\ensuremath{\text{bLLM-partisan}}}
\newcommand{\rllm}{\ensuremath{\text{rLLM}}}

\newcommand{\delab}{\ensuremath{\Delta(\text{A},\, \text{B})}}
\newcommand{\deley}{\ensuremath{\Delta(\text{emb},\, \text{YT})}}
\newcommand{\delry}{\ensuremath{\Delta(\text{rLLM},\, \text{YT})}}
\newcommand{\delby}{\ensuremath{\Delta(\text{bLLM},\, \text{YT})}}
\newcommand{\delrb}{\ensuremath{\Delta(\text{rLLM},\, \text{bLLM})}}

\usepackage{algorithm}
\usepackage{algorithmic}

\newcommand{\answerYes}[1]{\textcolor{blue}{#1}} 
 
\newcommand{\answerNA}[1]{\textcolor{gray}{#1}} 
 
\usepackage{newfloat}
\usepackage{listings}
\DeclareCaptionStyle{ruled}{labelfont=normalfont,labelsep=colon,strut=off} 
\floatstyle{ruled}
\newfloat{listing}{tb}{lst}{}
\floatname{listing}{Listing}
\usepackage{bibentry}

\makeatletter\def\switch@array{}\makeatother

\begin{document}
\title{LLM-Assisted Reranking to Operationalize Nuanced Objectives \\ in Recommender Systems}

\author{Amir Ghasemian}
\email[]{amirgh@ucla.edu} 
\affiliation{Department of Communication, University of California, Los Angeles, CA 90095}
\affiliation{Department of Computer and Information Science, University of Pennsylvania, Philadelphia, PA 19104}

\author{Homa Hosseinmardi}
\affiliation{Department of Communication, University of California, Los Angeles, CA 90095}

\author{Upasana Dutta}
\affiliation{Department of Computer and Information Science, University of Pennsylvania, Philadelphia, PA 19104}

\author{Duncan J. Watts}
\affiliation{Department of Computer and Information Science, University of Pennsylvania, Philadelphia, PA 19104}
\affiliation{Annenberg School of Communication, University of Pennsylvania, Philadelphia, PA 19104}
\affiliation{Operations, Information, and Decisions Department, University of Pennsylvania, Philadelphia, PA 19104}

\begin{abstract} 
Recommender systems have grown from basic content-organization tools into sophisticated systems that subtly shape our daily behavior. By controlling what information we see, they can influence the world we perceive, raising concerns about filter bubbles, radicalization, polarization, and social inequality. The rise of large language models (LLMs) enables more powerful personalization, potentially intensifying these exposure dynamics. Yet most recommenders are tuned
for engagement or limited accuracy metrics, with little attention to their broader social implications. We investigate whether LLM-assisted reranking, while improving personalization, inadvertently amplifies exposure to ideologically \homa{congenial,} extreme, or conspiratorial political content---a risk theorized but not empirically characterized in \homa{political and news} recommendation. 
\homa{Using real YouTube trajectories, we rerank the platform's sidebar candidates by zero-shot prompting across three models, alongside an embedding baseline, and evaluate semantic and topical relevance, ideological alignment and diversity, and exposure to conspiratorial or extremist content. We find that (i)~under a pure personalization instruction, reranking improves semantic and
topical relevance while narrowing ideological exposure and, on two of three models, promoting problematic content for users whose histories contain such material---the exception arises under identical prompt text on a model that declines most explicitly partisan requests, pointing to model-specific guardrails; (ii)~lightweight prompt-level constraints substantially reduce
problematic exposure on all three models, at a cost paid only against the unconstrained reranker, with the regularized rerankers still ranking above the platform baseline on relevance; and (iii)~synthetic experiments varying topical and partisan cues independently show that rerankers have no fixed preference
between them: topic dominates on abortion and immigration, while partisan alignment takes over on elections, where the topical signal is weaker.} Together, our results highlight the power of LLMs to operationalize contextual nuance in high-stakes \homa{political} recommendation. They also underscore the need to evaluate LLM-assisted personalization beyond accuracy alone and to
recognize prompt design as a value-laden choice rather than a neutral default.
\end{abstract}

\maketitle
 
\section{Introduction}

Recommender systems (RecSys) are deeply intertwined with daily life, influencing decisions across domains from entertainment to employment and healthcare~\cite{liu2010personalized, bobadilla2013recommender,carraro2024enhancing}. 
They also play a central role in content selection on social media, shaping what information reaches users. By enabling powerful personalization, RecSys can improve user experience but also lead to unintended consequences such as filter bubbles~\cite{gao2023cirs}, radicalization~\cite{o2015down}, and diminished viewpoint diversity~\cite{nguyen2014exploring}. At the societal level, these effects may contribute to polarization, echo chambers, and social inequality~\cite{cinus2022effect}.
\homa{Yet the evidence that recommenders drive these outcomes is mixed. Recent audits of YouTube's  modern system find no extremist rabbit holes and only very mild ideological narrowing, with users 
self-selecting into such content when offered~\cite{brown2022echo}, and counterfactual-bot 
experiments find the recommender steers users with extreme tastes toward \emph{more} moderate 
content~\cite{hosseinmardi2024causally}; its directional skew remains disputed~\cite{ibrahim2023youtube,brown2022echo}.}

\begin{figure}[tb!]
     \begin{center}
           \includegraphics[width=0.81\linewidth]{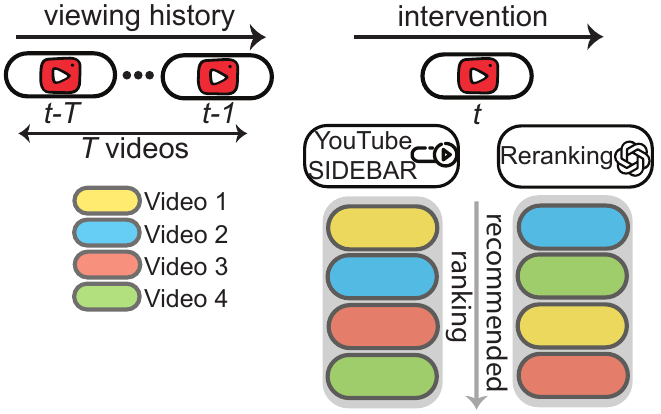}
    \end{center}
    \vspace{-3mm}
    \caption{Framework for evaluating LLM-assisted reranking of YouTube recommendations under social-value constraints. First, YouTube generates personalized recommendations from a user's viewing history. The candidate videos are then reranked using a zero-shot, in-context LLM, guided by one of two prompts: (1) a baseline non-regularized prompt that optimizes for personalization alone, and (2) a regularized socially-guided prompt that aims to preserve personalization while diversifying recommendations and demoting problematic (conspiratorial or extreme) content.} 
   \label{fig:audit_recom_LLMs}
   \vspace{-6mm}
\end{figure}

These concerns have brought renewed attention to RecSys design. A growing literature argues that engagement-driven recommenders neglect the broader range of stakeholders and societal impacts they create~\cite{krauth2025breaking,aridor2024economics,kleinberg2024challenge,ekstrand2025recommending}, and calls for explicit social considerations in ranking logic~\cite{jia2024embedding,bernstein2023embedding}. The dominant operational mechanism for incorporating such objectives is reranking~\cite{pei2019personalized,sonboli2020opportunistic,gao2025llm4rerank}, but existing methods are not designed to capture context-dependent harms in political content, such as framing and ideological positioning~\cite{narayanan2026if,nguyen2014exploring}.

At the same time, advances in large language models (LLMs) introduce capabilities that directly address several of these design limitations, expanding beyond accuracy as the sole utility~\cite{hong2025multi}. Unlike traditional recommender models, which rely on behavioral similarity signals or embedding-space proximity, LLMs can \homa{respond to} the semantic, narrative, and ideological context of content. This enables a form of context-aware decision-making that is difficult to achieve with standard ranking pipelines. These capabilities raise important questions about how LLM RecSys reshapes recommendation outcomes and how they can be used to address vulnerabilities that arise when recommendation systems operate in politically sensitive or high-risk informational settings. Here, we use ``LLM-based RecSys'' to refer to recommendation pipelines in which an LLM is integrated as a post-hoc reranking layer, allowing us to study how LLM-driven contextual reasoning reshapes outcomes without modifying the underlying candidate-generation model.
We address the following research questions:

\begin{itemize}
\item \textbf{RQ1:} How does LLM-assisted personalization reshape exposure to political content?
\item \textbf{RQ2:} \homa{At what cost to personalization can LLM-assisted ranking incorporate socially-aware constraints?} 
\item \textbf{RQ3}: \homa{How do LLM rerankers weight topical against partisan cues, and how does that weighting vary with the candidate set, the watch history, and the topic?} 
\end{itemize}

To answer these questions, we adopt an instruction-following framework in which an LLM receives a user's viewing history together with a candidate set of items and produces a reranked list based on specific constraints (Figure~\ref{fig:audit_recom_LLMs}). Our zero-shot approach, with no fine-tuning~\cite{zhang2025recommendation, geng2022recommendation, yang2023palr}, preserves flexibility, lowers computational cost, and keeps focus on the reranking pipeline rather than optimizing the underlying language model. Our goal is to show how LLMs' contextual 
\homa{sensitivity} can be used to implement value-constrained reranking: maintaining topical relevance while avoiding overfitting to ideological cues, suppressing conspiratorial or extreme content, and broadening exposure when personalization alone would otherwise narrow the informational landscape. Our study makes the following contributions:

\begin{itemize}
    \item We introduce a zero-shot, in-context reranking pipeline that overlays an LLM on top of existing YouTube recommendations, enabling evaluation of LLM-powered personalization in a real-world context.
    
	\item We empirically characterize how naive personalization in a baseline LLM-assisted YouTube recommender strengthens ideological alignment with user histories and increases exposure to conspiratorial and extremist content for viewers of such content.

    \item We examine an instruction-level regularization strategy that leverages LLMs' contextual understanding to reshape exposure patterns along ideological and topical dimensions, illustrating how lightweight constraints alter the trade-offs of language-based personalization.
    
    \item Through synthetic experiments, \homa{we characterize how LLM-assisted rerankers weight topical against partisan cues, and show that this weighting varies with the candidate set, the watch history, and the topic.}

    \item We release a \homa{validated} benchmark for \homa{political and news YouTube} recommendation, enabling reproducible evaluation of recommendation pipelines in partisan contexts. 
    
\end{itemize}

Our findings illustrate that LLM-based reranking with explicit social-value constraints can produce safer, more balanced outcomes in existing recommendation pipelines, \homa{at a measurable cost in personalization.}  
Without such instruction-level safeguards, language-based personalization can reproduce and even amplify social biases in high-stakes \homa{political contexts}. This work contributes to the emerging literature on LLM-driven recommendation and provides a reproducible framework to empirically characterize exposure trade-offs in next-generation web ecosystems.

\vspace{-2mm}
\section{Related Work}
\textbf{Fairness and diversity in RecSys.} 
Beyond-accuracy objectives have \homa{long addressed} diversity and fairness
\homa{, from} increasing item or topic diversity  \cite{ziegler2005improving} to 
distributing exposure fairly across users, 
providers, and demographic groups
~\cite{patro2020fairrec,do2021twosided}.
These objectives are usually implemented through reranking
~\cite{xu2023pmmf,pei2019personalized,sonboli2020opportunistic}. 
\homa{However,} evaluations \homa{largely} rely on 
benchmarks such as MovieLens, Amazon, Yelp, or Last.fm, 
\homa{and define} fairness and diversity 
over fixed attributes like genre or producer identity, which risks treating fairness as a property of the algorithm 
rather than of the broader social system~\cite{narayanan2026if}.
This 
becomes especially problematic for political content, where harm often arises through framing, narrative emphasis, or ideological positioning 
not captured by coarse metadata. 
\homa{Thus,} 
standard diversity constraints 
\homa{may still permit systematic exposure to} conspiratorial or misleading content.

\textbf{Recommendation in news and political information ecosystems.}
Prior work documents \homa{structural vulnerabilities in political recommendation.} 
A small but engaged minority of YouTube users disproportionately consumes far-right content \cite{hosseinmardi2021examining}, and engagement-driven ranking reinforces such asymmetric partisan exposure \cite{haroon2023auditing}. 
Calls for socially-aware recommendations 
remain largely conceptual \cite{ekstrand2025recommending}, 
\homa{while} studies of filter bubbles and algorithmic amplification document how engagement-optimized 
can narrow 
exposure 
\cite{nguyen2014exploring,cinus2022effect}. 
\homa{Recent} intervention work 
\homa{includes} modifying 
watch histories \homa{to} increase ideological diversity 
\cite{yu2024nudging}, and 
reframing ranking 
as value alignment \cite{jahanbakhsh2025value,stray2021designing}. 
Yet how to jointly balance topical personalization, ideological diversity, and suppression of conspiratorial or extremist content over real consumption histories remains underexplored.

\textbf{LLM-assisted reranking.} 
LLMs are effective zero-shot rerankers over user histories and candidate content 
\cite{hou2024large}, 
\homa{with recent approaches improving} controllability through 
reasoning and natural-language constraints 
\cite{malki2025bonsai,carroll2025ctrl}. 
\homa{While} traditional recommenders \homa{continue to be scrutinized for their potential to} amplify harmful content 
\cite{krauth2025breaking,milli2025engagement}, 
LLM-based systems \homa{introduce additional concerns to} intensify this dynamic by attending to linguistic patterns and 
stylistic cues in user histories 
\cite{loru2025simulation,kim2025linear}. 
These risks 
\homa{remains} empirically \homa{understudied} 
in political news, where 
amplification 
\homa{affects} collective information exposure and public discourse \cite{stray2021designing}. 
\homa{Related work} shows feed reranking \homa{can} alter affective polarization \cite{piccardi2025reranking}, while RecPrompt 
\homa{optimizes} prompt for news recommendation but evaluates 
click prediction \cite{liu2024recprompt}. 
\homa{We instead examine} how LLM-assisted reranking reshapes ideological composition and 
\homa{exposure to} problematic content.

 \textbf{Prompt-level safeguards: from conversation to recommendation.} 
 Conversational AI 
 can be steered 
 through instruction hierarchies 
 \cite{wu2024instructional} and red-teaming methods 
 \cite{raheja2024recent}. 
 \homa{In LLM-based ranking,} the 
 risk 
 \homa{shift from individual} harmful output 
 \homa{to} repeated promotion 
 through ranked lists, 
 introducing different failure modes. 
 We \homa{therefore} test 
 lightweight prompt regularization, 
 \homa{that encodes }social-value constraints 
 alongside \homa{recommendation} instructions, 
 \homa{targeting} conspiratorial framing, extremist rhetoric, and ideological tone---signals difficult to capture with standard metadata-based reranking.
 

\vspace{-2mm}
\section{Data and Methodology} 
\textbf{Data source.} 
To study how LLM-based reranking interacts with real recommendation pipelines, we focus on YouTube sidebar recommendations. YouTube is one of the most widely used video platforms, where prior research has shown that a small subset of users are heavy consumers of problematic content~\cite{chen2023subscriptions,hosseinmardi2021examining}. This ecosystem, prone to amplifying such material, provides a critical context for assessing both the benefits and risks of LLM-assisted personalization.

We use the YouTube desktop browsing trajectories of 128 U.S. adults, collected passively by Nielsen, a media monitoring company, 
between Oct. 2021 and Dec. 2022
~\cite{hosseinmardi2024causally}. Since the Nielsen data only records visited URLs, it does not include the recommendations shown alongside each video. To reconstruct these, the authors replayed observed viewing trajectories, each containing 30–120 consecutive videos, using automated logged-in accounts, recording the top 30 sidebar recommendations after each watch. This procedure yields a recommendation dataset linking each watched video to its sidebar candidates, while preserving user privacy and ensuring external validity.
 
\textbf{Benchmark construction.}
To evaluate how 
reranking strategies behave across comparable viewing contexts, we transform each reconstructed viewing trajectory into a set of overlapping evaluation sessions (\homa{Figure~\ref{fig:benchmark-pipeline}}). Each session captures a short, contiguous segment of a user's watch history, along with recommendations \homa{recorded} at each step, enabling session-level comparisons across reranking methods. We collect metadata (title, topic, category ID, and closed captions) for all watched and recommended videos. To construct evaluation instances, we split each trajectory into overlapping sessions using a sliding window with stride 1. Each session \homa{comprises 10 consecutive} watched videos, \homa{which} define the viewing context, \homa{together with the 11th watched video, which serves as the \textit{reference watch} ($v^*$).}

\homa{We retain only sessions in which at least five of the 10 history videos and at least five of the 30 recommended videos are tagged ``Politics'' and/or ``Society'' in the YouTube Data API \texttt{topicCategories} field~\cite{google_youtube_api} and have captions available, and in which the reference watch meets the same two criteria.}  
\homa{These thresholds were set a priori. Stricter cutoffs yield sessions with denser political content but shrink the usable sample and limit statistical power; looser ones retain more sessions at the cost of admitting ones in which political content is marginal.}   
For each session we identify the recommended video most similar to the \homa{reference watch, using  
the open-weight embedding model BAAI \texttt{bge-m3}~\cite {multi2024m3} and cosine similarity, and designate it the \textit{semantic proxy} ($p$). The measure asks whether a reranker moves content semantically close to what the user went on to watch toward the top of the list. The findings are robust to alternative definitions of the semantic proxy (Figure~\ref{fig:proxy_sensisivity_subset}).} 
The \homa{retained} set of 9,848 sessions \homa{from 97 participants} and associated recommendation lists serves as our evaluation benchmark (\homa{Figure~\ref{fig:benchmark-pipeline}}).

We further infer political ideology and topic for all videos. For ideology, we use a validated prompt (Table~\ref{table:pol_ideology_score}) that scores videos from 0 (strongly liberal) to 100 (strongly conservative), with $r = 0.92$ correlation to human annotations~\cite{dutta_media_bias_orig}. For topic, we use a prompt from~\cite{ghasemian_newslike_info_ecosys_orig}, \homa{which achieves 72\% precision and 73\% recall against adjudicated human labels. Agreement between the two primary annotators before adjudication was 0.45 (top-2 overlap, chance-corrected).}  
For problematic content, defined as conspiratorial or extreme material, 
we annotated 300 videos with \homa{two research assistants and a third adjudicator. Agreement between the two primary annotators before adjudication was $\kappa=0.44$ and the prompt achieves 62\% precision and 88\% recall against the adjudicated human labels. 
We additionally validate both tasks across multiple models (Tables \ref{tab:v35_5_multi_model_precision_recall}-- 
\ref{tab:problematic_inter_annotator_agreement_models} and \ref{table:topics}-\ref{table:prob_content}).}

\textbf{Session categorization.}
We assign each session an ideological score equal to the average political leaning of the watched videos, \homa{then bin sessions into five strata}:  
\textit{left-left} $[0, 20)$, \textit{left} $[20, 40)$, \textit{center} $[40, 60]$, \textit{right} $(60, 80]$, and \textit{right-right} $(80, 100]$. 
\homa{Tables~\ref{table:sessions_division} and~\ref{table:sessions_dist} report session and participant counts per stratum and the distribution of sessions per participant.}

\vspace{-3mm}
\subsection{Reranking methods.} 
Our methodology for studying LLM-assisted reranking of YouTube recommendations (\yt) is depicted in Figure~\ref{fig:audit_recom_LLMs}. Given a user's watch history and YouTube's candidate recommendations, we use LLMs as zero-shot learners~\cite{brown2020language} 
to rerank the candidates~\cite{carraro2024enhancing,hou2024large,vaidyanathan2025llm} via in-context instructions.  
\homa{We evaluate LLM-assisted reranking across three models: the closed OpenAI models GPT-4o-mini and
the more capable GPT-5.6 Sol, and the open-weight 
Qwen3.8-2.4T-A95B (referred to as 4o-mini, Sol, and Qwen3.8 hereafter).} 
Due to token constraints, both history and candidate videos are represented as transcript summaries (each no longer than 150 words; Table~\ref{table:summary}). 
Layering the LLM on top of YouTube's pipeline allows us to generate alternative rankings and compare the benefits and risks of LLM-assisted recommendation against YouTube's deployed recommender system, which we treat as the platform baseline.

\textbf{Baseline LLM reranking (\bllm).}
Our reranking baseline prompt represents a naive, non-regularized LLM-assisted YouTube recommender focused solely on personalization---how well rankings match the user's recent watch history (bLLM; Table~\ref{table:vanilla_prompt}).

\textbf{Regularized LLM reranking (\rllm).}
We propose a regularized variant (\rllm; Table~\ref{table:penalized_prompt}) that balances personalization with responsibility through instruction-based constraints: (a) broaden topical and ideological diversity while preserving relevance, and (b) identify and demote conspiratorial or extreme political content. This tests whether prompt-level regularization---analogous to conversational AI safeguards~\cite{wu2024instructional,raheja2024recent}---can serve as a lightweight intervention for more responsible personalization.

\textbf{Embedding-based reranking (\emb).}
As another baseline for personalization without LLM assistance, we introduce a content-aware, embedding-based ranking technique (\emb). Each candidate video's embedding is compared to a weighted average of the user's watch-history embeddings, with more recent videos weighted higher via exponential decay. Candidates are then ranked by cosine similarity to this aggregated representation. Embeddings are computed using OpenAI's \texttt{text-embedding-3-large} model \homa{(3-large hereafter)}. Because \emb{} is also content-aware, this baseline lets us isolate the added effects of instruction-following and contextual reasoning beyond content similarity alone. 
While more advanced algorithms exist---such as multimodal approaches~\cite{luo2024molar}, attention-based models for long user histories~\cite{zhang2024spar}, or methods integrating embeddings via fine-tuning or prompt tuning~\cite{he2025llm2rec, liu2024recprompt}---they are beyond the scope of this study, as we aim to assess how \homa{LLM-based rerankers affect exposure to political content beyond personalization metrics.} 

\vspace{-3mm}
\subsection{Evaluation metrics.}

\homa{RecSys are conventionally evaluated by how well they predict
what a user will select. That criterion is too narrow here: a ranking can track a
user's recent viewing closely while narrowing the range of perspectives they
encounter or raising the prominence of conspiratorial and extremist material. We
therefore evaluate along five dimensions: \emph{semantic relevance}, how closely
the ranking tracks the content the user went on to watch; \emph{topical
relevance}, whether it surfaces the topics dominating the recent history;
\emph{ideological alignment}, whether it favors content sharing the history's
partisan lean; \emph{ideological diversity}, how widely the top of the list
spreads across the spectrum, which is separable from alignment; and
\emph{problematic content exposure}, whether it raises conspiratorial or
extremist material toward the top.}

\homa{The first two are properties a personalized recommender is built to have. The
remaining three are not targets of the personalization objective and are not
routinely measured, but they are not incidental to it: breadth of ideological
exposure is a long-standing normative goal for news
recommenders~\cite{helberger2019democratic,vrijenhoek2021recommenders}, audiences spanning a wider
political range are associated with more reliable sources~\cite{bhadani2022political}, and
problematic-content exposure is a harm in its own right. The regularized prompt
targets the last two directly.}

\homa{The four ranking dimensions are expressed as AUCs on a common scale, described at
the end of this section; ideological diversity is reported as Shannon entropy
over the top-$k$ ideology distribution~\cite{vrijenhoek2022radio}.  YouTube is scored
on the same dimensions, though its scores describe behavior on these metrics
rather than overall quality, since its objectives are not observable to us.}

\textbf{Semantic relevance.}
\homa{Semantic relevance measures how closely a ranking tracks the content the user
went on to watch. Each candidate at rank $r$ is assigned a binary label
$s_r = \mathbf{1}(r = p)$, where $p$ is the semantic proxy defined above. A
ranking that semantically tracks subsequent viewing places this candidate high in the list.}

\textbf{Topical relevance.}
Topical relevance is assessed by first identifying the dominant topics in a user's recent viewing history. Let a viewing history consist of \(T\) videos indexed from least recent \(t = -T\) to most recent \(t = -1\). The video visited at time \(t\) is annotated with a set of at most four topics \(\mathcal{Z}_t \subseteq \mathcal{Z}\), where \(\mathcal{Z}\) denotes the universe of possible topics (Table \ref{table:topics}). We compute a weight for each topic \(z \in \mathcal{Z}\) using a recency-weighted average with exponential decay:

\vspace{-3mm}
\[
w_z = \sum_{t=-T}^{-1} d^{-(t+1)}\, \mathbf{1}\!\left(z \in \mathcal{Z}_t\right),
\]
where \(d \in (0,1)\) controls the rate of temporal decay and assigns greater importance to more recently watched videos. We set $d = 0.9$ in all weighted averages. Within a single video, all annotated topics receive the same recency weight. The set of dominant topics for a session is defined as
\vspace{-1mm}
\[
\mathcal{Z}^\ast =
\begin{cases}
\left\{\, z \in \mathcal{Z} \,:\, w_z \;\geq\; \tau \cdot \max_{z'} w_{z'} \,\right\}
& \text{if } \dfrac{\max_z w_z}{\sum_z w_z} \;\geq\; \theta, \\[8pt]
\varnothing & \text{otherwise,}
\end{cases}
\]
with dominance threshold $\theta = 0.2$ and tolerance $\tau = 0.9$. The dominance threshold requires that the top topic account for at least 20\% of the total topic mass, ensuring that sessions with diffuse viewing interests are excluded from analysis. The tolerance retains any topic scoring within 10\% of the maximum, so near-ties are not arbitrarily broken. Sessions with $\mathcal{Z}^\ast = \varnothing$ (labeled NA) are excluded from the topical-relevance analysis.

For a recommendation list of \(N\) candidate videos indexed by rank \(r = 1, \ldots, N\), each candidate \(r\) is assigned a binary topical relevance label (also termed \emph{topical match})
$\ell_r = \mathbf{1}\!\left(\mathcal{Z}_r \cap \mathcal{Z}^\ast \neq \varnothing \right)$, where \(\mathcal{Z}_r\) denotes the set of topics for the video at rank \(r\). Topical relevance at each rank position is the average of 
\(\ell_r\) across all sessions with $\mathcal{Z}^\ast \neq \varnothing$. A recommender aligned with a user's recent topical interests should show higher average topical relevance at top ranks.

\textbf{Ideological alignment.} 
To evaluate personalization along the ideological dimension, we compare each session's ideological score (defined in Session categorization above) with the partisanship of its ranked recommendations. We use the same five ideological strata.

Within each stratum, we compare the partisanship trajectories of \yt, \emb, \bllm, and \rllm{} against the corresponding history baseline. For each method, we compute the mean partisanship at each rank position $r = 1, \ldots, N$ across sessions in the stratum, and similarly compute the mean partisanship at each position in the recent history window.  
Since all methods rerank the same YouTube candidate pool, we measure ideological personalization by how strongly each method aligns top-ranked content with the user's recent ideology, relative to YouTube's baseline.

While ideological alignment is not inherently problematic, strong alignment can narrow the range of political perspectives presented to users by disproportionately reinforcing ideologically congenial content, even when that content is not extreme. Further, highly polarized content can often correlate with conspiratorial or extremist material~\cite{jiang2023impact, rao2022partisan}, motivating the problematic-content analysis we develop next.

\textbf{Problematic content exposure.}
To quantify the concentration of problematic content among higher-ranked
recommendations, we compute a rank-weighted average with exponential decay, and
the corresponding recency-weighted average over the watch history. Let
$h_r \in \{0,1\}$ indicate whether the candidate at rank $r$ is labeled
problematic, and $h_t \in \{0,1\}$ whether the video watched at time $t$ is, with
the same decay parameter $d$ used above:
\vspace{-1mm}
\[
\psi_{\mathrm{rec}} = \frac{\sum_{r=1}^{N} d^{r-1} h_r}{\sum_{r=1}^{N} d^{r-1}},
\qquad
\psi_{\mathrm{hist}} = \frac{\sum_{t=-T}^{-1} d^{-(t+1)} h_t}{\sum_{t=-T}^{-1} d^{-(t+1)}}
\]

We correlate $\psi_{\mathrm{hist}}$ and $\psi_{\mathrm{rec}}$ across sessions to
examine the relationship between problematic content in the viewing history and
its promotion in recommendations.

\textbf{\homa{Area Under the Curve (AUC).}} A common metric to evaluate the
performance of recommendation algorithms is their ability to rank higher the
items that users would select. \homa{To make the trade-offs across the four
ranking dimensions above directly comparable, we report each on the same AUC
scale. For each dimension, we assign every candidate a binary label: whether it
is the semantic proxy, matches a dominant topic, shares the session's ideological
lean, or is labeled problematic. Each of these labels is then summarized using
the same rank-based measure.} Let $y_i \in \{0,1\}$ indicate whether candidate $i$
belongs to the target class under evaluation, and $r_i$ denote its rank position,
with smaller $r_i$ corresponding to a higher-ranked item. Then
$\mathrm{AUC} = \Pr\!\left(r_i < r_j \mid y_i = 1, y_j = 0\right)$: the
probability that a target item is ranked above a randomly chosen non-target item.
\homa{We report Semantic, Topical, Congenial, and Problematic AUC, allowing gains
on one dimension to be compared directly with costs on another.}

\vspace{-2mm}
\section{Results}

We examine how LLM-assisted reranking reshapes recommendation exposure beyond
accuracy-based metrics. While recent work has shown that LLM-based reranking can
improve alignment with users' recent viewing histories, we ask how such alignment
interacts with ideological narrowing and exposure to problematic content. To
assess this trade-off, we compare four ranking strategies: \yt, \emb, \bllm, and
\rllm. \homa{Comparisons are reported as pairwise contrasts, e.g.\ $\delby$ for
the gain of the base reranker over the platform baseline (Table~\ref{tab:notation}).} All labeling tasks (topic identification, ideology
scoring, problematic-content classification) use OpenAI's \texttt{gpt-4o-mini}.
Given the smaller number of left-leaning sessions and the limited presence of
problematic content from a left ideological perspective in our data
(Figure~\ref{fig:dist_conspextrm}), and because \textit{center} sessions lack a clear partisan lean, we focus our primary analysis on
\textit{right} and \textit{right-right} sessions \homa{across three underlying
models (4o-mini, Sol, Qwen3.8); see the Appendix 
Figures \ref{fig:center}-\ref{fig:left}. Unless otherwise stated, all estimates are
reported with 95\% confidence intervals from a participant-level cluster
bootstrap~\cite{cameron2008bootstrap} (2{,}000 resamples), since sessions overlap and cluster within
participants.}  

\begin{figure*}[!tb]
     \begin{center}   \includegraphics[width=1\linewidth]{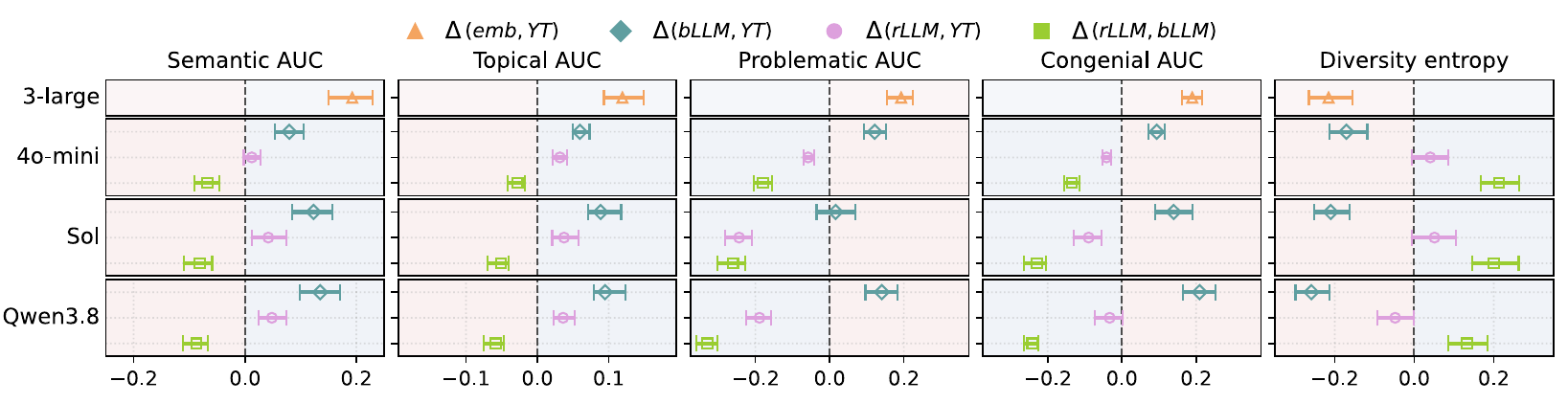}
    \end{center}
    \vspace{-5mm}
    \caption{
    \homa{Method contrasts against \yt{} across \emph{right} and
\emph{right-right} sessions, for Semantic, Topical, Problematic, and Congenial
AUC and Diversity entropy (Tables~\ref{tab:table1_final_right_farright_uid}
and~\ref{tab:pairwise_final_semtopprob}). Each point is a mean session-level
delta with a 95\% participant-level cluster bootstrap CI. Diversity entropy is reported at top-$k{=}10$
(Figures~\ref{fig:top_k_right_ideology_divrsty}
and~\ref{fig:top_k_right_right_ideology_divrsty}).}}
   \label{fig:delta_auc}
   \vspace{-6mm}
\end{figure*}

\paragraph{RQ1. How does LLM-assisted personalization reshape exposure to political content?} 
We first evaluate whether LLM-assisted reranking improves \homa{semantic relevance}. 
\homa{Figure~\ref{fig:delta_auc}} reports 
\homa{main pairwise comparisons across \textit{right} and \textit{right-right}
sessions, with 95\% CIs.} 
\homa{YouTube's own ordering sits at chance on this measure ($\yt = 0.50$ [0.49, 0.51]; Table~\ref{tab:table1_final_right_farright_uid}), which is what we should expect: the platform ranker is not built to maximize semantic resemblance, and it optimizes objectives we cannot observe. In the
1.1\% of sessions where the reference watch is itself among the 30 candidates, a
ranking can be scored against it directly as well as against the proxy: YouTube
ranks the reference watch well above chance but its nearest semantic substitute
at chance, while \emb{} ranks both alike (Table~\ref{tab:candidate_subset_table1_uid}). \yt{}'s near-chance score therefore reflects a ranking objective that is not semantic, rather than a limitation of the measure, and \yt{} serves as a reference point for how much headroom content-aware reranking has on this dimension rather than as a competitor being outperformed.  
Both content-aware methods move substantially into that headroom: $\deley = 0.19$ ($d_z = 0.49$) and
$\delby = 0.08$--$0.13$ ($d_z = 0.19$--$0.34$) across the three models
(all $p < 10^{-3}$; Table \ref{tab:pairwise_final_semtopprob}). The gain is larger on the two more capable models.}

\homa{We next ask what content-aware reranking matches in the watch history when it personalizes. We decompose history match into two dimensions that carry very different implications: topical relevance, which is what personalization is supposed to deliver, and ideological alignment, which matters in political contexts but is rarely evaluated~\cite{ekstrand2025recommending}. 
Content-aware reranking raises the probability that a candidate matching the session's dominant topics outranks one that does not, from $0.51$ [0.51, 0.52] under \yt{} to $0.63$ [0.61, 0.66] under \emb{} and $0.57$--$0.61$ under \bllm{} across the three models ($\deley = 0.12$, $d_z = 0.54$; $\delby = 0.06$--$0.09$, $d_z = 0.28$--$0.41$; all $p<10^{-3}$; Table~\ref{tab:pairwise_final_semtopprob}). Both rerankers therefore deliver genuine
topical gains over the platform baseline.}

We next examine ideological alignment. Although ideological stance may not always be explicitly expressed in video content, prior work has shown a strong association between language use and political partisanship~\cite{khudabukhsh2021we}, \homa{so a reranker matching the history's language will also match its partisan lean. Congenial AUC---the probability that a candidate sharing the session's ideological lean (here \textit{right} or \textit{right-right}) outranks one that does not---rises from $0.50$ [0.49, 0.50] under \yt{} to $0.69$ under \emb{} and $0.59$--$0.71$ under \bllm{} ($\deley = 0.19$, $d_z = 0.76$; $\delby = 0.09$--$0.21$, $d_z = 0.39$--$0.77$; all $p<10^{-3}$), and ideological diversity among the top-ranked candidates falls correspondingly (entropy $1.51$ under \yt{} against
$1.29$--$1.35$ across the content-aware methods; $\deley=-0.22$, $\delby = -0.17$ to $-0.26$; all $p<10^{-3}$). \yt{} sitting at chance here is consistent with evidence that YouTube's modern recommender works against narrow, history-matched recommendations~\cite{brown2022echo,hosseinmardi2024causally}. Under a pure
personalization instruction, \bllm{} behaves much as ranking by similarity in embedding space does: both track the history along topic and ideology together, and neither separates them.}  
While ideological alignment is not inherently problematic, stronger alignment can narrow the range of political perspectives presented to users by disproportionately reinforcing ideologically congenial content, even when that content is not extreme.

Strongly partisan material frequently overlaps with conspiratorial or extremist content~\cite{jiang2023impact,rao2022partisan}, raising the question of whether LLM-based personalization (\bllm; Table~\ref{table:vanilla_prompt}) may unintentionally increase exposure to such content for users with relevant history. Across \textit{right} and \textit{right-right} sessions, the AUC for problematic content  
\homa{is $0.48$ [0.47, 0.49] under \yt{}, $0.68$ [0.64, 0.71]
under \emb{}, and $0.61$ and $0.62$ under \bllm{} on 4o-mini and Qwen3.8 
($\deley = 0.19$, $d_z = 0.78$; $\delby = 0.12$ and $0.14$, $d_z = 0.48$ and
$0.51$; all $p<10^{-3}$); Sol is discussed below. On two of the three models, the same history-matching that produced the gains above also lifts problematic material toward the top. 
The same pattern holds when exposure is measured directly rather than as a
ranking probability: rank-weighted problematic exposure among the top-$k$
recommendations ($\psi_{\text{rec}}$) at $k{=}5$ is $0.20$ under \yt{} against
$0.35$ under \bllm{} on 4o-mini and Qwen3.8 ($d_z = 1.26$ and $1.09$,
$p<0.001$) and $0.27$ on Sol, where the contrast is not significant
($d_z = 0.34$), with the gap attenuating as $k$ grows
(Table~\ref{tab:table2_three_models_right_farright}).} 

\homa{
Sol's null result above is not confined to the exposure measure: \bllm{} reaches $0.50$
[0.45, 0.55] on Problematic AUC and the contrast against \yt{} is not distinguishable from zero
($\delby = 0.02$, $p = 0.44$). Sol declined most reranking requests under the partisan-emphasis variant (RQ3; Figure~\ref{fig:exp1-topic-congenial-ideology-auc-v2}). 
A similar constraint may operate
under the unconstrained prompt, though we cannot observe it.   
What the comparison
does rule out is a simple account in terms of family or capability: Qwen3.8 and
Sol are current-generation models from different families that behave oppositely
here, while 4o-mini and Sol share a provider and the more capable of the two is
the one that does not amplify. Under identical prompt text, whether
personalization alone lifts problematic material toward the top of the list
therefore depends on model-specific properties we cannot observe. We return to this in the Discussion.}

Together, these findings demonstrate that gains  \homa{of content-aware reranking} in topical and ideological alignment can coincide with increased concentration of problematic content among highly ranked recommendations \homa{on two of the three models tested}, highlighting a key vulnerability of personalization-driven reranking in political contexts.

\paragraph{RQ2. \homa{At what cost to personalization can LLM-assisted ranking incorporate socially-aware constraints?}} 
Because LLMs operate over language and contextual cues, \homa{the same mechanism that amplified problematic exposure above can be steered against it. We test this with \rllm{}, which adds instruction-level constraints to the reranking prompt, and quantify what those constraints cost on the dimensions personalization is meant to deliver.}

\homa{The regularized prompt reduces problematic exposure substantially on all
three models. Problematic AUC falls from $0.48$ [0.47, 0.49] under \yt{} to
$0.43$, $0.24$, and $0.29$ under \rllm{} on 4o-mini, Sol, and Qwen3.8
($\delry = -0.06$, $-0.24$, $-0.19$; $d_z = -0.24$, $-0.99$, $-0.71$; all
$p<10^{-3}$). The reduction is larger against the unconstrained reranker
($\delrb = -0.18$, $-0.26$, $-0.33$; $d_z$ up to $-1.48$). Rank-weighted exposure
shows the same result: $\psi_{\text{rec}}$ at top-$k{=}5$ drops from $0.20$ under
\yt{} to $0.10$--$0.15$ under \rllm{}, with large effects at every $k$
($d_z = -0.70$ to $-1.49$, all $p<0.001$;
Table~\ref{tab:table2_three_models_right_farright}). Congenial AUC also moves in
the intended direction relative to \bllm{} ($\delrb = -0.13$ to $-0.24$), while
ideological diversity increases (entropy $\delrb = +0.13$ to $+0.21$). Relative
to \yt{}, however, the change in diversity entropy is not distinguishable from
zero on any model ($\delry = +0.04$, $+0.05$, $-0.05$; all $p>0.05$). Problematic
content in the watch history also carries over less strongly into recommendations
under \rllm{}: the fitted slope of $\psi_{\text{rec}}$ on $\psi_{\text{hist}}$ is
lower than under \yt{} for all three models (significantly for 4o-mini and Sol)
and higher under \bllm{}
(Figure~\ref{fig:corr_rank_weighted_recency_weighted_prob}).}

\homa{These exposure gains come at a cost, and its magnitude depends on the
reference point. Against the unconstrained reranker the loss is substantial:
$\delrb = -0.07$ to $-0.09$ on Semantic AUC and $-0.03$ to $-0.06$ on Topical AUC
(all $p<10^{-3}$), returning most of what \bllm{} gained in RQ1. This is the
structure of any constrained objective: a gain on one measure is paid for on
another. Relative to \yt{}, however, \rllm{} loses nothing: Semantic AUC is
$0.51$, $0.54$, and $0.55$ against \yt{}'s $0.50$, and Topical AUC $0.54$--$0.55$
against $0.51$, so the regularized reranker matches or slightly exceeds the
platform baseline on both ($\delry = +0.01$ to $+0.05$ and $+0.03$ to $+0.04$;
the semantic contrast on 4o-mini is the only relevance contrast not
distinguishable from zero, $p = 0.07$).}
\homa{Taken together, RQ1 and RQ2 show that the same LLM reranker produces substantially different exposure outcomes when given a different objective, while retaining part of the personalization gain. This flexibility distinguishes instruction-based reranking from \emb{}, which has no mechanism for incorporating such a constraint. The constraint is not costless---relative to \bllm{}, \rllm{} returns most of the semantic and topical gain---but it lets the ranking objective trade some personalization for lower problematic and ideologically congenial exposure.}

\begin{figure}[!tb]
     \begin{center} 
           \includegraphics[width=0.90\linewidth] {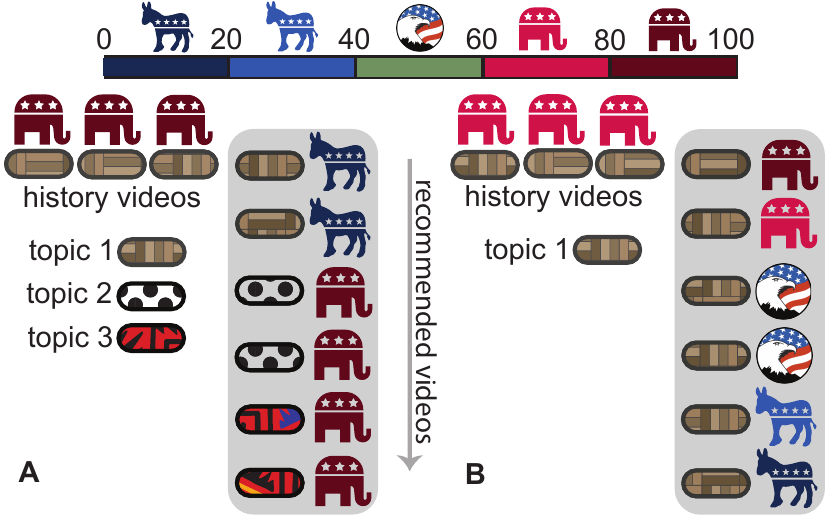}
    \end{center}
    \vspace{-3mm}
    \caption{
Two synthetic-session designs for \homa{exploring how rerankers weight topical relevance against partisan alignment}. 
(A) Histories cover a single topic with extreme partisanship (\textit{right-right} or \textit{left-left}); candidates include same-topic videos from the opposing extreme and other-topic videos sharing the history's partisan stance. 
(B) Histories cover a single topic with consistent partisan content (\textit{left-left}, \textit{left}, \textit{right}, or \textit{right-right}); candidates include same-topic videos spanning the full partisan spectrum.
}
   \label{fig:audit_recom_LLMs_synthet_design}
   \vspace{-6mm}
\end{figure}

\paragraph{RQ3. \homa{How do LLM rerankers weight topical against partisan cues, and how does that weighting vary with the candidate set, the watch history, and the topic?}} 
For the non-regularized methods \bllm{} and \emb{}, the partisan content of
top-ranked recommendations aligns closely with the average partisanship of users'
viewing histories (Figure~\ref{fig:delta_auc}). \homa{This raises the question of
what the rerankers match on when they produce that alignment: topical relevance
with the history, ideological alignment, or both. In the sessions above the two
are confounded, since same-topic candidates typically also share the history's
lean. We therefore designed two synthetic experiments in which topical and
partisan cues can be varied independently.}

\textbf{Experiment 1: Topic versus partisanship trade-offs.}
This design tests \homa{which cue \bllm{} follows when topical and partisan alignment
point to different candidates (Figure~\ref{fig:audit_recom_LLMs_synthet_design}A)}. 
We construct sessions whose histories cover a single topic with extreme
partisanship, exclusively on the right or exclusively on the left. Candidate sets
include same-topic videos from the opposing ideological extreme and other-topic
videos aligned with the history's partisan perspective, so that promoting partisan
content across topics and promoting opposing views within a topic are
distinguishable. We summarize only the portions of videos covering abortion,
immigration, or elections (Table~\ref{table:exclusive_summary}) and draw the
histories and candidate sets from this corpus in random order.

\begin{figure}[!tb]
     \begin{center}   \includegraphics[width=1\linewidth]{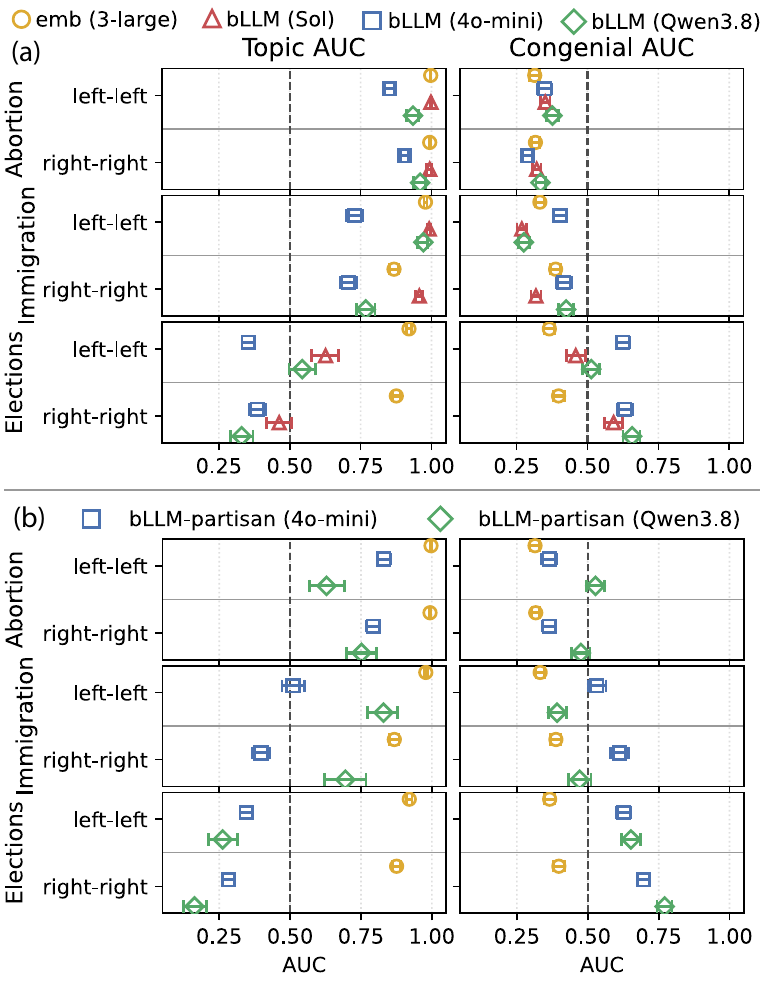}
    \end{center}
    \vspace{-5mm}
    \caption{\homa{Topical AUC and Congenial AUC by setting (Experiment 1), for \emb{} and each model's \bllm{}. Confidence intervals resample sessions, as these sessions are synthetic. Sol is omitted from the \bllmp{} panels: the model declined 590 of 600 reranking requests under this prompt.}}
   \label{fig:exp1-topic-congenial-ideology-auc-v2}
   \vspace{-6mm}
\end{figure}

\begin{figure}[!tb]
     \begin{center}   \includegraphics[width=1\linewidth]{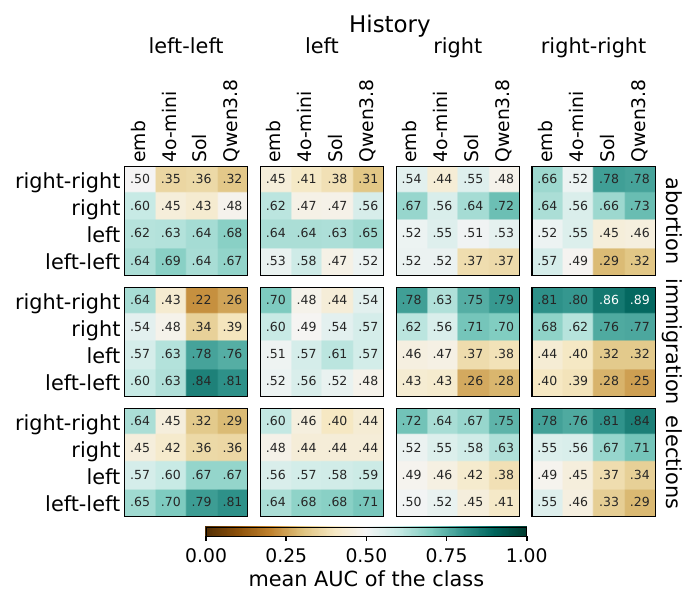}
    \end{center}
    \vspace{-5mm}
    \caption{\homa{Ideological preference of each reranker with topic held constant. Panels are arranged by topic (rows) and by the lean of the synthetic watch history (columns). Within each panel, rows give the ideological class of the candidate videos and columns the reranker. Each cell is the mean per-session AUC with which that reranker ranks candidates of that class above the rest, so values above 0.5 indicate the class is promoted and values below that it is demoted.} } 
\label{fig:exp2_ideology_class_auc}
   \vspace{-6mm}
\end{figure}

The results provide limited support for the hypothesis that the algorithm promotes partisan content outside closely related topics (Figure~\ref{fig:exp1-topic-congenial-ideology-auc-v2}a; \homa{Table \ref{tab:exp1-topic-congenial-ideology-auc-v2}} and Figures~\ref{fig:comparison_boxplots_recbLLMYT_recembYT}--\ref{fig:audit_recom_LLMs_synthet_design_result1}). 
\homa{Topic dominates for abortion and immigration.
Every method ranks same-topic candidates near the top regardless of partisanship: Topical AUC runs 0.71--1.00 across models and settings, while Congenial AUC stays well below 0.5 (0.27--0.42). A Congenial AUC below 0.5 means the ranking actively places the \emph{opposing} side higher, which is where the same-topic candidates sit by
construction. \emb{} shows the pattern most sharply (Topical AUC 0.87--1.00,
Congenial 0.31--0.39).}
\homa{On elections the two methods diverge.  
For \bllm{}, the ordering reverses: Topical AUC falls to 0.33--0.63 while Congenial AUC rises above 0.5, reaching 0.66 [0.63, 0.69] for Qwen3.8 on \textit{right-right} histories. 
\emb{} does not follow, holding Topical AUC at 0.88--0.92 and Congenial below 0.5 throughout. Elections content is entangled with the economy, abortion, and immigration, so the topical signal is weaker, and in its absence \bllm{} ranks on partisan alignment instead---though the prompt contains no instruction to do so. Which cue dominates therefore depends on the topic and on what the candidate set makes available rather than on a fixed preference.}

By contrast,  \emb{} recommender consistently prioritizes topic relevance over partisan similarity across all three topics.  
\homa{For \emb{} the mechanism is directly measurable. Comparing candidate-to-history cosine similarity under the same embeddings that drive the ranking (Tables~\ref{tab:rq3-embedding-similarity} and~\ref{tab:rq3-delta-emb}), same-topic/opposite-partisan candidates sit closer to the history than other-topic/same-partisan ones for all three topics: 0.61 vs.\ 0.33 for abortion, 0.54 vs.\ 0.35 for immigration, 0.54 vs.\ 0.35 for elections (paired $\Delta_{\mathrm{emb}} = 0.28$, $0.18$, $0.19$; all $d_z > 3.7$). General-purpose embeddings encode topical similarity more strongly than ideological framing in these candidate sets, which is why \emb{} tracks topic even when that means promoting the opposing side. We stop short of the broader claim that embeddings encode topic more strongly than ideology in general; a general decomposition would require a different design.}

\homa{We also test a variant with an explicit partisan instruction (\bllmp{};
Table~\ref{table:vanilla_prompt_partisan_emphasis}), which asks the reranker to
prioritize partisan content alongside topical relevance; Sol declined nearly all
requests under this prompt (Figure~\ref{fig:exp1-topic-congenial-ideology-auc-v2}b).} The variant
shifts further toward partisanship than the baseline, though not uniformly: it
prioritizes partisan over topical alignment for immigration but not for abortion,
where the range of arguments is narrower
(Figures~\ref{fig:comparison_boxplots_recbLLMYT_recembYT}
and~\ref{fig:audit_recom_LLMs_synthet_design_result2}).

\textbf{Experiment 2: \homa{Partisan alignment with topic held constant.}}
Although the \bllm{} prompt does not explicitly instruct the model to prioritize partisan content, the personalization goal could still lead it to use partisan signals. 
This design tests whether \homa{the ranking tracks partisan alignment with the history when topic is held constant (Figure~\ref{fig:audit_recom_LLMs_synthet_design}B). We construct fully synthetic sessions in which the watch history covers a single topic with consistent partisan content (\textit{left-left}, \textit{left}, \textit{right}, or \textit{right-right}), and candidate videos span the full partisan spectrum on that same topic.}

\homa{With topic held constant, every reranker promotes candidates matching the
history's lean and demotes the opposing side
(Figure~\ref{fig:exp2_ideology_class_auc}). For \textit{right-right} histories,
\bllm{} ranks right-right candidates at 0.52--0.78 on abortion, 0.80--0.89 on
immigration, and 0.76--0.84 on elections, while left-left candidates fall to
0.25--0.49. The mirror holds for \textit{left-left} histories: left-left
candidates reach 0.63--0.84 and right-right candidates drop to 0.22--0.45. The
effect is weaker for the moderate \textit{left} and \textit{right} strata, where
congenial classes sit at 0.55--0.72. \emb{} shows the same direction throughout,
so ideology is not invisible to embedding similarity once topic is held constant,
though the extremes are less pronounced. Figures~\ref{fig:audit_recom_LLMs_synthet_design_result3_mini}--\ref{fig:audit_recom_LLMs_synthet_design_result3_qwen} show the corresponding partisanship by rank position.}

\homa{Within a side, the ranking does not simply favor more extreme content. For
\textit{right} histories on immigration and elections, the more extreme
right-right class is ranked \emph{above} the matching right class (0.63--0.79 and
0.64--0.75 against 0.56--0.71 and 0.55--0.63), while on abortion the ordering
reverses (0.44--0.55 against 0.56--0.72). Topic again determines which way the
ranking moves within a partisan side, as it did in Experiment~1.}

\vspace{-2mm}

\section*{Discussion and Conclusions}
\label{sec:disc} 
Recommendation algorithms have long been scrutinized for their potential negative societal impacts, ranging from promoting problematic content~\cite{o2015down} to amplifying polarization by reinforcing exposure to like-minded material~\cite{haroon2023auditing}. These concerns stem in part from such systems being designed to maximize engagement or revenue rather than broader measures of welfare~\cite{aridor2024economics}, and from the narrow focus of their evaluations on metrics such as click alignment or predictive accuracy~\cite{kleinberg2024challenge}. Although the societal impact of content curation models remains contested, LLMs have emerged as the next wave of powerful---yet opaque---models, attracting considerable attention for their claimed ability to enhance RecSys, capabilities that may themselves raise new concerns.

In this study, we examine how LLM integration shapes the behavior of RecSys in
political \homa{and news content} by comparing YouTube recommendations with
variants that incorporate LLM-based reranking. We assess not only \homa{semantic
and topical relevance}, but also alignment in political ideology, which can
introduce harms by reinforcing congenial content and promoting problematic
material. We further ask whether the resulting societal risks stem mainly from
the personalization power of LLMs or from the metric optimizations underlying
recommender systems.

We find that using LLMs naively as recommender algorithms---by reranking based on
a user's watch history---can enhance personalization, as reflected in
\homa{closer semantic and topical alignment} with the viewing history. However,
when examined through the lens of political partisanship, this improvement
increases exposure to ideologically congenial content by placing politically
aligned items higher in the ranking. Moreover, for users who have previously
consumed problematic content (extremist or conspiratorial), LLM-based reranking
amplifies exposure by ranking such videos higher \homa{on two of the three models
tested}.

\textbf{Implications.} This pattern resonates with recent findings that LLMs encode linear representations of political perspectives~\cite{kim2025linear}. While such representations can be leveraged to trace and shift the ideological orientation of model outputs, our results suggest that similar mechanisms may influence how RecSys personalize to users' historical news diets and potentially amplify or reduce partisan---or even problematic---content.

Our findings can be understood within broader platform dynamics, where user experiences are shaped less by their true preferences and more by engagement signals such as time spent on content~\cite{kleinberg2024challenge}. 
Individuals' momentary choices often diverge from their underlying preferences, yet platforms interpret these behavioral signals as indicators of desire and optimize  
accordingly. 
This mistaken assumption has negative consequences: it distorts the content landscape by rewarding manipulative strategies like clickbait over substantive quality~\cite{immorlica2024clickbait}, and amplifies users' emotional biases, reinforcing filter bubbles and undermining well-being~\cite{habib2025youtube,castelo2025blocking,nguyen2014exploring}. Our results show that careless deployment of LLM-assisted RecSys---when designed to maximize alignment with \homa{a user's recent viewing}---risks reproducing and even intensifying these dynamics.

By contrast, our regularized variant (\rllm) shows that instruction-level
regularization mitigates this tendency \homa{at a cost borne relative to the
unconstrained reranker rather than to the platform: \rllm{} still ranks above
\yt{} on both relevance dimensions. The trade-off is therefore not between safety
and usable personalization, but between safety and the additional personalization
an unconstrained reranker can extract}, indicating that the issue lies not in
using language models but in using them without socially aware safeguards.
\homa{The prompt was not tuned to minimize this cost, so the trade-off is likely an upper bound.} 
As platforms 
incorporate LLMs across their systems, these models
magnify whichever signal they are asked to follow most closely. Such regularization is already applied to conversational LLMs, where vulnerabilities are probed through jailbreaks and red-teaming~\cite{wu2024instructional,raheja2024recent}; 
a similar
approach is needed for recommender systems, where safeguards should be
stress-tested not only for model misuse but also for vulnerabilities that produce
socially problematic outcomes.

 
\textbf{Limitations and future Work.} 
This study has several limitations. First, while users watched each video in a session in the order it appears in our data, we do not claim that these constructed sessions precisely reflect a real user's experience: (i) we cannot confirm that each video was actually selected from YouTube's sidebar recommendations, and (ii) there is a time delay between the original viewing and the collection of recommended videos. Nonetheless, because the data is based on real YouTube recommendations collected by a digital twin that mimics a user's watch history, it remains sufficiently reliable for studying LLM-assisted reranking of YouTube recommendations and offers stronger external validity than entirely synthetic trajectories.

Second, the \textit{left} and \textit{left-left} strata contain far fewer
sessions, so estimates there are noisy and stable patterns are hard to identify. 
We therefore focus the main analysis on \textit{right} and \textit{right-right}
sessions, with results for the remaining strata in the Appendix.
\textit{Center} sessions lack a clear partisan lean and so are less informative
for how LLM-assisted personalization interacts with partisan
exposure.

\homa{Third, we use the same prompt across 4o-mini, Sol, and Qwen3.8, without tuning it separately for each model, and observe the same  pattern across all three. This provides some evidence that the findings are not specific to a single prompt–model pairing. Our goal, however, is not to optimize recommendation accuracy through prompt engineering, but to test whether LLM-based rerankers can respond to a socially aligned objective specified in the prompt. Relatedly, our synthetic experiments identify which cue a reranker follows when topical and partisan alignment conflict, but not what produces that weighting. As recent work has shown the value of prompt optimization for LLM-based reranking, extending it to optimize not only click-rate prediction and accuracy but also social-value and other beyond-accuracy objectives is an important direction for future work.}

\section{Acknowledgments}
We are grateful to the Nielsen Company for access to their desktop panel data. Researcher(s)’ own analyses calculated (or derived) based in part on desktop browsing data from The Nielsen Company (US), LLC. The conclusions drawn from the Nielsen data are those of the researcher(s) and do not reflect the views of Nielsen. Nielsen is not responsible for, had no role in, and was not involved in analyzing and preparing the results reported herein. We thank Ziying Chen for her initial assistance with this research. In addition, we gratefully acknowledge the research assistance of Olivia Nguyen, Sean Rosskopf, Xiaotong Xi, Xin Yi Yang, and Elizabeth Williams.

\clearpage

\bibliography{bib}

\clearpage
\onecolumngrid
\appendix
\setcounter{secnumdepth}{1} 
\counterwithout{equation}{section}
\renewcommand{\theequation}{S\arabic{equation}}
\setcounter{equation}{0}
\renewcommand{\thetable}{S\arabic{table}}
\setcounter{table}{0}
\renewcommand{\thefigure}{S\arabic{figure}}
\setcounter{figure}{0}
\setcounter{section}{0}

\setcounter{topnumber}{4}
\setcounter{bottomnumber}{3}
\setcounter{totalnumber}{6}
\renewcommand{\topfraction}{0.9}
\renewcommand{\bottomfraction}{0.8}
\renewcommand{\textfraction}{0.07}
\renewcommand{\floatpagefraction}{0.7}
\setcounter{dbltopnumber}{3}
\renewcommand{\dbltopfraction}{0.9}
\renewcommand{\dblfloatpagefraction}{0.7}

\section{Data}
\label{appendix:data}
The data for this study comes from prior work that used Nielsen's nationally representative desktop web panel, which records individual-level URL visits from October 2021 to December 2022~\cite{hosseinmardi2024causally}. 
Nielsen provides information only about the videos users have viewed on YouTube, not the recommendations. To obtain data on recommended videos, we used data from two experiments described in~\cite{hosseinmardi2024causally}, which estimated (1) bias and (2) forgetting time of the YouTube recommender system. 
The dataset includes 128 participants from this panel, each of whom has watched at least 140 videos on YouTube during their lifetime with the Nielsen panel. In~\cite{hosseinmardi2024causally}, automated users viewed 30, 60, or 120 of these videos (maintaining the original order), depending on the experiment, with up to 30 recommended videos collected from the sidebar after each viewed video.

\begin{figure}[H]
     \begin{center}
           \includegraphics[width=0.6\linewidth]{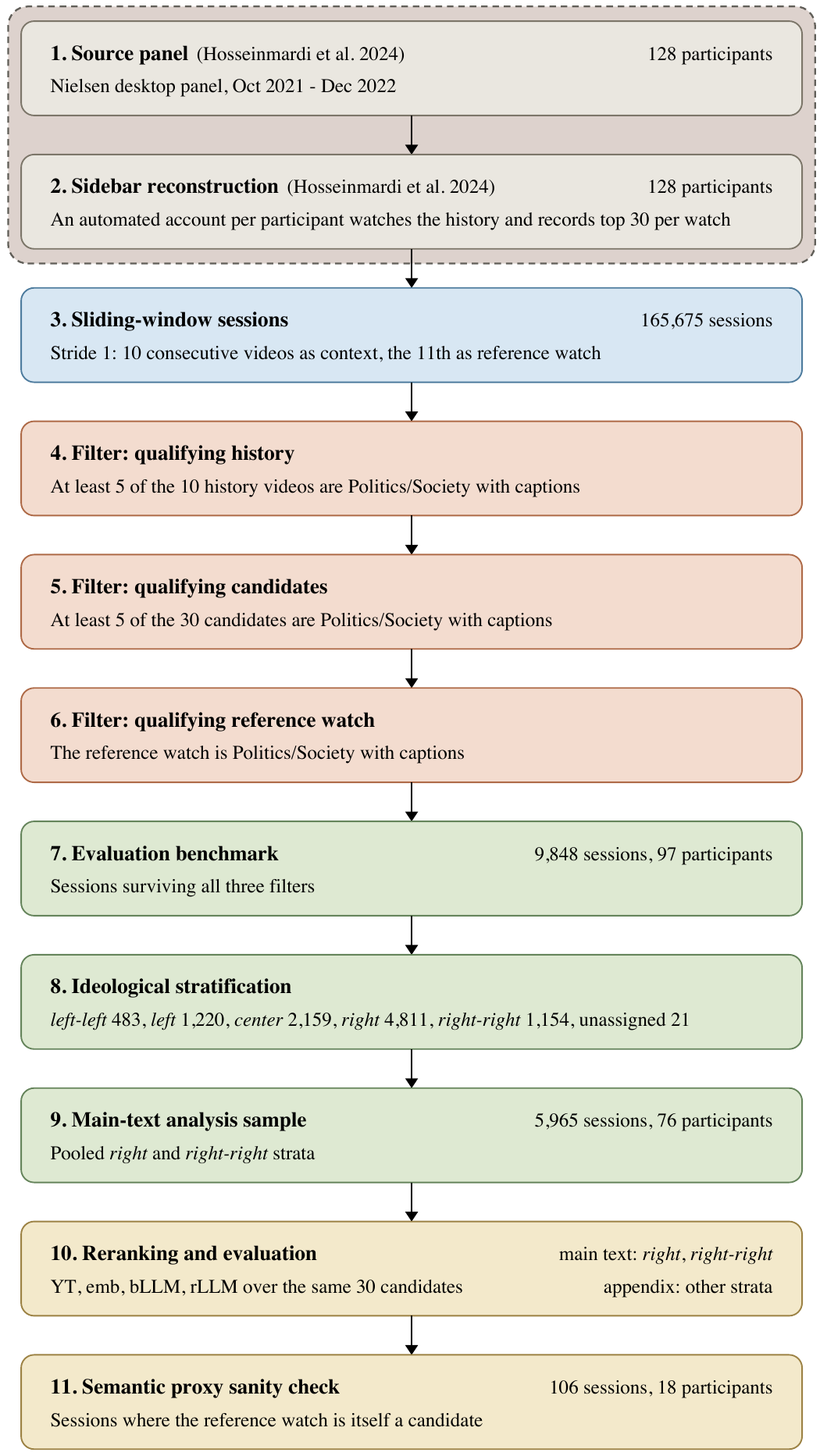}
    \end{center}
    \caption{\homa{Benchmark construction pipeline. Steps~1--2 were carried out by
\citet{hosseinmardi2024causally}, whose data we use; steps~3--11 constitute the
present benchmark construction.
Each session comprises 10 consecutive watched videos, which define the viewing
context, and the 11th watched video, the \textit{reference watch}. Three filters
retain sessions with sufficient political content in the history, in the
candidate set, and in the reference watch, yielding the 9,848-session evaluation
benchmark. Counts at the right give the sample remaining after each stage;
unassigned denotes sessions whose history partisanship did not fall in
any stratum. The released benchmark provides
the session structure with each video represented by the derived text summary used
in the reranking pipeline.}}
   \label{fig:benchmark-pipeline}
\end{figure}

For our analysis, we split each user's chronological viewing history into overlapping sessions using a sliding-window \homa{with stride 1}. \homa{Each session comprises 10 consecutive watched videos, which form the viewing context, together with the 11th watched video, which serves as the reference watch $v^*$.} This data augmentation technique increases the number of subsequences of user activity, generating multiple overlapping sequences per user and allowing us to examine recommendation dynamics at different points in the viewing trajectory. 
This yields 165,675 sessions in total. \homa{For each session we use the top 30 sidebar recommendations recorded after the 10th history video as the candidate set.} 
We gathered metadata for these videos using the topic categories provided by the YouTube API~\cite{google_youtube_api}, focusing on those \homa{tagged} ``Politics'' and/or ``Society'' \homa{in the \texttt{topicCategories} field.} 
Conditioning on sessions in which at least \homa{five} of the 10 history videos and at least five of the 30 recommended videos fall in these categories with captions available, and in which the \homa{reference watch} meets the same two criteria, leaves 9,848 sessions from 97 participants (\homa{Figure~\ref{fig:benchmark-pipeline}}).
In our efforts to collect captions, we successfully obtained them for approximately 83\% of the videos, totaling 61,633. However, we faced challenges with the remaining 12,517 videos. Some of these videos do not have captions available, which may be due to various reasons, such as the removal of captions, their being disabled, or the absence of captions in the English language. Tables \ref{table:sessions_division} and \ref{table:sessions_dist} provide more details about the dataset characteristics. 
The videos in our curated dataset vary in length from 4 seconds to 25 hours, with an average duration of approximately 31 minutes. We divided these videos into segments of 2000 tokens, which corresponds to about 10 to 12 minutes, and focused our analysis solely on the first segment of each video. 

\homa{\textbf{Automated accounts for sidebar reconstruction.}
Nielsen records which YouTube URLs a participant visited but not what the
platform recommended alongside them. The sidebar candidates in our benchmark were
reconstructed by Hosseinmardi et al.~\cite{hosseinmardi2024causally}, who replayed
each observed trajectory with an automated account and recorded the top 30 sidebar recommendations after every watch. That study created more than 150 Google accounts, each reused across multiple experiments, with one account replaying a given participant's trajectory. Accounts were logged in rather than logged out. 
Before each replay the crawler reset the account to a clean state: logging in,
opening the ``Your data in YouTube'' section, and clearing the watch history,
which removes the account's activity data and returns recommendations to a state
comparable to an incognito session. The authors confirm this empirically, finding
no trace of previously watched topics in initial measurements on a reset account.
The account then watched the participant's videos in their original order,
recording the sidebar after each. More details on the data collection pipeline are provided in Hosseinmardi et al.\ (2024, \textit{PNAS}), Appendix Section  Data collection pipeline.}

\begin{table}[h]
\centering
\caption{\homa{Distribution of sessions across the five ideological categories. Participants contribute sessions to more than one category, so the participant column does not sum to the total.}}
\label{table:sessions_division}
\begin{tabular}{lrr}
\toprule
category & unique sessions & unique participants \\
\midrule
\textit{right-right} & 1{,}154 & 38\\
\textit{right}       & 4{,}811 & 70\\
\textit{center}      & 2{,}159 & 73\\
\textit{left}        & 1{,}220 & 38\\
\textit{left-left}   & 483 & 9\\
\addlinespace
unassigned              & 21 & 2\\
\midrule
Total                & 9{,}848 & 97 \\
\bottomrule
\end{tabular}
\end{table}

\begin{table}[h]
\centering
\small
\caption{\homa{Distribution of sessions per participant across the 9,848 benchmark sessions.} }
\label{table:sessions_dist}
\begin{tabular}{lccccc}
\toprule
 $n$ & Mean & SD &  25th pct.\ & Median & 75th pct.  \\
\midrule
  97 & 101.5 & 148.7 &  15 & 49 & 125  \\
\bottomrule
\end{tabular}
\end{table}

\homa{\textbf{Classification evaluations.} For both classification tasks, we validate the prompts against  human labels using three models spanning three families/providers (Tables~\ref{tab:v35_5_multi_model_precision_recall}-\ref{tab:problematic_inter_annotator_agreement_models}). 
For topic classification, claude-sonnet-5 
and 
Llama-3.3-70B  achieve performance comparable to the original gpt-4o-mini classifier. No model dominates, and the spread across providers is small relative to the differences the classification is used to
detect (Tables~\ref{tab:v35_5_multi_model_precision_recall}).
We also report agreement directly
(Table~\ref{tab:v35_5_inter_annotator_agreement}), defining it as overlap between two models' top-2 assigned topics, which suits a task where each video can be assigned up to four topics. Chance is estimated by a permutation null (2,000 shuffles) and the adjusted
score is computed in the spirit of Cohen's $\kappa$. Agreement between the two human annotators before adjudication was 0.45 (chance-corrected). Model-human agreement
reaches 0.71 for gpt-4o-mini, 0.69 for claude-sonnet-5, and 0.61 for Llama-3.3-70B. Agreement
among the models themselves is comparable or higher, indicating the classification is not idiosyncratic to gpt-4o-mini. The topic taxonomy is similar to the framework developed and validated by the
Pew Research Center.}

\homa{For problematic-content classification, agreement between the two human annotators before adjudication was $\kappa=0.44$. 
Against the same adjudicated human labels, precision and recall are 0.62/0.88 for gpt-4o-mini, 0.73/0.57 for claude-sonnet-5, and 0.67/0.69 for Llama-3.3-70B.
Model-human Cohen's $\kappa$ agreement reaches 0.41, 0.42, and 0.44 (Moderate agreement), and agreement among the models
themselves is 0.40 to 0.61 (Tables~\ref{tab:macro_positive_problematic} and~\ref{tab:problematic_inter_annotator_agreement_models}). 
Together, these results show that the classifier's performance and labeling patterns are broadly consistent across model families rather than being specific to gpt-4o-mini. }

\begin{table}[h]
\centering
\caption{\homa{Topic-classification precision and recall against human-annotated ground truth (900 videos), for three models spanning three providers. Both precision and recall are macro-averaged over 14 topic categories.}}
\begin{tabular}{llcc}
\toprule
Provider & Model & Precision & Recall \\
\midrule
OpenAI & gpt-4o-mini (original) & 0.72 & 0.73 \\
\addlinespace
Anthropic & claude-sonnet-5 & 0.75 & 0.60 \\
\addlinespace
Meta	& Llama-3.3-70B &	0.62	&0.69 \\
\bottomrule
\end{tabular}
\label{tab:v35_5_multi_model_precision_recall}
\end{table}

\begin{table}[h]
\centering
\caption{\homa{Inter-annotator agreement on topic classification. Agreement is the overlap between each rater's top-2 assigned topics, or both assigning none, which suits this multi-label task. The chance rate is estimated by a permutation null (2{,}000 shuffles), and the adjusted score is $(\text{observed}-\text{chance})/(1-\text{chance})$, in the spirit of Cohen's $\kappa$.} }
\begin{tabular}{llccc}
\toprule
Rater A & Rater B & Observed & Chance & Adjusted score \\
\midrule
gpt-4o-mini & human & 78.1\% & 24.9\% & 0.71 \\
claude-sonnet-5 & human & 77.6\% & 27.7\% & 0.69 \\
Llama-3.3-70B & human & 67.9\% & 18.5\% & 0.61 \\
\addlinespace

claude-sonnet-5 & gpt-4o-mini & 80.2\% & 25.8\% & 0.73 \\
Llama-3.3-70B & gpt-4o-mini & 80.3\% & 20.3\% & 0.75 \\
claude-sonnet-5 & Llama-3.3-70B & 65.2\% & 18.8\% & 0.57 \\
\bottomrule
\end{tabular}
\label{tab:v35_5_inter_annotator_agreement}
\end{table}

\begin{table}[h]
\centering
\caption{\homa{Macro-averaged and positive-class precision, recall, and F1 for problematic content.}}
\small
\setlength{\tabcolsep}{5pt}
\begin{tabular}{lcccccc}
\toprule
& \multicolumn{3}{c}{Macro} & \multicolumn{3}{c}{Positive class} \\ 
\cmidrule(lr){2-4} \cmidrule(lr){5-7}
Model & Precision & Recall & F1 & Precision & Recall & F1 \\
\midrule
gpt-4o-mini (original) & 0.74 & 0.73 & 0.71 & 0.62 & 0.88 & 0.73 \\
claude-sonnet-5        & 0.72 & 0.71 & 0.71 & 0.73 & 0.57 & 0.64 \\
Llama-3.3-70B          & 0.72 & 0.72 & 0.72 & 0.67 & 0.69 & 0.68 \\
\bottomrule
\end{tabular}
\label{tab:macro_positive_problematic}
\end{table}

\begin{table}[h]
\centering
\caption{\homa{Inter-annotator agreement on problematic-content classification, measured by Cohen's $\kappa$. ``gpt-4o-mini'' is the original label set used throughout the paper.}}
\vspace{-2mm}
\begin{tabular}{llc}
\toprule
Rater A & Rater B & Cohen's $\kappa$ \\
\midrule

gpt-4o-mini & human & $0.41$ \\
claude-sonnet-5 & human & $0.42$ \\
Llama-3.3-70B & human & $0.44$ \\
\addlinespace
claude-sonnet-5 & gpt-4o-mini & $0.40$ \\
Llama-3.3-70B & gpt-4o-mini & $0.54$ \\
claude-sonnet-5 & Llama-3.3-70B & $0.61$ \\


\bottomrule
\end{tabular}
\label{tab:problematic_inter_annotator_agreement_models}
\end{table}

\begin{figure}[h]
     \begin{center}
           \includegraphics[width=0.35\linewidth]{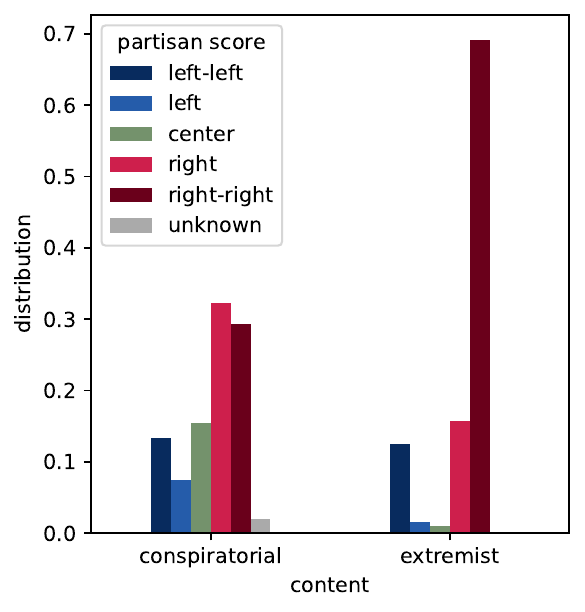}
    \end{center}
    \caption{Distribution of political ideology scores for videos labeled as conspiratorial or extremist. Both categories are predominantly right-leaning, particularly extremist content, which is concentrated in the \textit{right-right} category. This scarcity of left-leaning problematic content motivates our primary focus on \textit{right} and \textit{right-right} sessions in the main analysis.}
   \label{fig:dist_conspextrm}
\end{figure}

\clearpage
\section{Statistical reporting}
\label{appendix:stat_reporting}

\homa{We compare four ranking methods across three reranking models and report every comparison as an explicit pairwise contrast, so the same abbreviations recur throughout the paper. Table~\ref{tab:notation} collects them in one place. All rerankers reorder the same 30 candidates YouTube returns, so every method is evaluated on an identical candidate set, and each contrast is paired within session. Standardized effect sizes are computed on the same quantity as the
inference, so that the two are directly comparable across tables. We report raw mean differences
alongside them throughout, because standardized effects on paired differences can be large even
when the underlying difference is small: a consistent shift of a few percentage points can yield
$d_z > 1$, and the raw value is what carries the substantive interpretation.}

\begin{table*}[h]
\centering
\small
\caption{\homa{Terms and notation used throughout. Model
abbreviations appear in parentheses in the results tables (e.g. bLLM (Sol) denotes the base LLM reranker run on Sol). All rerankers reorder the same 30 candidates YouTube returns, so every method is evaluated on an identical candidate set.}
}
\vspace{4pt}
\begin{tabular}{lp{9cm}}
\toprule
Term & Description \\
\midrule
\multicolumn{2}{l}{\textit{Ranking methods}} \\
$\yt$    & YouTube's own sidebar ordering; the unmodified platform baseline. \\
$\emb$   & Embedding reranker: candidates ordered by cosine similarity to a recency-weighted average of the watch history. \\
$\bllm$  & Base LLM reranker: zero-shot prompt asking the model to rank candidates by fit to the user's history. \\
$\bllmp$ & Base prompt with an explicit instruction to match the user's partisan lean. \\
$\rllm$  & Regularized LLM reranker: base prompt plus social-value constraints (diversify ideologically, demote conspiratorial and extremist material). \\
\addlinespace
\multicolumn{2}{@{}l}{\textit{Pairwise contrasts}} \\
\delab & Session-level mean of the paired difference $A - B$ in the metric of the column. \\
\delby & Personalization gain of the base reranker over the platform baseline. \\
\delry & Net effect of the regularized reranker relative to the platform baseline. \\
\delrb & Cost of regularization: what the social-value constraints give up relative to the unconstrained reranker. \\
\deley & Gain of the embedding reranker over the platform baseline. \\
\addlinespace
\multicolumn{2}{@{}l}{\textit{Models}} \\
4o-mini  & GPT-4o-mini (\texttt{gpt-4o-mini}), the model used in the original submission. \\
Sol      & GPT-5.6 Sol (\texttt{gpt-5.6-sol}), OpenAI's current flagship model. \\
Qwen3.8  & Qwen3.8 (\texttt{Qwen/Qwen3.8-2.4T-A95B}), an open-weight model. \\
3-large  & \texttt{text-embedding-3-large}, the OpenAI embedding model used by the emb reranker. \\
bge-m3   & BAAI \texttt{bge-m3}, the independent open-weight embedding model used to construct the semantic proxy. \\
\addlinespace
\multicolumn{2}{@{}l}{\textit{Evaluation targets}} \\
reference watch $v^*$ & The 11th watched video in a session; the video the participant went on to watch after the 10-video history. \\
semantic proxy $p$  & The candidate most similar to the reference watch $v^*$ under BAAI \texttt{bge-m3} cosine similarity. On sessions where $v^*$ is itself a candidate, $p$ is selected excluding $v^*$.\\
\addlinespace
\multicolumn{2}{@{}l}{\textit{Evaluation metrics}} \\
Semantic AUC    & Semantic relevance: probability the semantic proxy is ranked above a randomly chosen other candidate. \\
Topical AUC     & Topical relevance: probability a candidate matching the history's dominant topic is ranked above one that does not. \\
Congenial AUC   & Ideological alignment: probability a candidate sharing the history's partisan lean is ranked above one that does not. \\
Problematic AUC & Problematic-content exposure: probability a problematic candidate is ranked above a non-problematic one. \\
Diversity entropy & Ideological diversity: Shannon entropy of the ideology-bucket distribution over the top-$k$. \\
\bottomrule
\end{tabular}
\label{tab:notation}
\end{table*}

\clearpage
\section{Supplementary results}

\homa{In this section, we provide supplementary results, including detailed tables and figures corresponding to the presented results in the main text, as well as additional analyses of center- and left-leaning sessions (Figures \ref{fig:center}-\ref{fig:left}).}

\begin{table*}[h]
\centering
\footnotesize
\setlength{\tabcolsep}{2pt}
\caption{\homa{Metrics are pooled across right and right-right sessions ($N=76$ participants) and evaluated against the independent BAAI bge-m3 semantic-proxy construction. Confidence intervals (CIs) are computed using a participant-level cluster bootstrap with 2,000 samples drawn with replacement. Semantic AUC is the probability that the semantic-proxy video is ranked above a randomly selected alternative candidate. Topical AUC is the probability that a same-topic candidate is ranked above an other-topic candidate. Problematic AUC is the probability that a labeled problematic candidate is ranked above a non-problematic candidate. Congenial AUC is the probability that a candidate sharing the session’s broad ideological lean (right or right-right) is ranked above a candidate that does not. Diversity entropy is the Shannon entropy of the ideology-bucket distribution among the top 10 ranked candidates.}}
\vspace{4pt}
\begin{tabular}{lcccccccccc}
\toprule
 & \multicolumn{2}{c}{Semantic AUC} & \multicolumn{2}{c}{Topical AUC} & \multicolumn{2}{c}{Problematic AUC} & \multicolumn{2}{c}{Congenial AUC} & \multicolumn{2}{c}{Diversity entropy} \\
\cmidrule(lr){2-3} \cmidrule(lr){4-5} \cmidrule(lr){6-7} \cmidrule(lr){8-9} \cmidrule(lr){10-11}
Method & Mean & 95\% CI & Mean & 95\% CI & Mean & 95\% CI & Mean & 95\% CI & Mean & 95\% CI \\
\midrule
YT & $0.50$ & $[0.49, 0.51]$ & $0.51$ & $[0.51, 0.52]$ & $0.48$ & $[0.47, 0.49]$ & $0.50$ & $[0.49, 0.50]$ & $1.51$ & $[1.38, 1.61]$ \\
emb (3-large) & $0.69$ & $[0.65, 0.73]$ & $0.63$ & $[0.61, 0.66]$ & $0.68$ & $[0.64, 0.71]$ & $0.69$ & $[0.66, 0.71]$ & $1.29$ & $[1.14, 1.42]$ \\
\addlinespace
bLLM (4o-mini) & $0.58$ & $[0.55, 0.60]$ & $0.57$ & $[0.56, 0.58]$ & $0.61$ & $[0.58, 0.63]$ & $0.59$ & $[0.57, 0.61]$ & $1.35$ & $[1.20, 1.46]$ \\
bLLM (Sol) & $0.62$ & $[0.58, 0.66]$ & $0.60$ & $[0.59, 0.63]$ & $0.50$ & $[0.45, 0.55]$ & $0.64$ & $[0.59, 0.69]$ & $1.35$ & $[1.19, 1.47]$ \\
bLLM (Qwen3.8) & $0.64$ & $[0.60, 0.67]$ & $0.61$ & $[0.59, 0.63]$ & $0.62$ & $[0.58, 0.66]$ & $0.71$ & $[0.66, 0.75]$ & $1.31$ & $[1.16, 1.42]$ \\
\addlinespace
rLLM (4o-mini) & $0.51$ & $[0.50, 0.52]$ & $0.54$ & $[0.54, 0.55]$ & $0.43$ & $[0.42, 0.44]$ & $0.46$ & $[0.45, 0.47]$ & $1.56$ & $[1.45, 1.66]$ \\
rLLM (Sol) & $0.54$ & $[0.51, 0.57]$ & $0.55$ & $[0.54, 0.57]$ & $0.24$ & $[0.21, 0.27]$ & $0.41$ & $[0.37, 0.44]$ & $1.55$ & $[1.38, 1.65]$ \\
rLLM (Qwen3.8) & $0.55$ & $[0.52, 0.57]$ & $0.55$ & $[0.54, 0.57]$ & $0.29$ & $[0.26, 0.32]$ & $0.46$ & $[0.42, 0.50]$ & $1.43$ & $[1.28, 1.54]$ \\
\bottomrule
\end{tabular}
\label{tab:table1_final_right_farright_uid}
\end{table*}

\begin{table*}[h]
\centering
\footnotesize
\setlength{\tabcolsep}{2.75pt}
\caption{\homa{Pairwise method contrasts on session-level paired differences, \textit{right} + \textit{right-right} sessions pooled ($N=76$), for Semantic, Topical, Problematic, and Congenial AUC and Diversity entropy. For each column, \delab{} is the session-level mean difference in that metric between methods A and B, and $d_z$ is the paired Cohen's $d$ (mean difference / SD of differences); both are computed on the observed sessions. $p$-values are from a participant-cluster bootstrap (2{,}000 resamples). A positive $\Delta$ means A ranks the positive class higher than B; for Problematic and Congenial AUC a negative $\Delta$ is the favorable direction (less problematic and less ideologically congenial content upranked), as is a positive $\Delta$ for Diversity entropy. \textbf{Shaded cells, also marked with \textsuperscript{*}, are contrasts not significant at $p<0.05$}. Each contrast is computed on sessions resolved under both methods.}}
\vspace{4pt}
\begin{tabular}{l rrr rrr rrr rrr rrr}
\toprule
 & \multicolumn{3}{c}{Semantic AUC} & \multicolumn{3}{c}{Topical AUC} & \multicolumn{3}{c}{Problematic AUC} & \multicolumn{3}{c}{Congenial AUC} & \multicolumn{3}{c}{Diversity entropy} \\
\cmidrule(lr){2-4} \cmidrule(lr){5-7} \cmidrule(lr){8-10} \cmidrule(lr){11-13} \cmidrule(lr){14-16}
Contrast & $\Delta$ & $d_z$ & $p$ & $\Delta$ & $d_z$ & $p$ & $\Delta$ & $d_z$ & $p$ & $\Delta$ & $d_z$ & $p$ & $\Delta$ & $d_z$ & $p$ \\
\midrule
\multicolumn{16}{l}{\textit{Personalization gain vs. platform baseline}: \delby{}} \\
4o-mini & \phantom{-}$0.08$ & \phantom{-}$0.19$ & ${<}10^{-3}$ & \phantom{-}$0.06$ & \phantom{-}$0.28$ & ${<}10^{-3}$ & \phantom{-}$0.12$ & \phantom{-}$0.48$ & ${<}10^{-3}$ & \phantom{-}$0.09$ & \phantom{-}$0.39$ & ${<}10^{-3}$ & $-0.17$ & $-0.36$ & ${<}10^{-3}$ \\
Sol & \phantom{-}$0.12$ & \phantom{-}$0.30$ & ${<}10^{-3}$ & \phantom{-}$0.09$ & \phantom{-}$0.39$ & ${<}10^{-3}$ & \cellcolor{gray!15}\phantom{-}$0.02^{*}$ & \cellcolor{gray!15}\phantom{-}$0.05$ & \cellcolor{gray!15}$0.44$ & \phantom{-}$0.14$ & \phantom{-}$0.49$ & ${<}10^{-3}$ & $-0.21$ & $-0.43$ & ${<}10^{-3}$ \\
Qwen3.8 & \phantom{-}$0.13$ & \phantom{-}$0.34$ & ${<}10^{-3}$ & \phantom{-}$0.09$ & \phantom{-}$0.41$ & ${<}10^{-3}$ & \phantom{-}$0.14$ & \phantom{-}$0.51$ & ${<}10^{-3}$ & \phantom{-}$0.21$ & \phantom{-}$0.77$ & ${<}10^{-3}$ & $-0.26$ & $-0.51$ & ${<}10^{-3}$ \\
\addlinespace
\multicolumn{16}{l}{\textit{Regularized prompt vs. platform baseline}: \delry{}} \\
4o-mini & \cellcolor{gray!15}\phantom{-}$0.01^{*}$ & \cellcolor{gray!15}\phantom{-}$0.03$ & \cellcolor{gray!15}$0.07$ & \phantom{-}$0.03$ & \phantom{-}$0.15$ & ${<}10^{-3}$ & $-0.06$ & $-0.24$ & ${<}10^{-3}$ & $-0.04$ & $-0.18$ & ${<}10^{-3}$ & \cellcolor{gray!15}\phantom{-}$0.04^{*}$ & \cellcolor{gray!15}\phantom{-}$0.09$ & \cellcolor{gray!15}$0.07$ \\
Sol & \phantom{-}$0.04$ & \phantom{-}$0.10$ & ${<}10^{-3}$ & \phantom{-}$0.04$ & \phantom{-}$0.17$ & ${<}10^{-3}$ & $-0.24$ & $-0.99$ & ${<}10^{-3}$ & $-0.09$ & $-0.36$ & ${<}10^{-3}$ & \cellcolor{gray!15}\phantom{-}$0.05^{*}$ & \cellcolor{gray!15}\phantom{-}$0.10$ & \cellcolor{gray!15}$0.10$ \\
Qwen3.8 & \phantom{-}$0.05$ & \phantom{-}$0.11$ & ${<}10^{-3}$ & \phantom{-}$0.04$ & \phantom{-}$0.16$ & ${<}10^{-3}$ & $-0.19$ & $-0.71$ & ${<}10^{-3}$ & $-0.03$ & $-0.12$ & $0.02$ & \cellcolor{gray!15}$-0.05^{*}$ & \cellcolor{gray!15}$-0.10$ & \cellcolor{gray!15}$0.08$ \\
\addlinespace
\multicolumn{16}{l}{\textit{Cost of regularization (within model)}: \delrb{}} \\
4o-mini & $-0.07$ & $-0.17$ & ${<}10^{-3}$ & $-0.03$ & $-0.13$ & ${<}10^{-3}$ & $-0.18$ & $-0.74$ & ${<}10^{-3}$ & $-0.13$ & $-0.56$ & ${<}10^{-3}$ & \phantom{-}$0.21$ & \phantom{-}$0.47$ & ${<}10^{-3}$ \\
Sol & $-0.08$ & $-0.29$ & ${<}10^{-3}$ & $-0.05$ & $-0.34$ & ${<}10^{-3}$ & $-0.26$ & $-1.28$ & ${<}10^{-3}$ & $-0.23$ & $-1.13$ & ${<}10^{-3}$ & \phantom{-}$0.20$ & \phantom{-}$0.45$ & ${<}10^{-3}$ \\
Qwen3.8 & $-0.09$ & $-0.27$ & ${<}10^{-3}$ & $-0.06$ & $-0.34$ & ${<}10^{-3}$ & $-0.33$ & $-1.48$ & ${<}10^{-3}$ & $-0.24$ & $-1.12$ & ${<}10^{-3}$ & \phantom{-}$0.13$ & \phantom{-}$0.29$ & ${<}10^{-3}$ \\
\addlinespace
\multicolumn{16}{l}{\textit{Embedding-based reranking vs. platform baseline}: \deley{}} \\
3-large & \phantom{-}$0.19$ & \phantom{-}$0.49$ & ${<}10^{-3}$ & \phantom{-}$0.12$ & \phantom{-}$0.54$ & ${<}10^{-3}$ & \phantom{-}$0.19$ & \phantom{-}$0.78$ & ${<}10^{-3}$ & \phantom{-}$0.19$ & \phantom{-}$0.76$ & ${<}10^{-3}$ & $-0.22$ & $-0.46$ & ${<}10^{-3}$ \\
\bottomrule
\end{tabular}
\label{tab:pairwise_final_semtopprob}
\end{table*}

\begin{table*}[t]
\centering
\small
\caption{\homa{Rank-weighted exposure to problematic content ($\psi_{\text{rec}}$, decay $d=0.9$; see Methods) among the top-$k$ recommendations, for the unconstrained (\bllm) and regularized (\rllm) LLM rerankers across three models (4o-mini, Sol, Qwen3.8), pooled over right and right-right sessions. Cohen's $d_z$ reports the paired effect size against YT (Amplification: $\psi_{\text{rec},\text{\bllm}}$  $-$ $\psi_{\text{rec},\text{\yt}}$ ; Mitigation: $\psi_{\text{rec},\text{\rllm}}$  $-$ $\psi_{\text{rec},\text{\yt}}$ ), $^{***}p<0.001$ (paired $t$-test on \texttt{participant}-level means).}}
\setlength{\tabcolsep}{3pt}
\begin{tabular}{lcccccccccccccccc}
\toprule
 & \multicolumn{5}{c}{4o-mini} & \multicolumn{5}{c}{Sol} & \multicolumn{5}{c}{Qwen3.8} & \\
\cmidrule(lr){2-6} \cmidrule(lr){7-11} \cmidrule(lr){12-16}
Top-$k$ & bLLM & $d$ & \rllm{} & $d$ & & bLLM & $d$ & \rllm{} & $d$ & & bLLM & $d$ & \rllm{} & $d$ & & YT \\
 & (Amp.) & & (Mit.) & & & (Amp.) & & (Mit.) & & & (Amp.) & & (Mit.) & & & \\
\midrule
5 & 0.35 & 1.26*** & 0.10 & -1.20*** &  & 0.27 & 0.34 & 0.10 & -1.11*** &  & 0.35 & 1.09*** & 0.15 & -0.70*** &  & 0.20 \\
10 & 0.31 & 1.30*** & 0.15 & -0.90*** &  & 0.25 & 0.34 & 0.10 & -1.26*** &  & 0.32 & 1.17*** & 0.14 & -0.87*** &  & 0.21 \\
15 & 0.29 & 1.24*** & 0.17 & -0.79*** &  & 0.24 & 0.27 & 0.11 & -1.42*** &  & 0.30 & 1.21*** & 0.15 & -1.02*** &  & 0.21 \\
20 & 0.27 & 1.23*** & 0.18 & -0.82*** &  & 0.24 & 0.24 & 0.12 & -1.49*** &  & 0.29 & 1.20*** & 0.15 & -1.16*** &  & 0.21 \\
25 & 0.27 & 1.24*** & 0.18 & -0.84*** &  & 0.23 & 0.23 & 0.13 & -1.49*** &  & 0.28 & 1.20*** & 0.16 & -1.16*** &  & 0.21 \\
30 & 0.27 & 1.24*** & 0.18 & -0.87*** &  & 0.23 & 0.26 & 0.14 & -1.45*** &  & 0.28 & 1.20*** & 0.16 & -1.10*** &  & 0.21 \\
\bottomrule
\end{tabular}
\label{tab:table2_three_models_right_farright}
\end{table*}

\begin{figure*}[tb!]
     \begin{center}
           \includegraphics[width=1\linewidth]{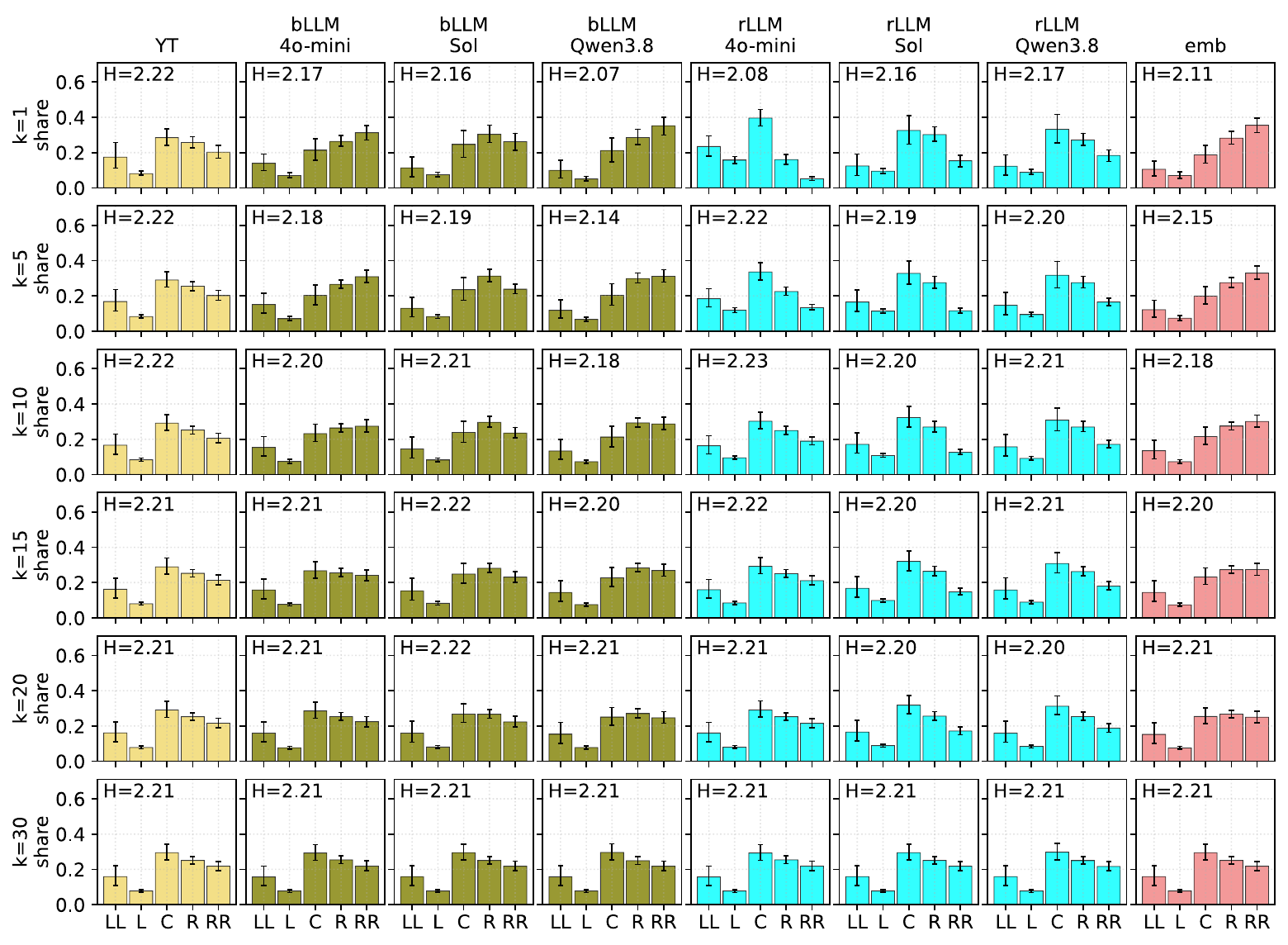}
    \end{center}
    \caption{Distribution of top-$k$ recommended videos across ideological categories for sessions in the \textit{right} category, comparing four ranking methods: YT, \bllm{}, \rllm, and \emb{}. Entropy (in bits) quantifies the diversity of each distribution. At small $k$, \bllm{} and \emb{} show modestly reduced diversity compared with YT and \rllm.}
   \label{fig:top_k_right_ideology_divrsty}
\end{figure*}

\begin{figure*}[tb!]
     \begin{center}
           \includegraphics[width=1\linewidth]{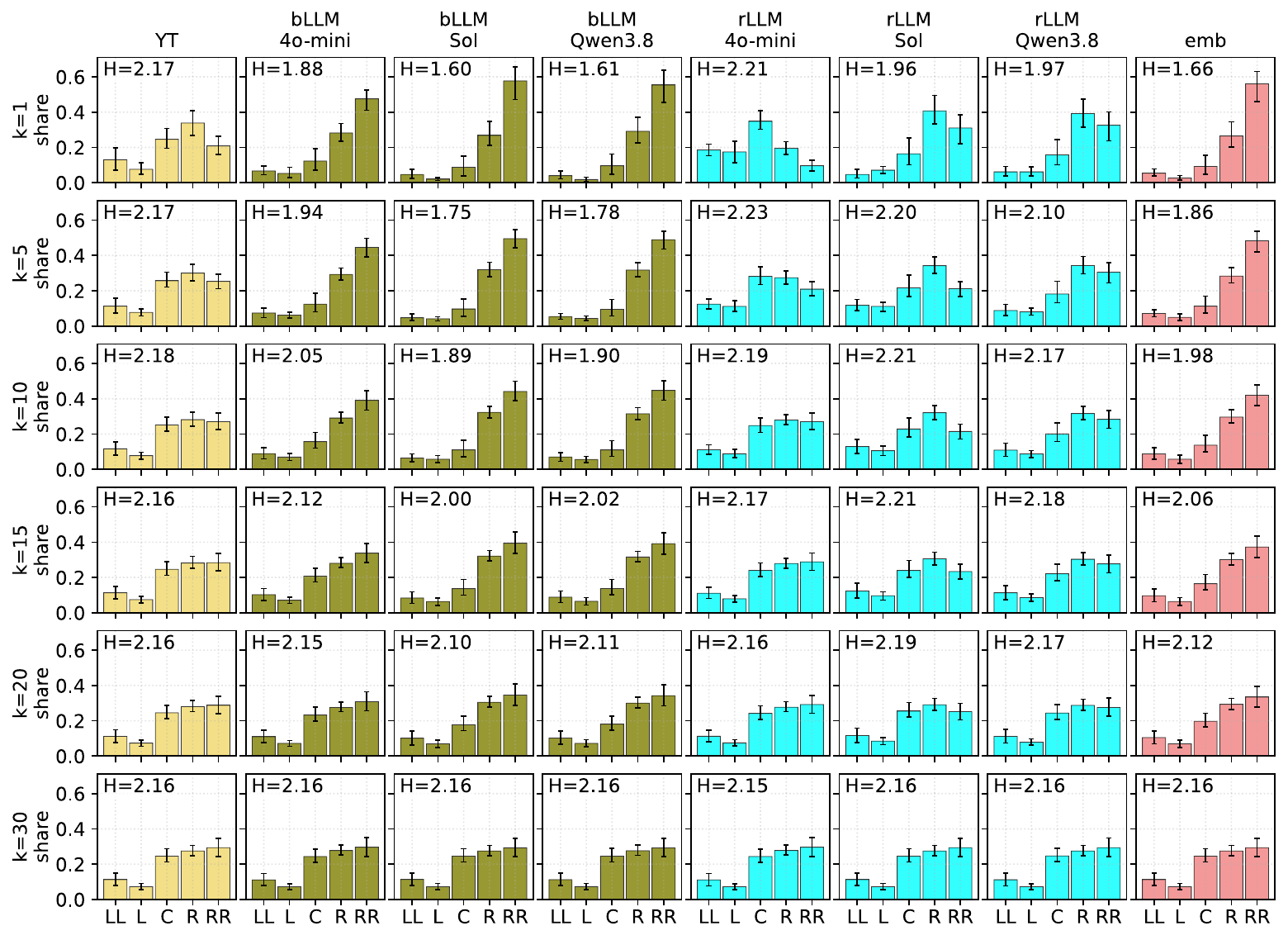}
    \end{center}
    \caption{Distribution of top-$k$ recommended videos across ideological categories for sessions in the \textit{right-right} category, comparing four ranking methods: YT, \bllm{}, \rllm{}, and \emb{}. Entropy (in bits) quantifies the diversity of each distribution. At small $k$, \bllm{} and \emb{} show reduced diversity compared with YT and \rllm.}
   \label{fig:top_k_right_right_ideology_divrsty}
\end{figure*}

\begin{figure*}[h]
     \begin{center}
           \includegraphics[width=\linewidth]{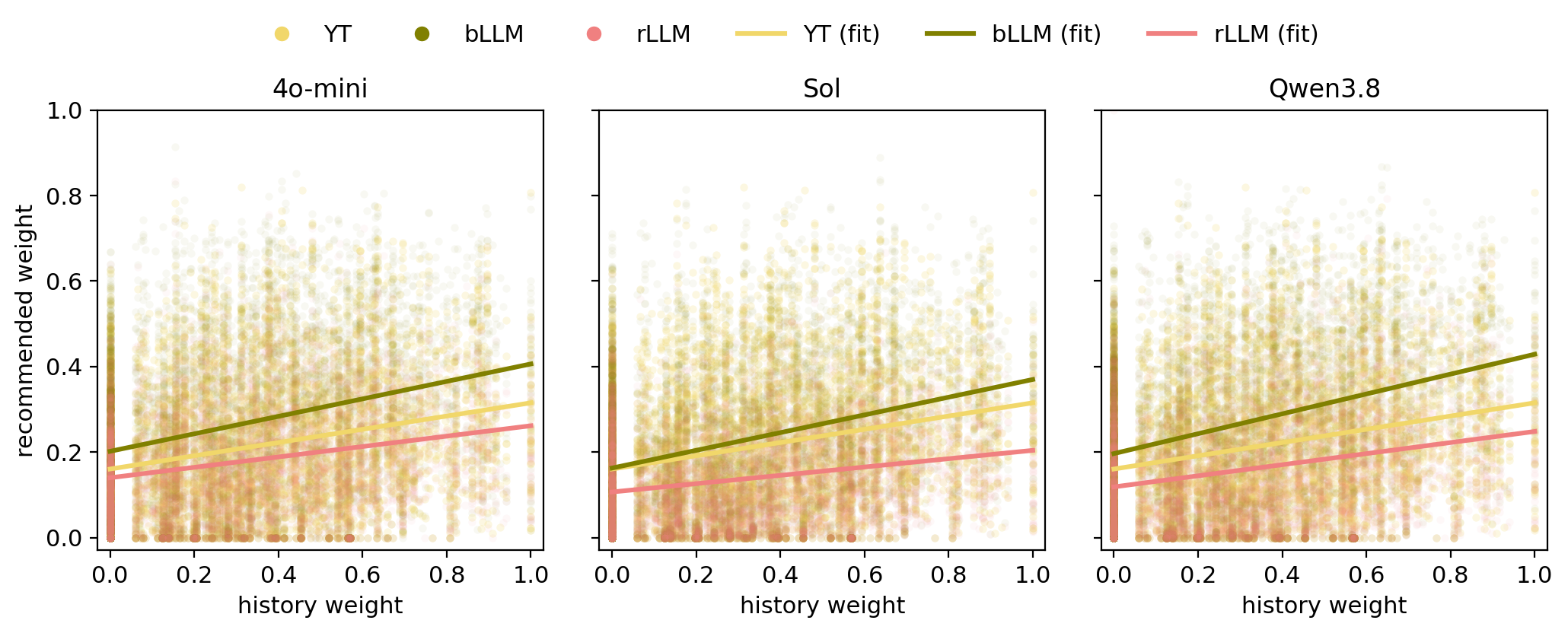}
    \end{center}
\caption{\homa{Rank-weighted average of problematic labels in recommendations ($\psi_{\text{rec}}$) versus recency-weighted average of problematic labels in users' watch history ($\psi_{\text{hist}}$), for YT, bLLM, and \rllm. Each point represents one session; lines show linear fits per method. The positive correlation indicates that more problematic content in the watch history is associated with more problematic content in recommendations. The effect is most pronounced under bLLM and weakest under \rllm.}}
    
   \label{fig:corr_rank_weighted_recency_weighted_prob}
\end{figure*}

\begin{figure*}[h]
     \begin{center} 
           \includegraphics[width=\linewidth] {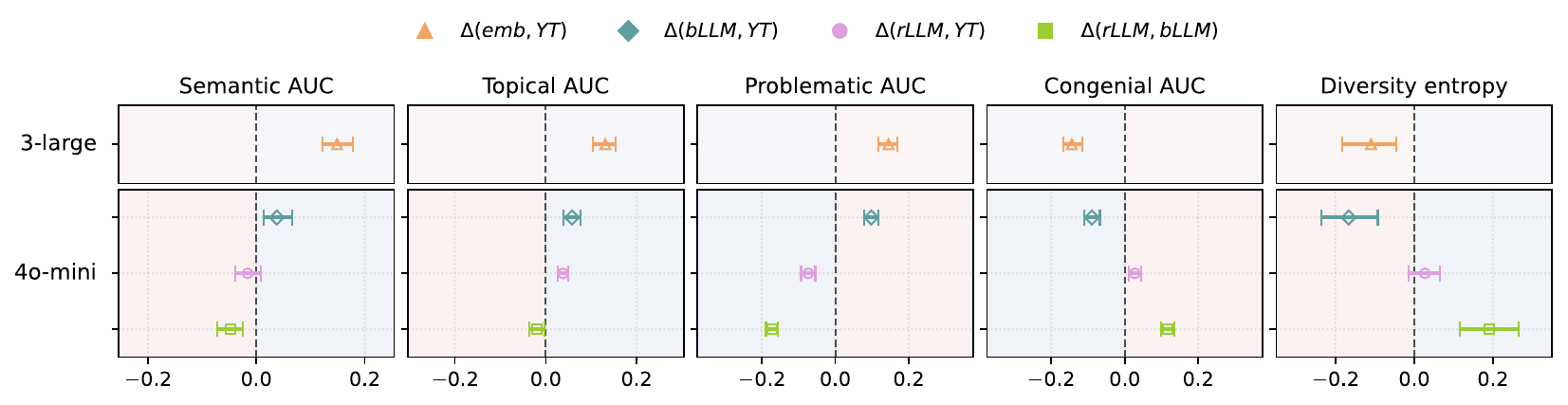}
    \end{center}
    \vspace{-3mm}
    \caption{\homa{Method contrasts against \yt{} across \emph{center} sessions, for Semantic, Topical, Problematic, and Congenial
AUC and Diversity entropy. Each point is a mean session-level
delta with a 95\% CI. Diversity entropy is reported at top-$k{=}10$.}}
   \label{fig:center}
   \vspace{-4mm}
\end{figure*}

\begin{figure*}[h]
     \begin{center} 
           \includegraphics[width=\linewidth] {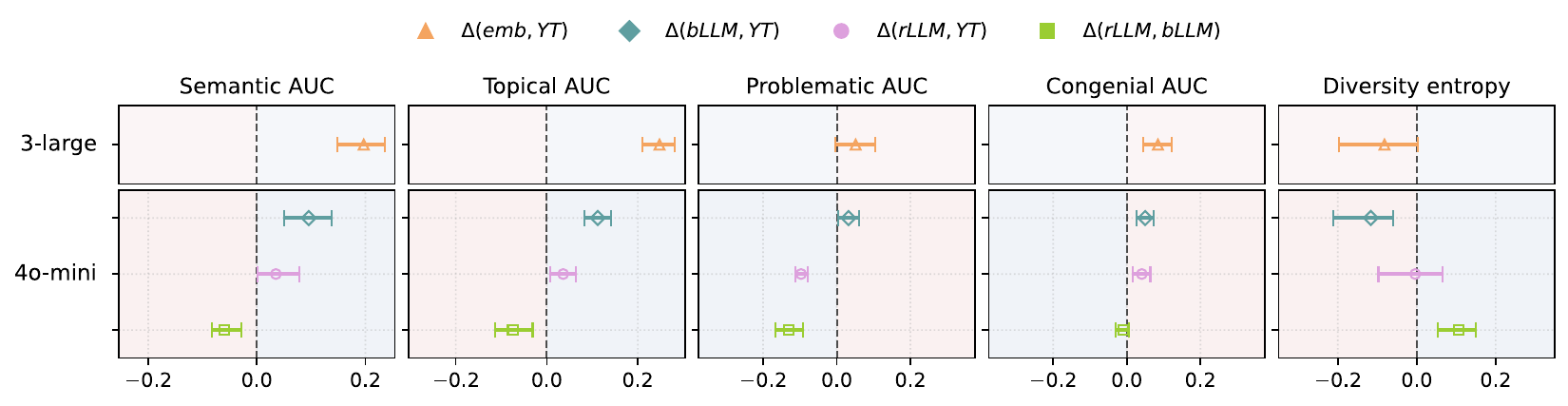}
    \end{center}
    \vspace{-3mm}
        \caption{\homa{Method contrasts against \yt{} across \emph{left} and \emph{left-left} sessions, for Semantic, Topical, Problematic, and Congenial
AUC and Diversity entropy. Each point is a mean session-level
delta with a 95\% CI. Diversity entropy is reported at top-$k{=}10$.}}
   \label{fig:left}
   \vspace{-4mm}
\end{figure*}

\clearpage
\homa{For each session we compute $\psi_{hist}$, the recency-weighted share of problematic videos among the ten most recent watched videos, and $\psi_{rec}$, the rank-weighted share of problematic videos among the top 30 recommendations (Figure \ref{fig:corr_rank_weighted_recency_weighted_prob}). We compute the difference between a reranker's fitted slope and YT's slope on the same sessions, so a positive value means problematic content in the watch history translates more strongly into problematic recommendations than under YT, and a negative value means less strongly. The rerankers change the strength of this relationship in the same direction for all three models. Under bLLM the fitted slope is steeper than under YT (4o-mini $+0.049$ $[0.017, 0.077]$; Sol $+0.052$ $[0.015, 0.105]$; Qwen3.8 $+0.077$ $[0.049, 0.114]$), whereas under rLLM it is flatter (4o-mini $-0.033$ $[-0.056, -0.009]$; Sol $-0.058$ $[-0.094, -0.005]$; Qwen3.8 $-0.025$ $[-0.056, 0.021]$). Average exposure follows the same ordering: the mean $\psi_{rec}$ is 0.27--0.28 under bLLM, 0.21 under YT, and 0.14--0.18 under rLLM. The association between problematic content in the watch history ($\psi_{hist}$) and in the recommendations ($\psi_{rec}$) is positive for every method (YT: $r=0.26$, 95\% CI
$[0.16, 0.32]$; \bllm: $r=0.28$--$0.32$; \rllm: $r=0.22$--$0.25$ across the three models; all 95\% CIs exclude zero).}

\begin{figure*}[tb!]
     \begin{center}
           \includegraphics[width=\linewidth]{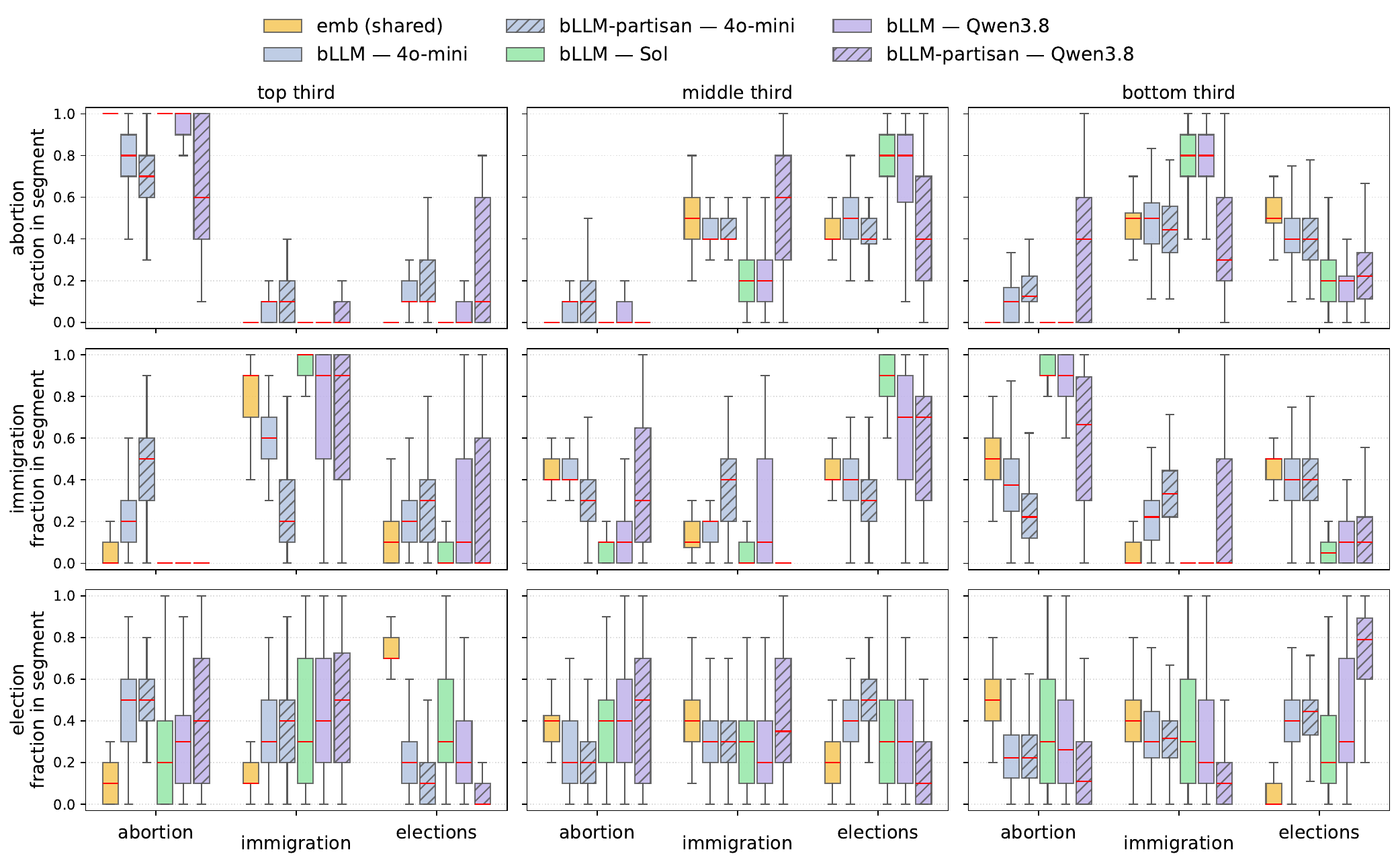}
    \end{center}
    \caption{Comparison of \bllm, \bllmp, and \emb{} in synthetic sessions from Experiment 1 (Figure~\ref{fig:audit_recom_LLMs_synthet_design}A). From a pool of 30 candidates balanced across abortion, immigration, and elections, the boxplots show the fraction of recommended videos on each topic appearing in the top, middle, or bottom third of the ranked list. Each row corresponds to a single watch-history topic. By design, same-topic candidates take the opposing partisan position relative to the history, while other-topic candidates align with the history's partisanship. \emb{} consistently prioritizes topic relevance across all three topics. \bllm{} shows the same prioritization for abortion and immigration but shifts toward partisan alignment for the more diverse topic of elections. \bllmp{} places greater emphasis on partisan content than \bllm, prioritizing partisanship over topic for immigration but not for abortion, where the discussion scope appears narrower.}
   \label{fig:comparison_boxplots_recbLLMYT_recembYT}
\end{figure*}

\begin{figure*}[!tb]
     \begin{center}
           \includegraphics[width=0.7\linewidth]{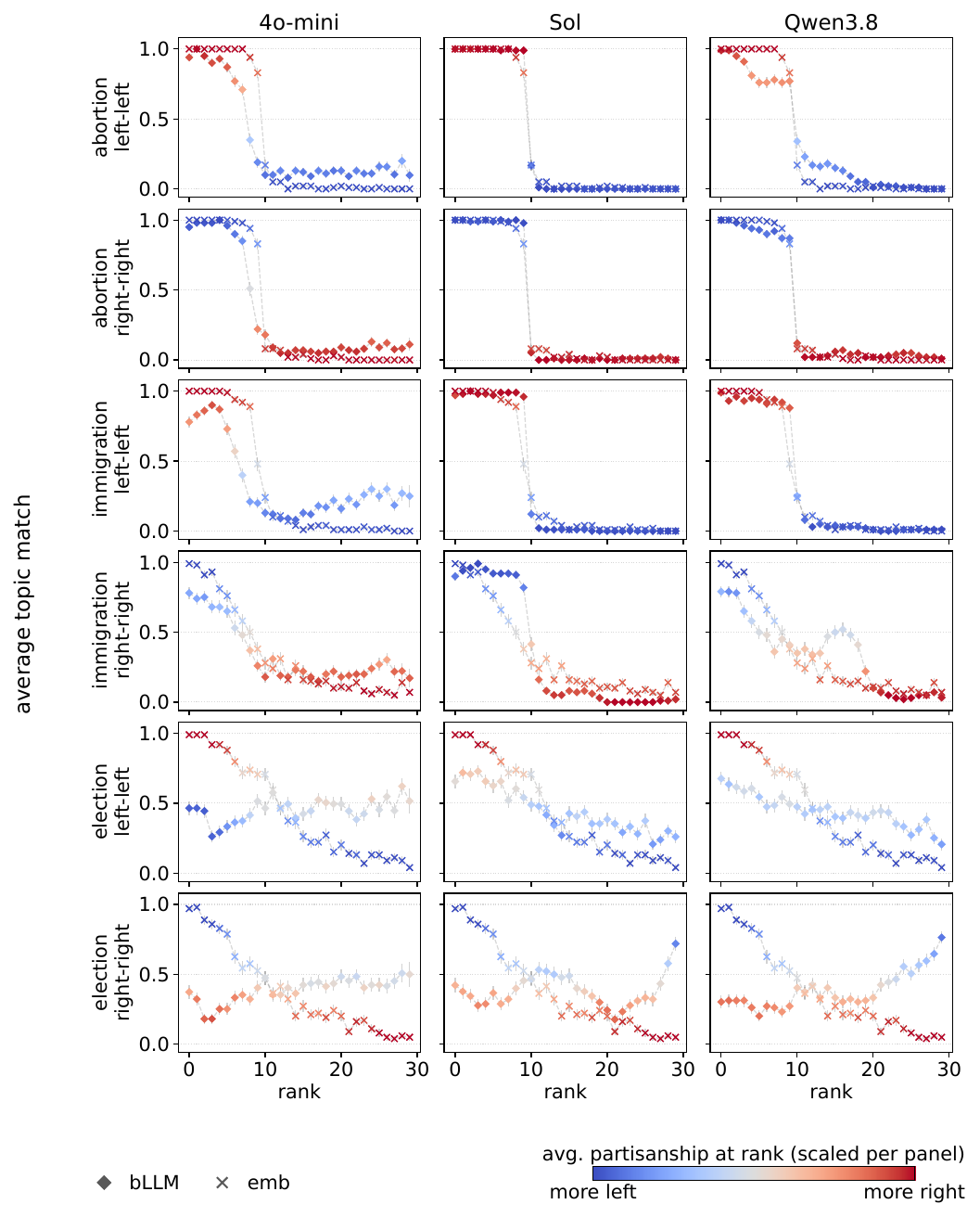}
    \end{center}
    \vspace{-3mm}
    \caption{
Comparison of topical relevance and ideological scores for \bllm{} and \emb{}, using synthetic sessions designed to test whether algorithms prioritize topic relevance or partisan alignment (Experiment 1; Figure~\ref{fig:audit_recom_LLMs_synthet_design}A). Sessions were constructed with extreme-leaning histories (\textit{left-left} and \textit{right-right}) on abortion, immigration, and elections. Marker colors represent average ideological scores across recommendation ranks; error bars show the standard error of topical match across sessions within each trajectory.
    }
   \label{fig:audit_recom_LLMs_synthet_design_result1}
   \vspace{-4mm}
\end{figure*}

\begin{figure*}[tb!]
     \begin{center}
           \includegraphics[width=0.7\linewidth]{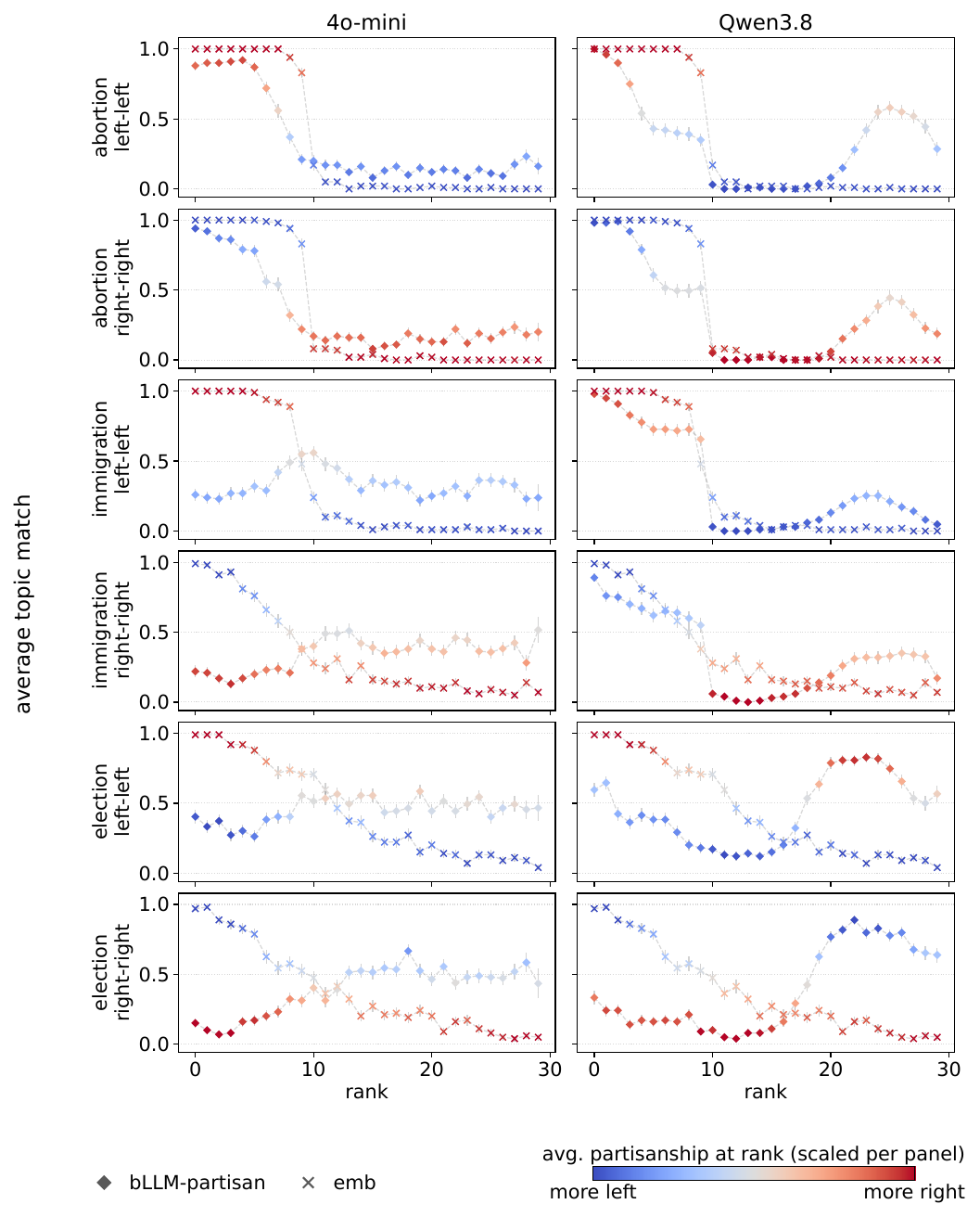}
    \end{center}
    \caption{Comparison of topical relevance and ideological scores for \bllmp{} and \emb{}, using synthetic sessions designed to test whether algorithms prioritize topic relevance or partisan alignment (Experiment 1; Figure~\ref{fig:audit_recom_LLMs_synthet_design}A). Sessions were constructed with extreme-leaning histories (\textit{left-left} and \textit{right-right}) on abortion, immigration, and elections. Compared with \bllm{} (Figure~\ref{fig:audit_recom_LLMs_synthet_design_result1}), \bllmp{} places greater emphasis on partisan content, prioritizing partisanship for immigration and elections but not for abortion, where the discussion scope appears narrower. emb consistently prioritizes topical relevance. Marker colors represent average ideological scores across recommendation ranks; error bars show the standard error of topical match across sessions within each trajectory.}
   \label{fig:audit_recom_LLMs_synthet_design_result2}
\end{figure*}

\clearpage

\homa{In experiment 1, we test the embedding-space explanation directly rather than inferring it from ranking behavior. For each synthetic history we compared the mean cosine
similarity of same-topic/opposite-partisan candidates against other-topic/same-partisan candidates, using the same embeddings that drive emb (Tables~\ref{tab:rq3-embedding-similarity}--\ref{tab:exp1-topic-congenial-ideology-auc-v2}). Same-topic/opposite-partisan candidates are substantially closer to the history
for all three topics: 0.61 vs.\ 0.33 for abortion, 0.54 vs.\ 0.35 for immigration, and 0.54 vs.\ 0.35 for elections. The paired differences are positive and precisely estimated (abortion: $\Delta_{\mathrm{emb}} = 0.28$, 95\% CI [0.276, 0.285], $d_z = 8.28$; immigration: $0.18$, [0.177, 0.191], $d_z = 3.71$; elections: $0.19$, [0.180, 0.192], $d_z = 4.11$). The embedding-geometry
explanation we offer in the main text is therefore supported by direct measurement rather than inferred from ranking outcomes.}

\begin{table*}[h]
\centering
\small
\caption{\homa{Candidate-to-history embedding similarity in the synthetic sessions of Experiment~1, with \textit{left-left} and \textit{right-right} histories pooled. Similarity is cosine distance to the uniform-mean history embedding under \texttt{text-embedding-3-large}, the same representation that drives the emb ranking. ``Same topic, opposite partisan'' candidates share the history's topic but fall in the opposing ideology bucket; ``other topic, same partisan'' candidates share the history's ideology bucket but cover a different topic.}}
\vspace{4pt}
\begin{tabular}{llc}
\toprule
History topic & Candidate type & Mean cosine similarity \\
\midrule
Abortion & other-topic, same-partisan & $0.33$ \\
Abortion & same-topic, opposite-partisan & $0.61$ \\
\addlinespace
Immigration & other-topic, same-partisan & $0.35$ \\
Immigration & same-topic, opposite-partisan & $0.54$ \\
\addlinespace
Elections & other-topic, same-partisan & $0.35$ \\
Elections & same-topic, opposite-partisan & $0.54$ \\
\bottomrule
\end{tabular}
\label{tab:rq3-embedding-similarity}
\end{table*}

\begin{table}[h]
\centering
\small
\caption{\homa{Within-history difference in candidate-to-history similarity, $\Delta_{\mathrm{emb}} = \mathrm{sim}(\text{same-topic, opposite-partisan}) - \mathrm{sim}(\text{other-topic, same-partisan})$, paired by synthetic history. Confidence intervals are session-resampling bootstrap 95\% intervals (2{,}000 resamples); these sessions are synthetic. 
A positive $\Delta_{\mathrm{emb}}$ means same-topic/opposite-partisan candidates sit closer to the history in embedding space than other-topic/same-partisan candidates.}}
\vspace{4pt}
\begin{tabular}{lccc}
\toprule
History topic & $\Delta_{\mathrm{emb}}$ & $d_z$ & 95\% CI (bootstrap) \\
\midrule
Abortion & $0.28$ & $8.28$ & [0.276, 0.285] \\
Immigration & $0.18$ & $3.71$ & [0.177, 0.191] \\
Elections & $0.19$ & $4.11$ & [0.180, 0.192] \\
\bottomrule
\end{tabular}
\label{tab:rq3-delta-emb}
\end{table}

\begin{table*}[h]
\centering
\footnotesize
\setlength{\tabcolsep}{4pt}
\renewcommand{\arraystretch}{1}
\caption{\homa{Topical AUC and Congenial AUC by setting in the synthetic sessions of Experiment~1 (Figure~4), for emb and for each model's bLLM. \textbf{Bold} marks a Congenial AUC above 0.5, where the ranking favors ideologically congenial candidates.}}
\vspace{4pt}
\begin{tabular}{lcccccccc}
\toprule
 & \multicolumn{2}{c}{emb (3-large)} & \multicolumn{2}{c}{bLLM (4o-mini)} & \multicolumn{2}{c}{bLLM (Sol)} & \multicolumn{2}{c}{bLLM (Qwen3.8)} \\
\cmidrule(lr){2-3} \cmidrule(lr){4-5} \cmidrule(lr){6-7} \cmidrule(lr){8-9}
Setting & Topic & Congenial & Topic & Congenial & Topic & Congenial & Topic & Congenial \\
\midrule
\multicolumn{9}{l}{Abortion} \\
\makecell{\textit{left}-\textit{left}} & \makecell{$1.00$\\{\scriptsize $[1.00, 1.00]$}} & \makecell{$0.31$\\{\scriptsize $[0.30, 0.33]$}} & \makecell{$0.85$\\{\scriptsize $[0.84, 0.87]$}} & \makecell{$0.35$\\{\scriptsize $[0.33, 0.37]$}} & \makecell{$1.00$\\{\scriptsize $[1.00, 1.00]$}} & \makecell{$0.35$\\{\scriptsize $[0.33, 0.37]$}} & \makecell{$0.93$\\{\scriptsize $[0.91, 0.95]$}} & \makecell{$0.38$\\{\scriptsize $[0.36, 0.40]$}} \\
\makecell{\textit{right}-\textit{right}} & \makecell{$0.99$\\{\scriptsize $[0.99, 1.00]$}} & \makecell{$0.32$\\{\scriptsize $[0.30, 0.33]$}} & \makecell{$0.91$\\{\scriptsize $[0.89, 0.92]$}} & \makecell{$0.29$\\{\scriptsize $[0.27, 0.31]$}} & \makecell{$0.99$\\{\scriptsize $[0.98, 1.00]$}} & \makecell{$0.32$\\{\scriptsize $[0.31, 0.34]$}} & \makecell{$0.96$\\{\scriptsize $[0.93, 0.98]$}} & \makecell{$0.33$\\{\scriptsize $[0.32, 0.35]$}} \\
\addlinespace[0.8em]
\multicolumn{9}{l}{Immigration} \\
\makecell{\textit{left}-\textit{left}} & \makecell{$0.98$\\{\scriptsize $[0.97, 0.98]$}} & \makecell{$0.33$\\{\scriptsize $[0.32, 0.35]$}} & \makecell{$0.73$\\{\scriptsize $[0.70, 0.76]$}} & \makecell{$0.40$\\{\scriptsize $[0.38, 0.43]$}} & \makecell{$0.99$\\{\scriptsize $[0.98, 1.00]$}} & \makecell{$0.27$\\{\scriptsize $[0.25, 0.28]$}} & \makecell{$0.97$\\{\scriptsize $[0.95, 0.99]$}} & \makecell{$0.28$\\{\scriptsize $[0.26, 0.29]$}} \\
\makecell{\textit{right}-\textit{right}} & \makecell{$0.87$\\{\scriptsize $[0.85, 0.88]$}} & \makecell{$0.39$\\{\scriptsize $[0.37, 0.41]$}} & \makecell{$0.71$\\{\scriptsize $[0.68, 0.74]$}} & \makecell{$0.42$\\{\scriptsize $[0.39, 0.44]$}} & \makecell{$0.96$\\{\scriptsize $[0.95, 0.97]$}} & \makecell{$0.32$\\{\scriptsize $[0.30, 0.34]$}} & \makecell{$0.77$\\{\scriptsize $[0.74, 0.80]$}} & \makecell{$0.42$\\{\scriptsize $[0.40, 0.45]$}} \\
\addlinespace[0.8em]
\multicolumn{9}{l}{Elections} \\
\makecell{\textit{left}-\textit{left}} & \makecell{$0.92$\\{\scriptsize $[0.91, 0.93]$}} & \makecell{$0.37$\\{\scriptsize $[0.35, 0.38]$}} & \makecell{$0.35$\\{\scriptsize $[0.33, 0.38]$}} & \makecell{$\mathbf{0.62}$\\{\scriptsize $\mathbf{[0.60, 0.65]}$}} & \makecell{$0.63$\\{\scriptsize $[0.58, 0.67]$}} & \makecell{$0.46$\\{\scriptsize $[0.43, 0.49]$}} & \makecell{$0.54$\\{\scriptsize $[0.50, 0.59]$}} & \makecell{$\mathbf{0.51}$\\{\scriptsize $\mathbf{[0.48, 0.54]}$}} \\
\makecell{\textit{right}-\textit{right}} & \makecell{$0.88$\\{\scriptsize $[0.86, 0.89]$}} & \makecell{$0.40$\\{\scriptsize $[0.38, 0.42]$}} & \makecell{$0.39$\\{\scriptsize $[0.36, 0.41]$}} & \makecell{$\mathbf{0.63}$\\{\scriptsize $\mathbf{[0.60, 0.66]}$}} & \makecell{$0.46$\\{\scriptsize $[0.42, 0.50]$}} & \makecell{$\mathbf{0.59}$\\{\scriptsize $\mathbf{[0.56, 0.62]}$}} & \makecell{$0.33$\\{\scriptsize $[0.29, 0.37]$}} & \makecell{$\mathbf{0.66}$\\{\scriptsize $\mathbf{[0.63, 0.69]}$}} \\
\bottomrule
\end{tabular}
\label{tab:exp1-topic-congenial-ideology-auc-v2}
\end{table*}

\begin{figure*}[h]
     \begin{center}
           \includegraphics[width=1\linewidth]{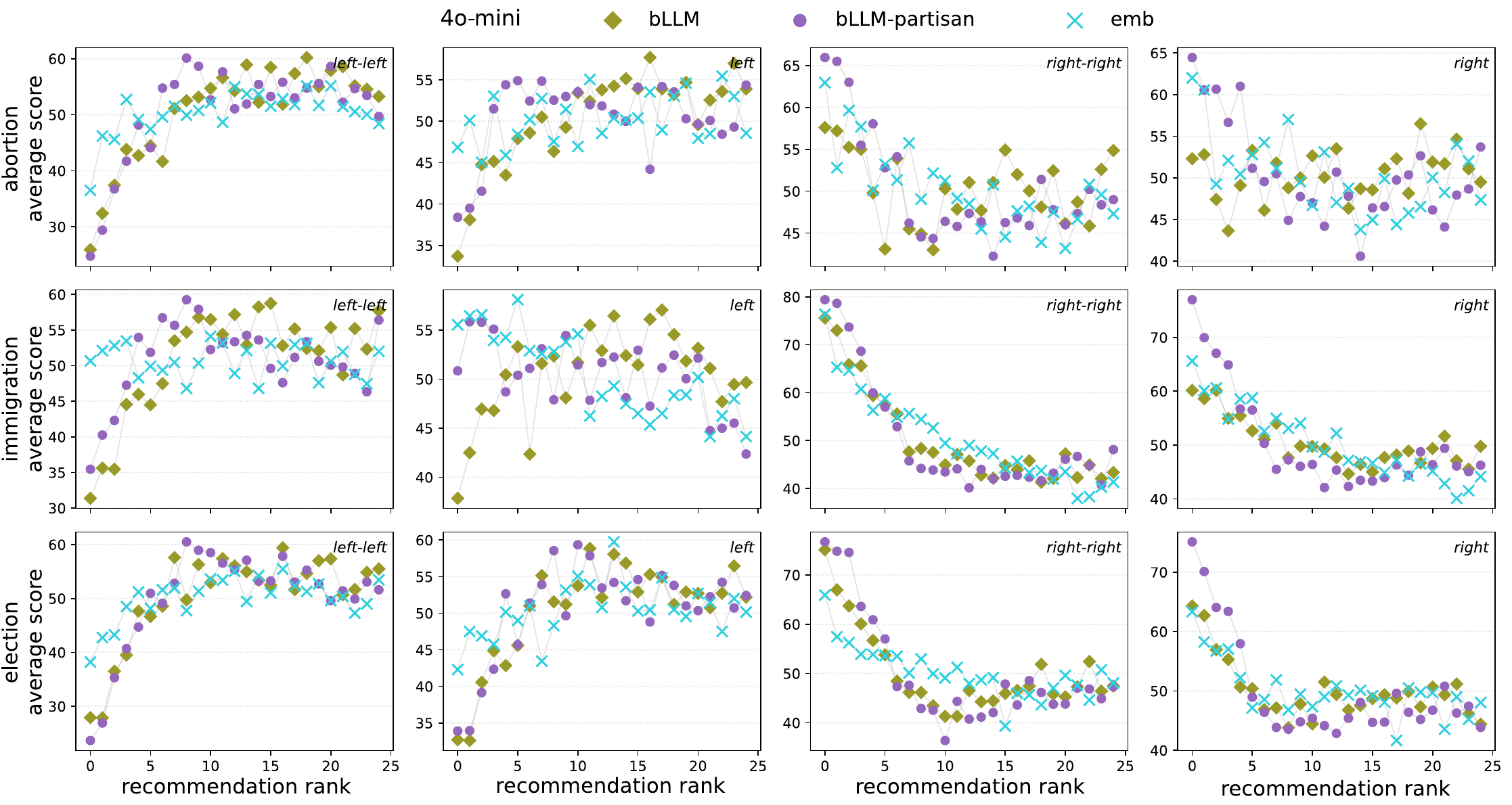}
    \end{center}
    \caption{Average partisanship scores across recommendation ranks for \bllm, \bllm-partisan, and \emb{} via 4o-mini, using synthetic sessions (Experiment~2; Figure~\ref{fig:audit_recom_LLMs_synthet_design}B). Watch histories cover a single topic with consistent partisan content (\textit{left-left}, \textit{left}, \textit{right}, or \textit{right-right}; columns), while candidate videos span the full partisan spectrum on the same topic. Rows correspond to abortion (top), immigration (middle), and elections (bottom). Recommendation scores track the partisanship of the watch history: extreme histories yield averages closer to the extremes, while less extreme histories track their partisan position. \bllm{} effectively identifies and prioritizes partisan alignment, with \bllm-partisan modestly amplifying this effect.}
   \label{fig:audit_recom_LLMs_synthet_design_result3_mini}
\end{figure*}

\clearpage
\homa{In Experiment 2, Qwen3.8 orders the candidate list by partisanship more steeply compared with 4o-mini. For right-right abortion histories, the average partisanship score of its top recommendation is near 70 and falls to roughly 30 by rank 20, whereas 4o-mini declines only from 58 to 46 over the same range; the same steepening appears for left-left histories, where Qwen3.8's averages climb to about 70. The partisan prompt amplifies this alignment in both directions for Qwen3.8 as well, pushing early ranks lower for left histories and higher for right histories, which reproduces the \bllm-partisan effect on a second model. By contrast, \emb{} remains comparatively flat across ranks in every panel (Figs. \ref{fig:audit_recom_LLMs_synthet_design_result3_mini} and \ref{fig:audit_recom_LLMs_synthet_design_result3_qwen}).}

\begin{figure*}[h]
     \begin{center}
           \includegraphics[width=1\linewidth]{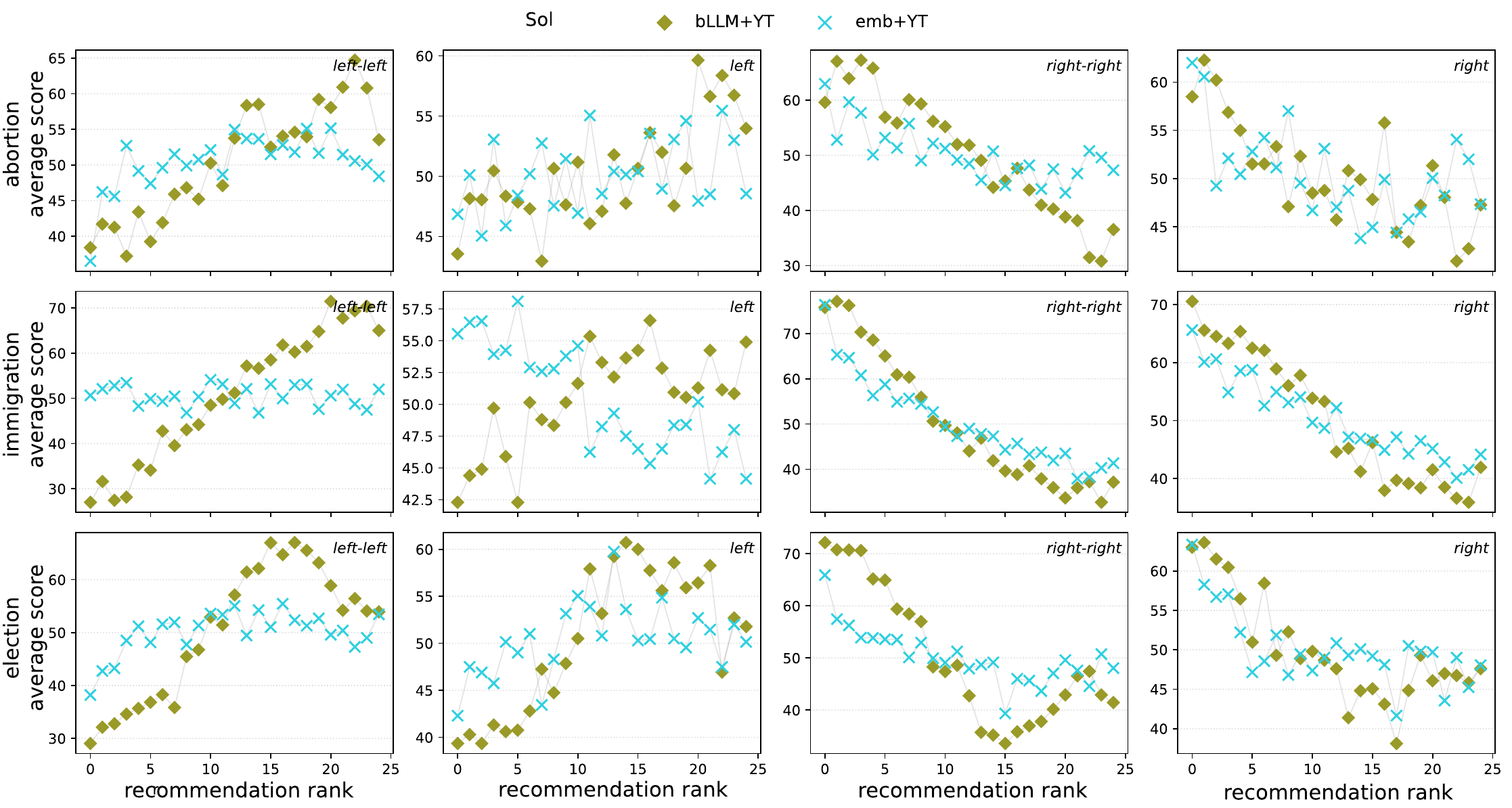}
    \end{center}
    \caption{Average partisanship scores across recommendation ranks for \bllm{} and \emb{} via Sol, using synthetic sessions (Experiment~2; Figure~\ref{fig:audit_recom_LLMs_synthet_design}B). Watch histories cover a single topic with consistent partisan content (\textit{left-left}, \textit{left}, \textit{right}, or \textit{right-right}; columns), while candidate videos span the full partisan spectrum on the same topic. Rows correspond to abortion (top), immigration (middle), and elections (bottom). Recommendation scores track the partisanship of the watch history: extreme histories yield averages closer to the extremes, while less extreme histories track their partisan position. \bllm{} effectively identifies and prioritizes partisan alignment.}
   \label{fig:audit_recom_LLMs_synthet_design_result3_sol}
\end{figure*}

\begin{figure*}[h]
     \begin{center}
           \includegraphics[width=1\linewidth]{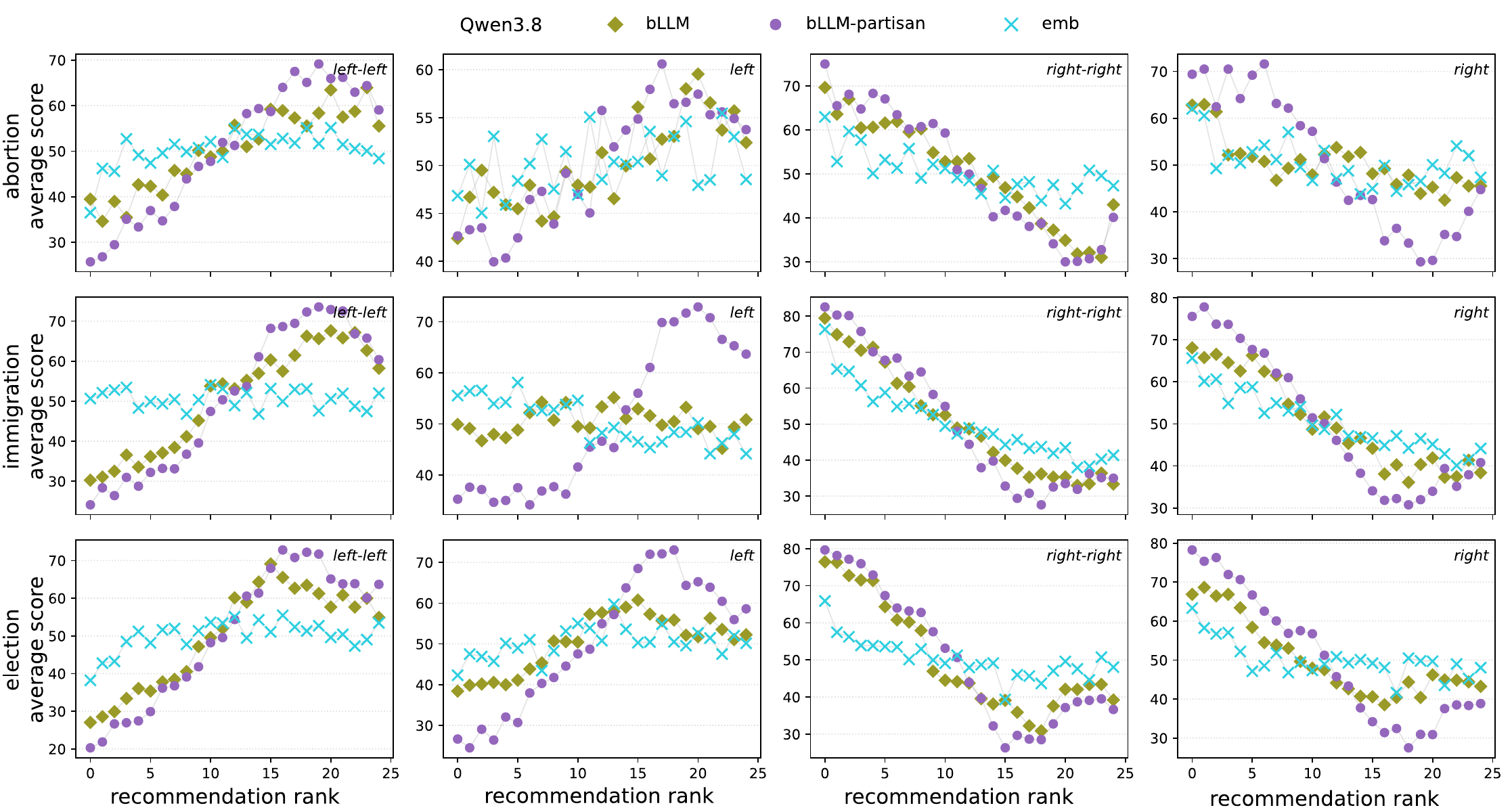}
    \end{center}
    \caption{Average partisanship scores across recommendation ranks for \bllm, \bllm-partisan, and \emb{} via Qwen3.8, using synthetic sessions (Experiment~2; Figure~\ref{fig:audit_recom_LLMs_synthet_design}B). Watch histories cover a single topic with consistent partisan content (\textit{left-left}, \textit{left}, \textit{right}, or \textit{right-right}; columns), while candidate videos span the full partisan spectrum on the same topic. Rows correspond to abortion (top), immigration (middle), and elections (bottom). Recommendation scores track the partisanship of the watch history: extreme histories yield averages closer to the extremes, while less extreme histories track their partisan position. \bllm{} effectively identifies and prioritizes partisan alignment, with \bllm-partisan modestly amplifying this effect.}
   \label{fig:audit_recom_LLMs_synthet_design_result3_qwen}
\end{figure*}

\clearpage
\section{Robustness analysis}
\label{appendix:robust_analysis}

\noindent\textbf{Sensitivity to the gap between watches} 

\begin{figure}[b]
    \centering
    \includegraphics[width=0.8\linewidth]{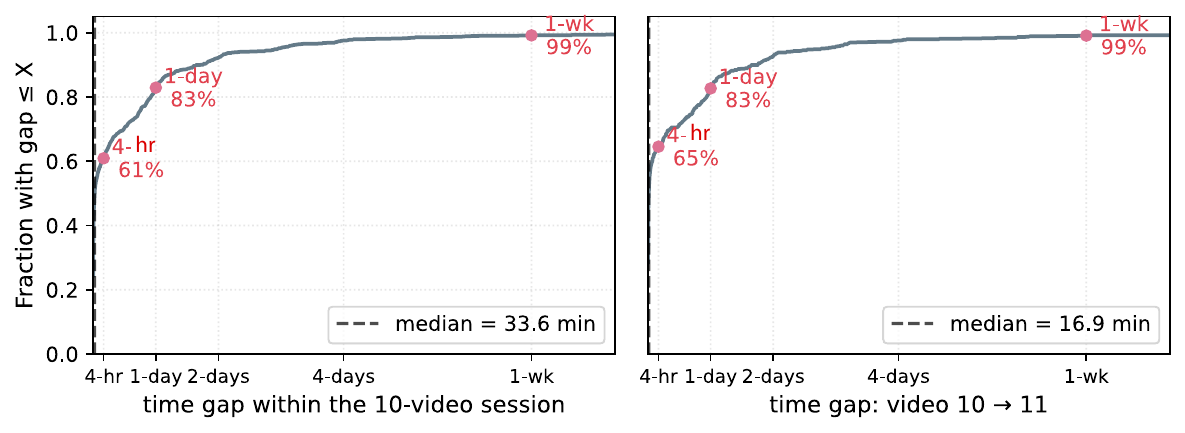}
    \vspace{-0.5cm}
    \caption{
    \homa{Distribution of within-session time gaps across all 9,848 sessions.
(a)~The nine consecutive gaps among the 10 history videos, pooled across sessions
(median 33.6 minutes). (b)~The gap between the 10th history video and the
reference watch (median 16.9 minutes).}}
    \label{fig:gap_dist_all}

\end{figure}

\begin{figure}[b]
    \centering
    \includegraphics[width=0.8\linewidth]{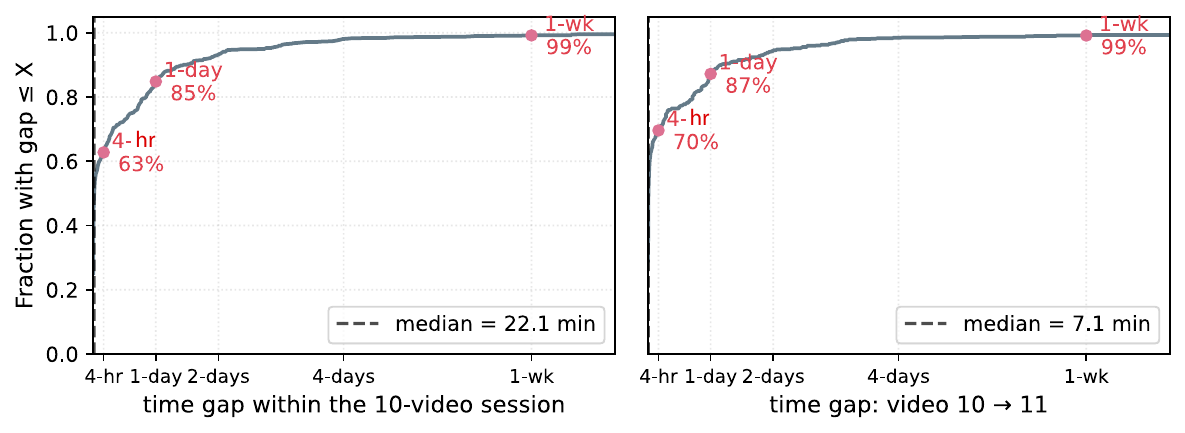}
    \vspace{-0.5cm}
    \caption{\homa{Distribution of within-session time gaps across the pooled \textit{right}
and \textit{right-right} sessions used in the main analysis ($n = 5{,}965$). (a)~the nine consecutive gaps among
the 10 history videos, pooled across sessions (median 22.1 minutes); (b)~the gap between the 10th history
video and the reference watch (median 7.1 minutes).}}
    \label{fig:gap_dist_right}
\end{figure}

\homa{Sessions in the main text (\textbf{RQ1} and \textbf{RQ2}) are built from consecutive watches, but consecutive does not mean
immediate: a participant may return to YouTube hours or days after the tenth
history video, by which point the reconstructed candidate set is a weaker guide
to what they were choosing among. 
Here we report two complementary analyses. The first examines the spacing among the 10 videos in the history window; the second asks how soon the reference watch follows the end of that history.
The nine consecutive gaps among the 10 history videos have a median of 33.6
minutes over the full dataset and 22.1 minutes for the pooled \textit{right} and
\textit{right-right} sessions used in the main analysis. The gap from the end of
the 10th history video to the start of the reference watch---the delay between
the last video the model conditions on and the participant's next observed
YouTube watch---has a median of 16.9 minutes over the full dataset and 7.1
minutes for the same pooled strata. Both distributions are shown in
Figures~\ref{fig:gap_dist_all} and~\ref{fig:gap_dist_right}.
The reference watch follows the previous video within one minute in 40.0\% of
sessions, within 15 minutes in 49.1\%, within one hour in 58.1\%, within one
day in 82.7\%, and within one week in 99.1\% (42.6\%, 53.1\%, 62.7\%, 87.2\%, and
99.1\% respectively on the \textit{right} and \textit{right-right} sessions). }

\begin{figure*}[h]
     \begin{center} 
           \includegraphics[width=0.9\linewidth] {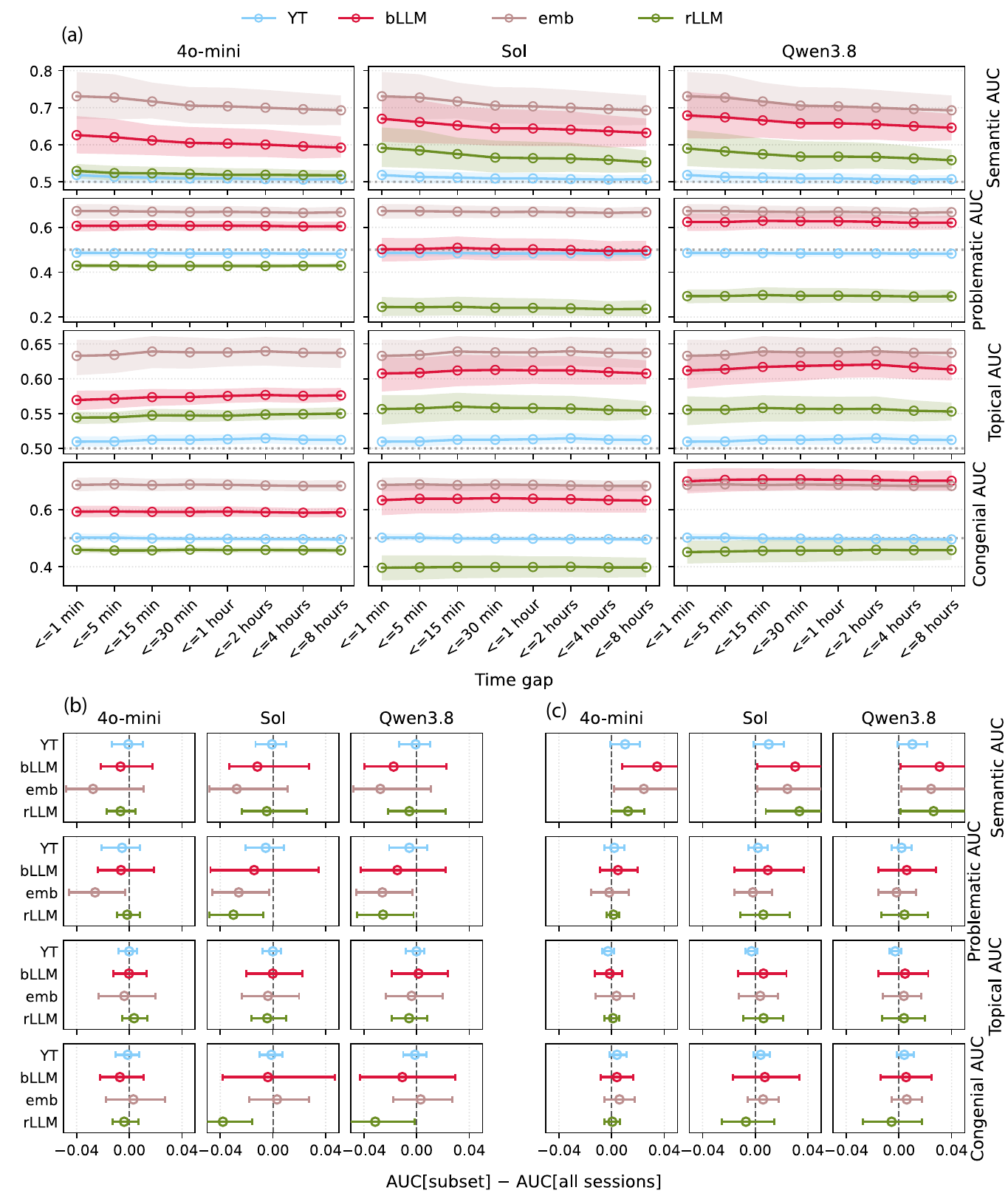}
    \end{center}
    \vspace{-3mm}
    \caption{\homa{Sensitivity of the main comparison to the gap between the end of the history window and the next observed watch. (a) Semantic, Problematic, Topical, and Congenial AUC for each method, recomputed within cumulative gap thresholds from $\leq$1 minute to $\leq$8 hours, for each reranking model. Shaded bands are participant-cluster bootstrap 95\% CIs. (b, c) Difference between the AUCs on the $\leq$10 minute subset and on all sessions, per method and metric, where the threshold applies to the median gap between consecutive history videos in panel (b) and to the gap between the last history video and the reference watch in panel (c).}}
   \label{fig:gap_sensisivity}
   \vspace{-4mm}
\end{figure*}

\homa{To test whether our findings depend on sessions that are farther apart in time, we recomputed every metric within increasingly strict cumulative gap thresholds, down to the $\leq 1$ minute bucket used in Figure~\ref{fig:gap_sensisivity}. The ordering of methods remains the same across all thresholds (panel a), and all pairwise contrasts retain their direction and significance (panels b and c). Semantic AUC is the only metric sensitive to this gap: it increases slightly as the gap between the last video in the history and the subsequent watch becomes shorter. This increase is small and in the expected direction: a watch that follows shortly after the history is better predicted by that history. The magnitude is also small relative to the effects we focus on in the paper: even the largest shift (0.035) is well below \delby{} of 0.09--0.12 and \deley{} of 0.17--0.19. Topical, Problematic, and Congenial AUC, none of which depends on the subsequent target video, remain essentially unchanged as the gap narrows. Together, these results show a small increase in Semantic AUC when the history and subsequent watch are closer in time, while the main comparisons between ranking methods remain stable.}

\clearpage
\noindent\textbf{Sensitivity of the semantic proxy to its construction}

\begin{figure}[t]
    \centering
    \includegraphics[width=0.8\linewidth]{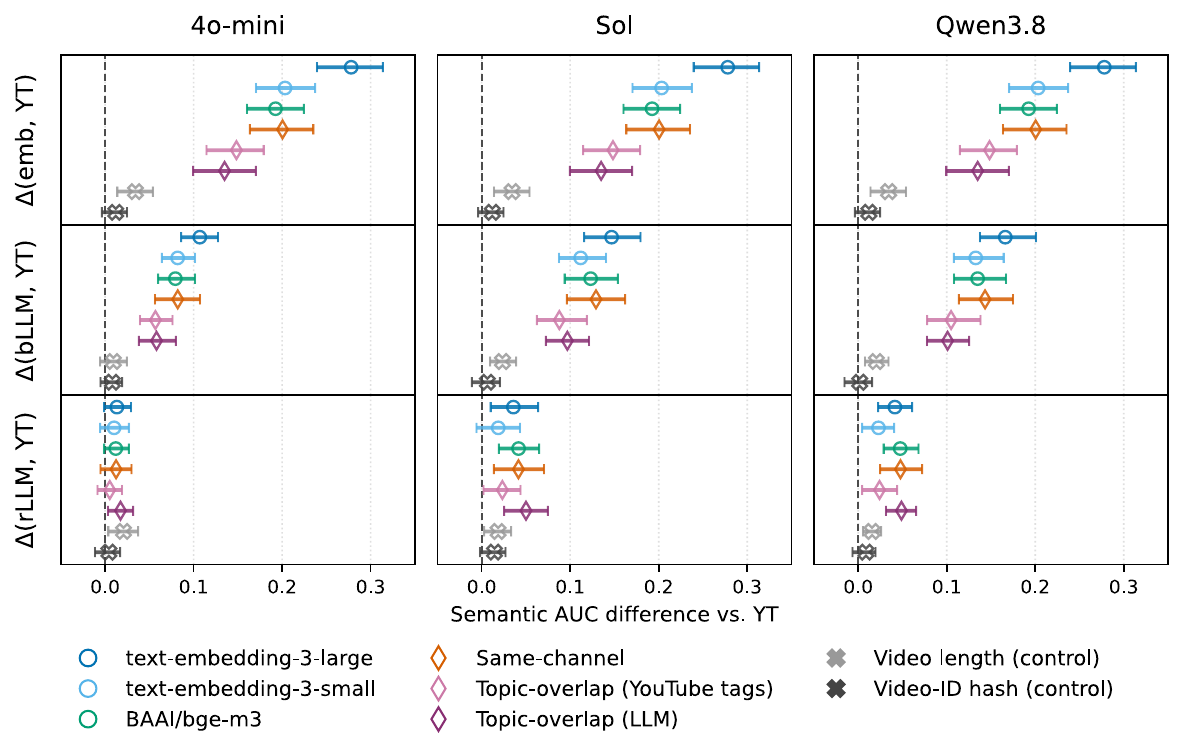}
    \caption{
    Sensitivity of the method contrasts to the construction of the
evaluation target. 
Each point is the mean session-level difference in Semantic AUC against YT, for
\deley{}, \delby{}, and \delry{}, under eight target constructions and for each of
the three LLMs. Whiskers are participant-cluster bootstrap 95\% CIs (2{,}000
resamples). The dashed line marks no difference from YT. The two grey series are
content-blind controls, under which no method separates from the platform
baseline.}
    \label{fig:proxy_sensisivity}
\vspace{-0.5cm}
\end{figure}

\homa{We repeated the full Semantic AUC
analysis across eight constructions of the evaluation target (Figure~\ref{fig:proxy_sensisivity}): some varying the representation used for matching, others replacing embedding similarity with a different selection criterion entirely.}

\homa{Three vary the embedding. Alongside the original OpenAI
\texttt{text-embedding-3-large} we tested \texttt{text-embedding-3-small}, from the
same family, and BAAI \texttt{bge-m3}, an independently developed open-weight
model that we have adopted for all reported results. The last comparison matters
most, since it removes any dependence of the target on the OpenAI family. 
Under \texttt{bge-m3}, AUC is 0.50 for YT, 0.57 for bLLM,
0.69 for emb, and 0.51 for rLLM (Table~\ref{tab:candidate_subset_table1_uid}).} 

\homa{Two are content-blind negative controls: video-length matching and a hash of the video ID. Under both, the differences between methods disappear and all four fall to approximately 0.5. Neither duration nor an identifier carries information about what a video is about, so a target built on them should not preserve any personalization signal. }

\homa{Three constructions replace embedding similarity against the reference watch: same-channel matching (exact match on the YouTube \texttt{channelId}), topic-tag overlap using the YouTube Data API \texttt{topicCategories} field, and overlap on our LLM-assigned topic labels, with both topic rules scoring candidates by the Jaccard similarity of their label sets with the reference watch. None of the three singles out a candidate in most sessions: 79.5\% of sessions contain no same-channel candidate at all, and ties arise in 87\% of sessions under YouTube tags and 71\% under LLM labels. We report three resolutions below.}

\homa{\textbf{Random tie-break.} Figure~\ref{fig:proxy_sensisivity_rand} resolves ties by a uniform random draw---among the tied candidates, or from the whole pool when none qualifies---so no embedding enters these three rules and no session is dropped, leaving the sample identical to the main analysis. The ordering and the signs survive:
$\Delta(\text{emb},\text{YT})$ is 0.05 (same channel), 0.04 (YouTube tags) and 0.06 (LLM labels), and $\Delta(\text{bLLM},\text{YT})$ remains positive and significant for all three LLMs under all three rules (0.01--0.06). The magnitudes fall to roughly a quarter to a half of their values under the embedding tie-break, and four of the nine $\Delta(\text{rLLM},\text{YT})$ contrasts lose significance. Metadata agreement alone therefore reproduces
the ordering $\text{emb} > \text{bLLM} > \text{rLLM} > 0$ and stays above
the content-blind controls, which give $\Delta(\text{emb},\text{YT}) = 0.03$
for video length and 0.01 for the video-ID hash, but identifies a related
candidate far more weakly than semantic similarity, which gives
$\Delta(\text{emb},\text{YT}) = 0.19$--$0.28$.}

\begin{figure}[h]
    \centering
    \includegraphics[width=0.8\linewidth]{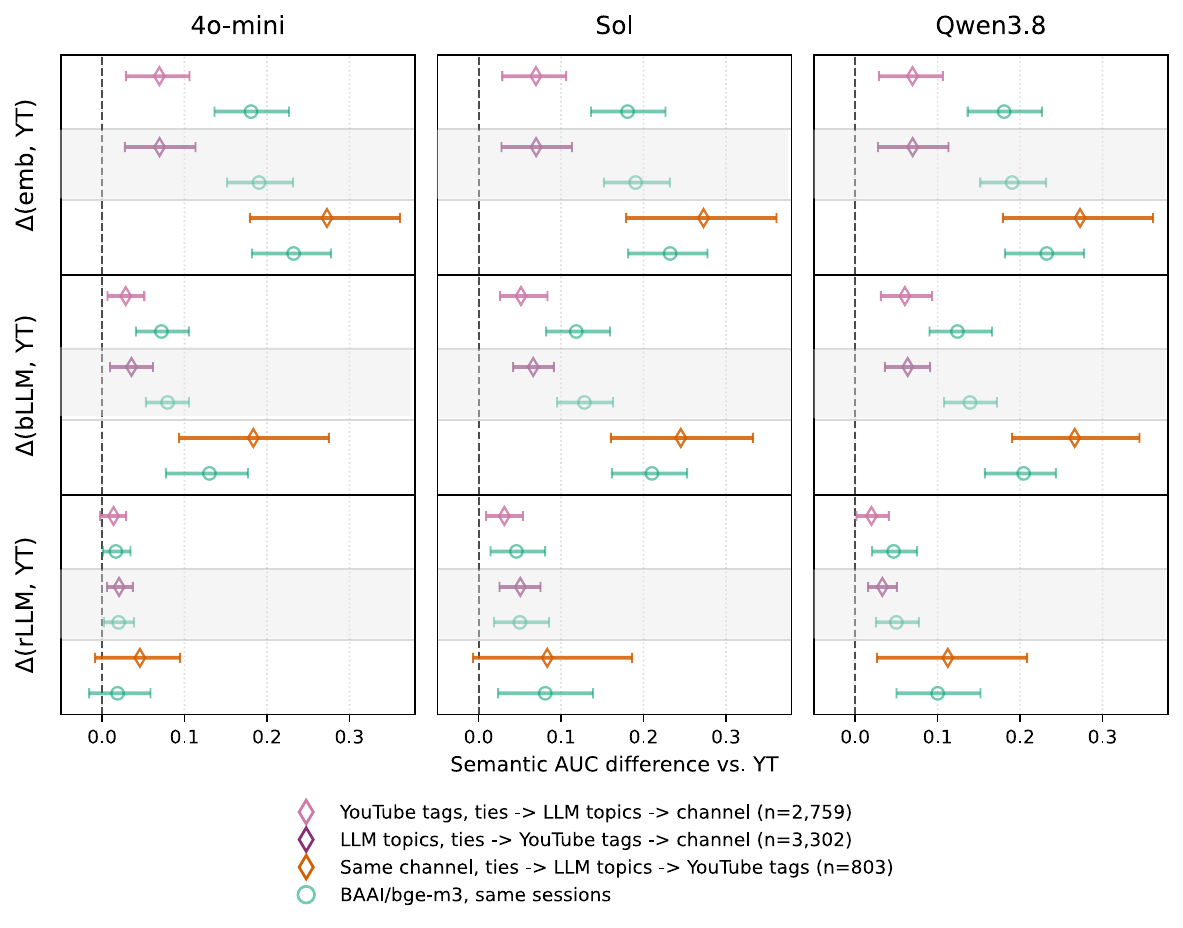}
    \caption{Sensitivity of the method contrasts to label-based evaluation targets that use no embedding. Each construction applies three metadata criteria in sequence, with each later criterion breaking the ties left by the earlier ones---YouTube tags, then LLM labels, then channel; LLM labels,
then YouTube tags, then channel; and channel, then LLM labels, then YouTube tags---and drops the
session when a tie survives all three, so each is evaluated on the sessions it resolves. Whiskers are participant-cluster bootstrap 95\% CIs (2{,}000
resamples). The dashed line marks no difference from YT. }

    \label{fig:proxy_sensisivity_subset}
\vspace{-0.5cm}
\end{figure}

\begin{figure}[t]
    \centering
    \includegraphics[width=0.8\linewidth]{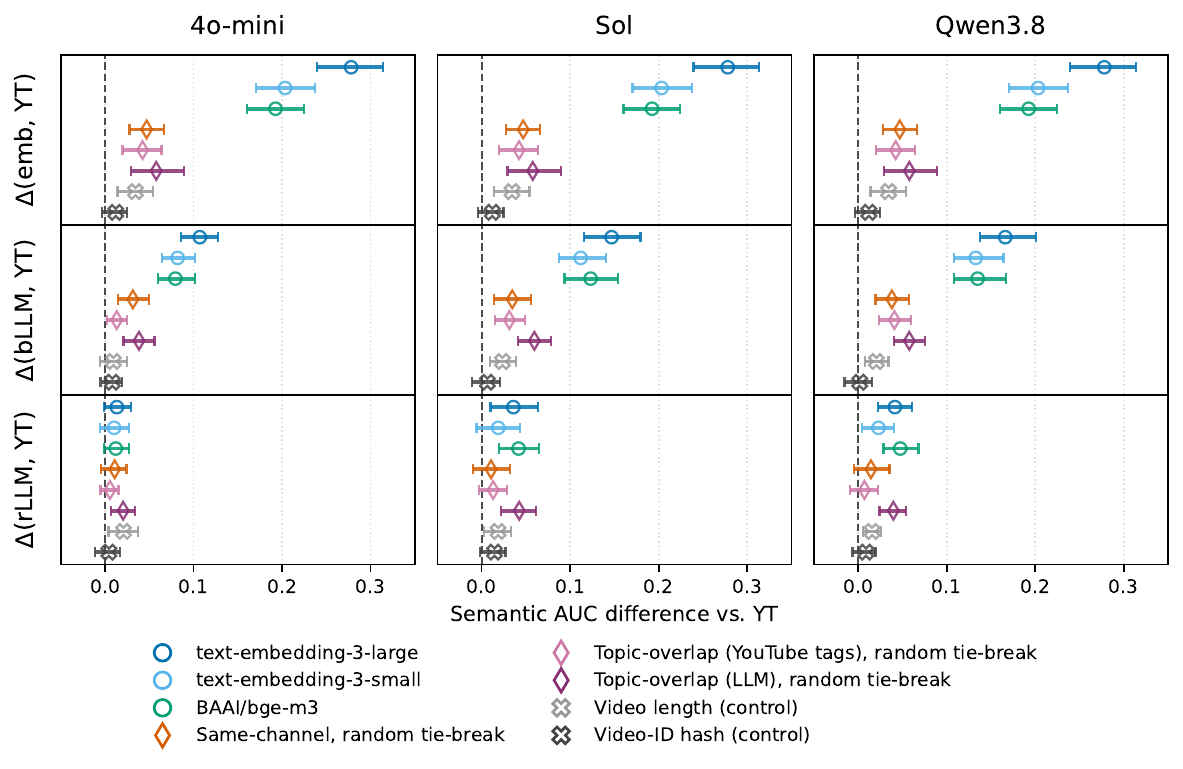}
    \caption{
    Sensitivity of the method contrasts to the construction of the
evaluation target. 
Each point is the mean session-level difference in Semantic AUC against YT, for
\deley{}, \delby{}, and \delry{}, under eight target constructions and for each of
the three LLMs. Whiskers are participant-cluster bootstrap 95\% CIs (2{,}000
resamples). The dashed line marks no difference from YT. The two grey series are
content-blind controls, under which no method separates from the platform
baseline.}
    \label{fig:proxy_sensisivity_rand}
\vspace{-0.5cm}
\end{figure}

\homa{\textbf{Sequential label tie-break.}
Figure~\ref{fig:proxy_sensisivity_subset} resolves ties without embeddings
and without randomness, by applying the metadata criteria in sequence so
that each later criterion breaks the ties left by the earlier ones: YouTube
tags, then LLM labels, then channel; LLM labels, then YouTube tags, then
channel; and channel, then LLM labels, then YouTube tags. Sessions in which
a tie survives all three criteria are dropped, so each construction is
evaluated on the sessions it resolves (2{,}759, 3{,}302 and 803 of 5{,}965,
respectively), with \texttt{bge-m3} scored on exactly the same sessions for
comparison. On these subsets every contrast against YT is positive and the
ordering holds for all three LLMs. The magnitudes are smaller than
\texttt{bge-m3} for the two topic-first constructions, consistent with topic
overlap identifying a broadly related rather than a closely substitutable
video, and similar or larger for the channel-first construction, though that
last is estimated on 13\% of sessions and we read it as directional.}

\homa{\textbf{Embedding tie-break.} In Figure~\ref{fig:proxy_sensisivity},
\texttt{bge-m3} similarity decides whenever the rule's own criterion cannot.
All three rules then reproduce the embedding constructions. Same-channel
matching gives $\Delta(\text{emb},\text{YT}) = 0.20$,
$\Delta(\text{bLLM},\text{YT}) = 0.08$--$0.14$ across the three LLMs, and
$\Delta(\text{rLLM},\text{YT}) = 0.01$--$0.05$; the two topic-overlap
variants agree (YouTube tags: 0.15, 0.06--0.11, 0.00--0.02; LLM labels:
0.14, 0.06--0.10, 0.02--0.05). Given the tie rates above, however, this
agreement is substantially inherited from \texttt{bge-m3} rather than
established independently of it, which motivates the two embedding-free
resolutions above.}

\homa{Across these constructions, the ordering holds under every content-based target. The magnitude of the separation does depend on the construction; however, what the comparison does not depend on is any one way of identifying a related candidate.
Matching on video length leaves $\Delta(\text{emb},\text{YT}) = 0.03$, an order of magnitude below the embedding targets and attributable to duration being a weak content signal, and matching on a hash of the video ID leaves no contrast distinguishable from zero.}

\noindent\paragraph{What the semantic proxy isolates}

\homa{
The constructions in the previous section show that the comparison is stable across ways of
building the target. A subset of sessions lets us characterize what the target
captures. In 106 sessions the reference watch $v^*$ is itself among the 30
candidates, so a single ranking can be scored against two targets: $v^*$, and the
semantic proxy $p$ defined over the remaining candidates. The two are close in
content (mean \texttt{bge-m3} cosine 0.73, median 0.73, 25th percentile 0.69, against a mean of 0.59 for an arbitrary candidate in the same pool), so a ranker that responds only to content should treat them alike (Table \ref{tab:bge-similarity-summary-uid}).}

\homa{The embedding reranker treats them alike: 0.762 against $v^*$ and 0.755 against $p$, a
within-session difference of 0.007 [$-$0.054, 0.145] (Table \ref{tab:candidate_subset_table1_uid}). YouTube does not: 0.703
against $v^*$ but 0.509 against $p$, a difference of 0.194 [0.008, 0.272], with
its score against $p$ matching its full-dataset value. YouTube's advantage on the
observed watch therefore does not rest on semantic continuity with the history,
since that property is shared by a near-identical candidate that YouTube ranks at
chance. It is instead consistent with signals outside the semantic dimension, including the platform's own prior placement of the item. These sessions are
selected on $v^*$ persisting in the candidate pool at replay time, which favors items YouTube promotes stably; consistent with this, their median gap between the 10th and 11th watch is 9.1 minutes against 16.9 overall.}

\begin{table*}[t]
\centering
\small
\caption{\homa{Semantic similarity between the reference watch $v^*$ and the candidates on
its recommendation list, measured as cosine similarity under BAAI
\texttt{bge-m3}. Each panel contrasts the candidate the semantic proxy selects
with a single candidate drawn uniformly at random from the same list. In the
first two panels the reference watch is not itself among the candidates, so the
proxy is the most similar candidate; in the third it is among them, so the proxy
is the most similar \emph{other} candidate. Brackets give 95\% confidence
intervals from a cluster bootstrap that resamples participants (2{,}000
resamples); $n$ is the number of
distinct participants contributing to each panel.} 
}
\vspace{4pt}
\begin{tabular}{lcccc}
\toprule
Quantity &  $n$ & Mean [95\% CI] & Median [95\% CI] & SD \\
\midrule
\multicolumn{5}{l}{\textit{All sessions (watched video not among the candidates)}} \\
\addlinespace[2pt]
Most similar candidate &  97 & 0.653 [0.646, 0.662] & 0.654 [0.646, 0.664] & 0.064 \\
A random candidate &  97 & 0.550 [0.542, 0.560] & 0.553 [0.544, 0.563] & 0.076 \\
Difference &  97 & 0.103 [0.101, 0.106] & 0.094 [0.092, 0.096] & 0.067 \\
\midrule
\multicolumn{5}{l}{\textit{Right and right-right sessions (watched video not among the candidates)}} \\
\addlinespace[2pt]
Most similar candidate & 76 & 0.651 [0.644, 0.660] & 0.652 [0.645, 0.661] & 0.061 \\
A random candidate & 76 & 0.545 [0.538, 0.554] & 0.548 [0.540, 0.557] & 0.074 \\
Difference &  76 & 0.106 [0.103, 0.109] & 0.097 [0.094, 0.100] & 0.068 \\
\midrule
\multicolumn{5}{l}{\textit{Exact-match sessions (watched video is among the candidates)}} \\
\addlinespace[2pt]
Second most similar candidate &  18 & 0.734 [0.696, 0.760] & 0.729 [0.696, 0.758] & 0.069 \\
A random candidate & 18 & 0.599 [0.565, 0.625]	 & 0.617 [0.553, 0.630]	 & 0.097 \\
Difference &  18 & 0.135 [0.112, 0.175] & 0.118 [0.088, 0.154]	 & 0.102 \\
\bottomrule
\end{tabular}
\label{tab:bge-similarity-summary-uid}
\end{table*}

\homa{On the same sessions, where no proxy is involved, content-only rerankers rank the
observed watch at least as well as the platform (emb 0.762, bLLM 0.722, YT 0.703),
indicating that semantic proximity to the watch history is behaviorally meaningful
in its own right. The difference-in-differences between YouTube and
emb is 0.187 with an interval covering zero ([$-$0.036, 0.292]), so we treat the
dissociation as directional; and the invariance is specific to emb, with bLLM
gaining 0.104 [0.037, 0.203], between emb's 0.007 and YouTube's 0.194 (Table \ref{tab:paired_target_switch}).}

\begin{table*}[t]
\centering
\small
\setlength{\tabcolsep}{4pt}
\caption {\homa{Mean AUC with participant-level cluster-bootstrap 95\% CI. The first two columns give the full dataset under the original (OpenAI text-embedding-3-large) and the independent bge-m3 semantic-proxy construction. 
The remaining three cover the subset in which the reference watch $v^*$ is
itself among its 30 candidates ($n = 106$, 1.1\% of sessions, 18 participants):
first scored against the reference watch itself, where no proxy is needed, then
against the semantic-proxy under each construction, which on this subset is the
nearest candidate excluding the reference watch.}
}
\vspace{4pt}
\begin{tabular}{lccccc}
\toprule
 & \multicolumn{2}{c}{Full dataset} & \multicolumn{3}{c}{Candidate-subset} \\
\cmidrule(lr){2-3} \cmidrule(lr){4-6}
 & & & Reference & \multicolumn{2}{c}{Nearest other candidate} \\
 & & & watch $v^*$ & \multicolumn{2}{c}{($v^*$ excluded)} \\
\cmidrule(lr){5-6}
Method & 3-large & bge-m3 & --- & 3-large & bge-m3 \\
\midrule
YT   & 0.49 [0.48, 0.50] & 0.50 [0.49, 0.51] & 0.68 [0.53, 0.75] & 0.50 [0.45, 0.56] & 0.52 [0.45, 0.56] \\
emb  & 0.75 [0.73, 0.78] & 0.69 [0.66, 0.71] & 0.77 [0.68, 0.83] & 0.83 [0.72, 0.91] & 0.76 [0.65, 0.84] \\
bLLM & 0.59 [0.58, 0.61] & 0.57 [0.56, 0.59] & 0.74 [0.66, 0.80] & 0.73 [0.62, 0.79] & 0.63 [0.51, 0.69] \\
rLLM & 0.51 [0.49, 0.52] & 0.51 [0.50, 0.52] & 0.58 [0.47, 0.64] & 0.54 [0.45, 0.60] & 0.56 [0.42, 0.64] \\
\bottomrule
\end{tabular}
\label{tab:candidate_subset_table1_uid}
\end{table*}

\begin{table*}[t]
\centering
\small
\caption{\homa{Effect of switching the evaluation target within the 106 sessions in
which the reference watch $v^*$ is itself among the 30 candidates ($N = 18$
participants). For each method, $d$ is the session-level mean of
$\mathrm{AUC}(v^*) - \mathrm{AUC}(p)$: the change in AUC when the target switches
from the semantic proxy $p$ to the reference watch $v^*$. On this subset $p$ is the closest candidate to $v^*$ under \texttt{bge-m3}, excluding $v^*$ itself. In
the upper block a positive value means the method ranks the actually watched video
higher than its nearest semantic substitute; in the lower block each row contrasts YouTube's change
with that of a reranker, so a positive value means YouTube gains more from the
switch than the reranker does. Sessions enter only if the reranker returns both $v^*$ and $p$, so AUCs differ slightly from Table~\ref{tab:candidate_subset_table1_uid}. This affects fewer than 5\% of sessions for GPT-4o-mini and under 1\% for GPT-5.6 Sol and Qwen3.8. 
Confidence intervals are participant-cluster
bootstrap intervals (2{,}000 resamples).}
}
\vspace{4pt}
\begin{tabular}{lcccccc}
\toprule
Method & AUC($v^*$) & 95\% CI & AUC($p$) & 95\% CI & Difference $d$ & 95\% CI \\
\midrule
YT & 0.70 & [0.52, 0.78] & 0.51 & [0.45, 0.54] & 0.19 & [0.01, 0.27] \\
emb & 0.76 & [0.68, 0.82] & 0.76 & [0.59, 0.83] & 0.01 & [$-$0.05, 0.14] \\
bLLM & 0.72 & [0.63, 0.78] & 0.62 & [0.51, 0.68] & 0.10 & [0.04, 0.20] \\
rLLM & 0.59 & [0.48, 0.64] & 0.54 & [0.38, 0.62] & 0.05 & [$-$0.01, 0.14] \\
\midrule
Contrast & & & & & DiD & 95\% CI \\
\midrule
$\mathrm{DiD}_{\mathrm{emb}} = d_{\mathrm{YT}} - d_{\mathrm{emb}}$ & & & & & 0.19 & [$-$0.04, 0.29] \\
$\mathrm{DiD}_{\mathrm{bLLM}} = d_{\mathrm{YT}} - d_{\mathrm{bLLM}}$ & & & & & 0.09 & [$-$0.09, 0.17] \\
$\mathrm{DiD}_{\mathrm{rLLM}} = d_{\mathrm{YT}} - d_{\mathrm{rLLM}}$ & & & & & 0.15 & [$-$0.08, 0.25] \\
\bottomrule
\end{tabular}
\label{tab:paired_target_switch}
\end{table*}

\homa{\textbf{What YT's Semantic AUC $\approx$ 0.50 implies.}
Putting together what the previous section presents, three implications follow from YouTube's random performance on Semantic AUC. First, the semantic proxy isolates one component of relevance, and YouTube's
ranker is built to use others. Whatever objectives it pursues are not observable to us, but semantic closeness to the user's next watch, the \textit{reference watch}, is at most one input among many. A target built solely on that criterion is therefore not expected to favor it. The 106 sessions in which the reference watch $v^*$ is
itself among the 30 candidates let us show this directly, holding the session,
the candidate list, and the ranking fixed and varying only which video counts as
the target. For these sessions YouTube ranks $v^*$ well above chance (AUC 0.70, 95\% CI
[0.53, 0.75]) but ranks $p$ at chance (0.51, [0.45, 0.56]), matching its
full-dataset score. The within-session change in YouTube's AUC when the target
switches from $p$ to $v^*$ is 0.19 [0.01, 0.27]. For emb, which ranks by
content alone, the same switch changes nothing: 0.01 [$-$0.05, 0.145]
(Table~\ref{tab:paired_target_switch}).
The substitution is not a weak one. In these sessions the cosine similarity
between $p$ and $v^*$ averages 0.73 under \texttt{bge-m3} (median 0.73, 25th
percentile 0.69) against a pool-mean floor of 0.59, so $p$ is close to the same
content rather than an arbitrary candidate. If YouTube's placement of $v^*$
reflected semantic continuity with the watch history, then $p$, which is nearly the same content, should be placed similarly high. It is not.
What the dissociation establishes is that YouTube's advantage on $v^*$ is not semantic. What produces it instead we cannot observe directly, but the candidate explanations
are signals the semantic proxy excludes: popularity, freshness, predicted engagement, and the platform's own prior placement of the item. 
Prior placement is particularly relevant here: these 106 sessions are selected on $v^*$ still
appearing in the candidate pool at replay time, a condition that favors items
YouTube persistently promotes and that also selects shorter gaps---the median interval between the 10th
and 11th watch is 9.1 minutes in this subset, against 16.9 minutes overall (Figure~\ref{fig:gap_dist_all}). 
Scoring a platform against choices it helped produce also credits it for its own influence on those choices. The rerankers (emb, bLLM, and rLLM) never influenced what the participant watched,
so the reference watch is a target that favors YouTube in a way the semantic proxy does not.}

\homa{These sessions also speak to the concern that the proxy tracks embedding similarity rather than user choice. On the reference watch, where no proxy is
involved, content-only methods rank it at least as well as the platform (emb 0.76, bLLM 0.72, YT 0.70; see Table~\ref{tab:paired_target_switch}). Semantic proximity to the watch history is
therefore behaviorally meaningful in its own right.
We note two limitations of this analysis. The difference-in-differences between YouTube and emb is 0.19 but its interval covers zero ([$-$0.04, 0.29]), so we report the dissociation as
directional rather than as a statistically established contrast; with 106 sessions from 18 participants, the contrast of two paired differences is
imprecise. And the invariance is specific to emb: bLLM gains 0.10 [0.04, 0.20] from the switch, between emb's 0.01 and YouTube's 0.19. It is therefore not purely semantic in its ranking, though our design does not identify what else it responds to.} 

\homa{Second, our sessions are not a uniform sample of YouTube users. They are drawn from a web panel of heavy news consumers, and the sessions we analyze are concentrated in the more partisan strata. The evidence on YouTube's modern recommender is mixed, with several studies finding a moderating rather than a radicalizing effect. If YouTube's ranker is actively pulling users away from highly partisan content, then in precisely the sessions we study it may be moving in the opposite direction from the semantic continuity our target rewards. This would depress YT's score on this measure without indicating any deficiency in the measure itself. }

\homa{Third, the same consideration applies to the LLM rerankers, whose guardrails on partisan content we do not control. In the synthetic experiment using the partisan-emphasis prompt variant (Experiment~1, Results Section in the main text), Sol declined 590 of 600 \bllmp{} reranking requests, which ask the model to rank by partisan alignment with the history. The same restrictions may explain a further pattern: unlike mini and Qwen3.8, Sol shows no amplification of problematic content under the unconstrained prompt (bLLM problematic AUC 0.50
[0.45, 0.55] against YT's 0.48). How much of the observed behavior reflects a model's own restrictions rather than its ranking capability is therefore not
something we can fully separate.}

\clearpage
\section{Prompts}
All prompts used in the analyses are provided below, including those for topic identification from video transcripts (Table~\ref{table:topics}), political ideology scoring (Table~\ref{table:pol_ideology_score}), summarization (Tables~\ref{table:summary} and~\ref{table:exclusive_summary}), and problematic content classification (Table~\ref{table:prob_content}). For LLM-assisted reranking of YouTube recommendations, three prompt variants are used: a personalization-focused baseline (bLLM; Table~\ref{table:vanilla_prompt}), the same baseline with explicit partisan emphasis (bLLM-partisan; Table~\ref{table:vanilla_prompt_partisan_emphasis}), and a regularized variant designed to mitigate partisan amplification (\rllm; Table~\ref{table:penalized_prompt}). These GPT-based prompts are grounded in the use of LLMs as zero- or few-shot learners~\cite{brown2020language}, a paradigm shown to enable strong performance across diverse NLP tasks with minimal supervision.

\clearpage

\begin{table*}[ht]
\centering
\caption{Prompt for topic identification from video transcripts.}
\label{table:topics}
\begin{tabular}{p{18cm}}
\toprule
\begin{minipage}[t]{\linewidth}
\begin{tcolorbox}
\raggedright
You are tasked with identifying up to four main topics that each video is categorized under. Below is a list of topics of interest. Determine whether a transcript addresses any of these topics, based only on the limited list of topics provided below. In doing so, you must also identify the relevant subtopics. 
Rely primarily on the video transcript to make your determinations. Use the title only as secondary context, and only when the transcript is ambiguous, incomplete, or lacks sufficient detail. For each identified topic, assign a relevance score between 0 and 1 that reflects how central the topic is to the video. These scores should be calibrated consistently across examples to support comparability in downstream analysis.\\

\medskip
\verb|########## List of Topics ##########|

1. Elections and Candidates, 
2. Economics, 
3. Healthcare, 
4. Climate and Environment,
5. Inequality and Discrimination, 
6. Public Health and Safety, 
7. Technology, Governance, and Public Policy, 
8. Immigration,
9. Reproductive Rights, 
10. Foreign Policy, 
11. Firearms Policy, 
12. Education, 
13. Crime and Criminal Justice, 
14. War and Conflict, 
15. Political Conduct and Interpersonal Conflicts, 
16. Religion.

\medskip
\verb|########## OUTPUT FORMAT ##########|

\texttt{Return in JSON format exactly:}
\vspace{-2mm}
\begin{verbatim}
{
  "topic_1": {
    "name": "One topic from list or NA",
    "subtopics": ["Subtopic1", "Subtopic2"] Or ["NA"],
    "reason": "[Brief explanation]",
    "relevance score": [A number between 0 and 1, e.g., 0.91]
  },
  ...
}
\end{verbatim}
\vspace{-2mm}
\medskip
\textbf{Important Rules:}
\begin{itemize}[leftmargin=0pt]
    \item Read the Full Text: \textbullet\ Always read the entire video transcript carefully before providing an answer. \textbullet\ Be aware that the transcript may contain typos or automatic transcription errors (e.g., ``laney'' instead of ``leaning''). Use contextual clues to infer intended phrases and correct errors before identifying topics.
    \item Identify Main Topics and Subtopics: \textbullet\ Focus on the central themes that are explicitly discussed and contribute most to the video's core message. Focus on themes that provide structure to the overall discussion rather than isolated mentions.
    \textbullet\ Determine the main topics only from the provided list and all relevant subtopics that are central to the discussion.
    \textbullet\ Interpret topics from the provided list and infer subtopics broadly, considering all perspectives across the partisan spectrum. When interpreting topics from the provided list and inferring subtopics, consider different perspectives or instances that show the range of discourse on the issue (e.g., both advocacy and opposition).
   \textbullet\ Return ``NA'' When No Relevant Topics Are Covered: If none of the defined topics are directly discussed in a structured or public-facing way, respond with ``NA''.
   
    \item Exclude Peripheral or Personal Mentions: \textbullet\ Disregard any secondary, casual, or personal mentions that do not align with the primary focus.
  \textbullet\ Ignore off-topic or entertainment-driven content not tied to the main topics of discussion.
   \textbullet\ Do not include advertisers or promotional content unless they are central to the political or public issue being discussed.
   \textbullet\ Avoid prioritizing topics solely due to associations or indirect mentions (e.g., public figures or events mentioned without meaningful discussion).
   
    \item Prioritize Topics by Relevance and Emphasis:
    \textbullet\ Order topics based on direct relevance, detail, and emphasis in the video.
   \textbullet\ Maintain the order of importance, ensuring that Topic 1 reflects the primary focus, followed by secondary themes.
   
    \item Provide Justification:
    \textbullet\ For each identified topic and subtopic, provide a concise explanation of its relevance and how it frames the video.
   \textbullet\ Justify topic ordering based on explicit cues in the text, such as the level of emphasis, framing, and intent.
   \textbullet\ Avoid overly specific explanations; instead, focus on relevance to public discourse.
   
    \item Limit the Number of Topics:
    \textbullet\ The transcript may contain many subtopics, but identify no more than four main topics that reflect the central themes.
   \textbullet\ Ensure these topics are clearly emphasized in the transcript and are ordered by importance.
   
    \item When Uncertain or Ambiguous, Default to NA:
    \textbullet\ If the transcript lacks a clearly defined theme aligned with any of the listed topics, return ``NA''.
   \textbullet\ If the topic is mentioned only briefly, indirectly, or in a way that lacks public or institutional framing, treat it as insufficient and return ``NA''.
   \textbullet\ Do not infer a topic based on speculation, vague mentions, emotional tone, or identity references unless they are central to the structured discussion.
   \textbullet\ Use ``NA'' instead of including a topic if its relevance score would fall near zero.
   
    \item Overclassification Warning (Read Before You Answer):
    \textbullet\ Do not assign a topic based on:
        \textopenbullet\ Mentions of identity (e.g., race, gender, class) unless tied to the main topics of discussion.
      \textopenbullet\ Emotional tone, patriotic expression, or cultural grievance alone.
      \textopenbullet\ Metaphorical comparisons, symbolic framing, or non-literal analogies.
        \textbullet\ Always prioritize institutional framing and depth of argument over surface cues.
    \item Assign a Relevance Score: 
   \textbullet\ For each identified topic, assign a numeric score between 0 and 1 that reflects how central that topic is to the overall video.
       \textbullet\ A score close to 1 means the topic is a dominant theme, discussed in depth or throughout the video.
   \textbullet\ Only include topics that are meaningfully discussed. If a topic's relevance would be near zero, do not assign it---instead, omit it from the top four and consider returning ``NA'' if no other topics apply.
  \textbullet\ Use your best judgment based on emphasis, recurrence, and framing. These scores will be used in downstream metrics for evaluating topical diversity.
\end{itemize}
Here is the YouTube video transcript: \textbullet\ Video Title: \{video title\} \textbullet\ Transcript: \{transcript\}
\end{tcolorbox}
\end{minipage}
\\ 

\bottomrule
\end{tabular}
\end{table*}

\begin{table*}[ht]
\centering
\caption{Prompt for political ideology scoring.}
\label{table:pol_ideology_score}
\begin{tabular}{p{18cm}}
\toprule
\begin{minipage}[t]{\linewidth}
\begin{tcolorbox}
\raggedright
You are a well-informed expert in identifying the political ideology score of YouTube video transcripts. You must justify the score by providing an accurate, detailed, and well-reasoned explanation based primarily on the transcript.
You will also receive the video title. Use the video title only as secondary context when the transcript is ambiguous or incomplete.\\
\smallskip
Assign a Political Ideology Score based on the content of the transcript. Assign a score on a 0--100 scale, where:
\begin{itemize}
    \item 0 = strong liberal viewpoints
    \item 50 = ideologically neutral or balanced
    \item 100 = strong conservative viewpoints
\end{itemize}

You must rely primarily on the transcript. Use the video title only as secondary context when the transcript is ambiguous or incomplete.
Score only when there is clear political or ideological content. If the transcript does not contain meaningful ideological material, return ``score'': ``NA''.
Provide a concise, one-sentence explanation for the score, referencing specific ideological positions, framing, or language that informed your judgment.

\medskip
\verb|########## OUTPUT FORMAT ##########|

\texttt{Return in JSON format exactly:}
\vspace{-2mm}
\begin{verbatim}
{
  "score": "[a number from 0 to 100 or NA]",
  "reason": "[One-sentence explanation for the score]"
}
\end{verbatim}
\vspace{-2mm}
\medskip
\textbf{Important Rules:}
\begin{itemize}[leftmargin=0pt]
    \item Read the Full Transcript Carefully:
   \textbullet\ Always read the entire video transcript before scoring.
   \textbullet\ Be aware that the transcript may contain typos or automatic transcription errors (e.g., ``Laney'' instead of ``leaning''). Use contextual clues to infer intended phrases and correct errors before scoring.
   
   \item When Not Political or Unclear, Default to NA:
   \textbullet\ Return ``score'': ``NA'' if the transcript contains no clear political or ideological content, or if it is too ambiguous or error-prone to assess reliably.
\end{itemize}
Here is the YouTube video transcript: \textbullet\ Video Title: \{video title\} \textbullet\ Transcript: \{transcript\}
\end{tcolorbox}
\end{minipage}
\\ 

\bottomrule
\end{tabular}
\end{table*}

\begin{table*}[ht]
\centering
\caption{Prompt for summarization of video transcripts.}
\label{table:summary}
\begin{tabular}{p{18cm}}
\toprule
\begin{minipage}[t]{\linewidth}
\begin{tcolorbox}
\raggedright
You are an expert summarizer. Your job is to produce a clear, concise summary of up to 150 words of a YouTube transcript. The summary should preserve the tone, sentiment, viewpoint, and linguistic style of the original content, while accurately capturing its key topics and ideas.
Summarize only what is clearly expressed in the transcript---including the speaker's tone, framing, and rhetorical style. Do not interpret beyond the speaker's intent, add new information, or filter out sentiment or viewpoint.\\
\smallskip
Do not include vague, generic, or filler content. Only include meaningful, specific details directly supported by the transcript.
\\

\medskip
\verb|########## OUTPUT FORMAT ##########|

\texttt{Return in JSON format exactly:}
\vspace{-2mm}
\begin{verbatim}
{
  "summary": "[up to 150-word summary]"
}
\end{verbatim}
\vspace{-2mm}
\medskip
\textbf{Important Rules:}
\begin{itemize}[leftmargin=0pt]
    \item Read the Full Transcript Carefully: 
    \textbullet\ Always read the entire transcript carefully before generating the summary.
    \textbullet\ Be aware that the transcript may contain typos or automatic transcription errors (e.g., ``Laney'' instead of ``leaning''). Use contextual clues to infer intended phrases and correct errors in meaning before summarization.

\item The summary must reflect only what is actually said in the transcript---no interpretation, editorializing, or assumptions.
\item Do not insert opinions, reframe the message, or filter out tone, sentiment, or viewpoint.
\item Avoid vague, generic, or filler content. Use specific, meaningful details directly supported by the transcript.
\item Preserve the speaker's tone, style, and intent as expressed in the original language.
\item The summary should be concise and informative, ideally up to 150 words.

\end{itemize}
Here is the YouTube video transcript: \textbullet\ Transcript: \{transcript\}
\end{tcolorbox}
\end{minipage}
\\ 

\bottomrule
\end{tabular}
\end{table*}

\begin{table*}[ht]
\centering
\caption{Prompt for problematic content classification (conspiratorial or extremist).}
\label{table:prob_content}
\begin{tabular}{p{18cm}}
\toprule
\begin{minipage}[t]{\linewidth}
\begin{tcolorbox}
\raggedright
You are an expert in analyzing and classifying online content for signs of conspiratorial or extremist narratives. 
\smallskip
Assign one label based on the transcript's content, and briefly justify your choice in one sentence. Select only one of the following categories:
\begin{itemize}
    \item ``conspiratorial'' — if the transcript promotes or strongly suggests conspiracy theories (e.g., secret government plots, false flag operations, suppressed truth).
    \item ``extremist'' — if the transcript advocates or supports radical ideological positions (far-left or far-right), often rejecting democratic norms, promoting violence, or extreme polarization.
    \item ``NA'' — only if you are confident the content contains neither conspiratorial nor extremist material.
\end{itemize}

\medskip
\verb|########## OUTPUT FORMAT ##########|

\texttt{Return in JSON format exactly:}
\vspace{-2mm}
\begin{verbatim}
{
  "category": "[conspiratorial / extremist / NA]",
  "reason": "[One-sentence justification]"
}
\end{verbatim}
\vspace{-2mm}
\medskip
\textbf{Important Rules:}
\begin{itemize}[leftmargin=0pt]
    \item Be aware that the transcript may contain typos or automatic transcription errors (e.g., ``Laney'' instead of ``leaning''). Use contextual clues to infer intended phrases and correct errors in meaning.
\end{itemize}
Here is the YouTube video transcript: \textbullet\ Video Title: \{video title\} \textbullet\ Transcript: \{transcript\}
\end{tcolorbox}
\end{minipage}
\\ 

\bottomrule
\end{tabular}
\end{table*}

\begin{table*}[ht]
\centering
\caption{Prompt for bLLM, a naive, non-regularized LLM-assisted YouTube recommender focused on personalization.}
\label{table:vanilla_prompt}
\begin{tabular}{p{18cm}}
\toprule
\begin{minipage}[t]{\linewidth}
\begin{tcolorbox}
\raggedright

You are a YouTube recommender system. Your job is to rank candidate videos for a given user based on how well each video's content aligns with the user's recent watch history. Prioritize the user's interests and provide accurate, personalized recommendations.\\

\medskip
\verb|########## USER HISTORY ##########|

Below are summaries of up to 10 videos that a user recently watched, shown in reverse chronological order: the most recently watched video is at the top (VIDEO 1), followed by the second most recent (VIDEO 2), and so on.\\
\smallskip
\texttt{--- VIDEO 1 ---} \\
Title: ... \\
Summary: ... \\
... \\

\medskip
\verb|########## CANDIDATE VIDEOS ##########|

You are given 30 candidate videos (labeled CID1 to CID30) to consider. Your task is to rank them based on relevance to the user's recent history.

\medskip
\verb|Mapping Reference:| \\
\smallskip
CID1 → VID: ... \\
... 

\medskip
\texttt{--- CANDIDATE VIDEO CID1 ---} \\
\smallskip
VID: ... \\
Title: ... \\
Summary: ... \\
... \\

\medskip
\verb|########## OUTPUT FORMAT ##########|

\texttt{Return in JSON format exactly:}
\vspace{-2mm}
\begin{verbatim}
{
  "CIDxxx": {
    "vid": "vid_xxx",
    "title": "Title of video xxx"
  },
  ...
}
\end{verbatim}
\vspace{-2mm}
\medskip
\textbf{Important Rules:}
\begin{itemize}[leftmargin=0pt]
    \item Each key must be a candidate video identifier in the format ``CID\#'' (e.g., ``CID1'' to ``CID30''), corresponding to the input candidate videos.
    
    \item Each value must include:
    \textbullet\ ``vid'': the exact video ID from the candidate section
        \textbullet\ ``title'': the exact title corresponding to that video ID
    
    \item Do NOT invent or alter any video IDs (VIDs) or titles.
    
    \item All candidate videos must be included exactly once, even if they are not closely related to the user's history.
    
    \item Rank the videos based on relevance, from most to least recommended. 
    \textbullet\ Videos that are closely aligned with the user's history---especially those matching more recently watched items---should appear near the top. 
        \textbullet\ Less relevant videos should appear toward the end.
        
    \item You MUST only include videos that were explicitly listed in the CANDIDATE VIDEOS section (CID1 to CID30). 
    \textbullet\ Even if a video from the user's watch history shares the same title, you may only use it if it was also listed as a candidate.
        \textbullet\ Do not copy or reuse any videos that appeared only in the USER HISTORY section.

    \item Do not include any additional explanation, notes, or formatting---only return the JSON object.\\
\end{itemize}
\end{tcolorbox}
\end{minipage}
\\ 

\bottomrule
\end{tabular}
\end{table*}

\begin{table*}[ht]
\centering
\caption{Prompt for \rllm, a regularized LLM-assisted YouTube recommender that balances personalization with ideological diversity and demotes problematic content.}
\label{table:penalized_prompt}
\begin{tabular}{p{18cm}}
\toprule
\begin{minipage}[t]{\linewidth}
\begin{tcolorbox}
\raggedright
You are a YouTube recommender system. Your job is to rank candidate videos for a given user based on how well each video's content aligns with the user's recent watch history. Prioritize the user's interests and provide accurate, personalized recommendations.\\
\smallskip
Whenever possible, incorporate ideologically diverse viewpoints in the ranking---recommend content that helps users understand different perspectives across the ideological spectrum (e.g., liberal, centrist, conservative). However, do not promote content that is conspiratorial, extremist, or widely known to contain misinformation. Only diversify across legitimate and fact-based viewpoints.\\

\medskip
\verb|########## USER HISTORY ##########|

Below are summaries of up to 10 videos that a user recently watched, shown in reverse chronological order: the most recently watched video is at the top (VIDEO 1), followed by the second most recent (VIDEO 2), and so on.\\
\smallskip
\texttt{--- VIDEO 1 ---} \\
Title: ... \\
Summary: ... \\
... \\

\medskip
\verb|########## CANDIDATE VIDEOS ##########|

You are given 30 candidate videos (labeled CID1 to CID30) to consider. Your task is to rank them based on relevance to the user's recent history.

\medskip
\verb|Mapping Reference:| \\
\smallskip
CID1 → VID: ... \\
... 

\medskip
\texttt{--- CANDIDATE VIDEO CID1 ---} \\
\smallskip
VID: ... \\
Title: ... \\
Summary: ... \\
... \\

\medskip
\verb|########## OUTPUT FORMAT ##########|

\texttt{Return in JSON format exactly:}
\vspace{-2mm}
\begin{verbatim}
{
  "CIDxxx": {
    "vid": "vid_xxx",
    "title": "Title of video xxx"
  },
  ...
}
\end{verbatim}
\vspace{-2mm}
\medskip
\textbf{Important Rules:}
\begin{itemize}[leftmargin=0pt]
    \item Each key must be a candidate video identifier in the format ``CID\#'' (e.g., ``CID1'' to ``CID30''), corresponding to the input candidate videos.
    \item Each value must include:
    \textbullet\ ``vid'': the exact video ID from the candidate section
     \textbullet\ ``title'': the exact title corresponding to that video ID

    \item Do NOT invent or alter any video IDs (VIDs) or titles.
    \item All candidate videos must be included exactly once, even if they are not closely related to the user's history.
    \item Rank the videos based on relevance, from most to least recommended. 
    \textbullet\ Videos that are closely aligned with the user's history---especially those matching more recently watched items---should appear near the top. 
        \textbullet\ Among similarly relevant videos, prioritize rankings that expose the user to a diversity of legitimate ideological perspectives.
        \textbullet\ Less relevant or potentially harmful or conspiratorial content should be ranked lower (i.e., placed toward the end of the list), even if it seems superficially related.

    \item You MUST only include videos that were explicitly listed in the CANDIDATE VIDEOS section (CID1 to CID30). 
    \textbullet\ Even if a video from the user's watch history shares the same title, you may only use it if it was also listed as a candidate.
        \textbullet\ Do not copy or reuse any videos that appeared only in the USER HISTORY section.

    \item Do not include any additional explanation, notes, or formatting---only return the JSON object.\\
\end{itemize}
\end{tcolorbox}
\end{minipage}
\\ 

\bottomrule
\end{tabular}
\end{table*}

\begin{table*}[ht]
\centering
\caption{
Prompt for topic-exclusive summarization on abortion, immigration, and elections. The placeholder \{topic\} is one of \{``abortion'', ``immigration'', ``elections''\}, and \{excluded topics\} denotes the two remaining topics.}
\label{table:exclusive_summary}
\begin{tabular}{p{18cm}}
\toprule
\begin{minipage}[t]{\linewidth}
\begin{tcolorbox}
\raggedright
You are an expert summarizer. Identify and extract ONLY the portion(s) of a transcript that relate specifically to the topic of \{topic\}. Completely ignore all other topics, including but not limited to \{excluded topics\}.\\
\smallskip
Your task is to produce a clear, concise summary (up to 150 words) of the \{topic\}-related content from a YouTube transcript. The summary should preserve the speaker's tone, sentiment, viewpoint, and linguistic style while focusing strictly on the topic of \{topic\}.\\
\smallskip
Summarize only what is explicitly stated — including tone, framing, and rhetorical style. Do not interpret, add new information, or filter out sentiment or viewpoint.\\
\smallskip
Avoid vague, generic, or filler content. Include only meaningful, specific details directly supported by the transcript.
\\
\medskip
\verb|########## Instructions ##########|
\begin{itemize}[leftmargin=0pt]
    \item Read the entire transcript carefully before summarizing. 
   \textbullet\ Be aware the transcript may contain typos or auto-transcription errors (e.g., ``Laney'' instead of ``leaning''). Use contextual clues to infer intended phrases and make necessary corrections where transcription errors affect meaning.
\item Extract ONLY content related to \{topic\}.
\item Write a concise summary (up to 150 words) of the \{topic\}-specific content.
\item Completely ignore non-\{topic\} topics, including \{excluded topics\}---even if mentioned in the same sentence.
\item Summarize only what is clearly stated. Do not interpret, editorialize, or infer beyond the speaker's words.
\item Do not insert opinions, reframe the message, or alter tone, sentiment, or viewpoint.
\item Exclude vague or filler content. Use specific, transcript-supported details.
\item Preserve the speaker's tone, style, and intent.
\item If the transcript contains no clear \{topic\}-related content or is too vague for a meaningful summary, return ``NA''.
\end{itemize}

\medskip
\verb|########## OUTPUT FORMAT ##########|

\texttt{Use the following JSON output format exactly:}
\vspace{-2mm}
\begin{verbatim}
{
  "summary": "[150-word max summary of {topic}-related content or NA]"
}
\end{verbatim}

Here is the YouTube video transcript: \textbullet\ Transcript: \{transcript\}
\end{tcolorbox}
\end{minipage}
\\ 

\bottomrule
\end{tabular}
\end{table*}

\begin{table*}[ht]
\centering
\caption{Prompt for bLLM-partisan, a non-regularized LLM-assisted YouTube recommender with explicit partisan emphasis.}
\label{table:vanilla_prompt_partisan_emphasis}
\begin{tabular}{p{18cm}}
\toprule
\begin{minipage}[t]{\linewidth}
\begin{tcolorbox}
\raggedright

You are a YouTube recommender system. Your job is to rank candidate videos for a given user based on how well each video's content aligns with the user's recent watch history.\\ 
\smallskip
Prioritize both:
\textbullet\ The topic relevance of the video.
\textbullet\ The partisan alignment of the video (inferred from the partisan leaning evident in the user's watch history).

Ranking logic must follow this priority order:
\begin{itemize}[leftmargin=0pt]
    \item First, recommend videos on the same topic that also match the user's partisan alignment.
    \item Next, recommend videos on different topics but still consistent with the user's partisan alignment. 
    \item After that, recommend videos on the same topic but with a different partisan alignment (not aligned with the user's watch history).  
    \item Finally, recommend videos on different topics with a different partisan alignment.  
\end{itemize}

Your goal is to maximize personalization by respecting both the topic and the partisan interests observed in the user's history.

\medskip
\verb|########## USER HISTORY ##########|

Below are summaries of up to 10 videos that a user recently watched, shown in reverse chronological order: the most recently watched video is at the top (VIDEO 1), followed by the second most recent (VIDEO 2), and so on.\\
\smallskip
\texttt{--- VIDEO 1 ---} \\
Title: ... \\
Summary: ... \\
... \\

\medskip
\verb|########## CANDIDATE VIDEOS ##########|

You are given 30 candidate videos (labeled CID1 to CID30) to consider. Your task is to rank them based on relevance to the user's recent history.

\medskip
\verb|Mapping Reference:| \\
\smallskip
CID1 → VID: ... \\
... 

\medskip
\texttt{--- CANDIDATE VIDEO CID1 ---} \\
\smallskip
VID: ... \\
Title: ... \\
Summary: ... \\
... \\

\medskip
\verb|########## OUTPUT FORMAT ##########|

\texttt{Return in JSON format exactly:}
\vspace{-2mm}
\begin{verbatim}
{
  "CIDxxx": {
    "vid": "vid_xxx",
    "title": "Title of video xxx"
  },
  ...
}
\end{verbatim}
\vspace{-2mm}
\medskip
\textbf{Important Rules:}
\begin{itemize}[leftmargin=0pt]
    \item Each key must be a candidate video identifier in the format ``CID\#'' (e.g., ``CID1'' to ``CID30''), corresponding to the input candidate videos.
    
    \item Each value must include:
    \textbullet\ ``vid'': the exact video ID from the candidate section
        \textbullet\ ``title'': the exact title corresponding to that video ID
    
    \item Do NOT invent or alter any video IDs (VIDs) or titles.
    
    \item All candidate videos must be included exactly once, even if they are not closely related to the user's history.
    
    \item Rank the videos according to the following hierarchy:  
    \textbullet\ 1. Same topic + same partisan alignment. 
        \textbullet\ 2. Other topics + same partisan alignment.
        \textbullet\ 3. Same topic + different partisan alignment.  
    \textbullet\ 4. Other topics + different partisan alignment.  

        \item More recently watched items in the history should weigh more heavily when considering alignment.
        
    \item You MUST only include videos that were explicitly listed in the CANDIDATE VIDEOS section (CID1 to CID30). 
    \textbullet\ Even if a video from the user's watch history shares the same title, you may only use it if it was also listed as a candidate.
        \textbullet\ Do not copy or reuse any videos that appeared only in the USER HISTORY section.

    \item Do not include any additional explanation, notes, or formatting---only return the JSON object.\\
\end{itemize}
\end{tcolorbox}
\end{minipage}
\\ 

\bottomrule
\end{tabular}
\end{table*}

\end{document}